\documentclass[preprint]{elsarticle}

\usepackage{multirow}

\usepackage{color}

\usepackage{booktabs}

\usepackage{algorithm}
\usepackage{algorithmicx}
\usepackage{amsmath,amsfonts}
\usepackage{dsfont}
\usepackage{algpseudocode}
\usepackage{xcolor}
\usepackage{lipsum}
\usepackage{chngcntr}
\usepackage{tocloft}
\usepackage{pifont}
\usepackage{amssymb}
\usepackage[margin=2.90cm]{geometry}
\usepackage{rotating}
\usepackage{tikz}
\usepackage{soul}
\usepackage{graphicx}

\usepackage{csquotes}
\MakeOuterQuote{"}
\usepackage{hyperref}

\usepackage[justification=centering]{caption}
\usepackage{blindtext}

\usepackage{balance}

\usepackage{color}

\makeatletter
\def\ps@pprintTitle{%
   \let\@oddhead\@empty
   \let\@evenhead\@empty
   \def\@oddfoot{\reset@font\hfil\thepage\hfil}
   \let\@evenfoot\@oddfoot
}
\makeatother

\usepackage{adjustbox}
\DeclareMathOperator*{\argmin}{argmin}
\usepackage{subfig}

\usepackage{makecell}

\usepackage{float}

\makeatletter
\def\blfootnote{\xdef\@thefnmark{}\@footnotetext}
\makeatother

\journal{}

\begin{document}

\begin{frontmatter}

\title{Feature Engineering for Mid-Price Prediction with Deep Learning}

\author[first]{Adamantios Ntakaris\corref{mycorrespondingauthor}}
\cortext[mycorrespondingauthor]{Corresponding authors}
\ead{\string\href{mailto:adamantios.ntakaris@tut.fi}{adamantios.ntakaris@tuni.fi}}

\author[second]{Giorgio Mirone\corref{mycorrespondingauthor}}
\ead{\string\href{mailto:gmirone@econ.au.dk}{gmirone@econ.au.dk}}

\author[first]{Juho Kanniainen}

\author[first]{Moncef Gabbouj}

\author[third]{Alexandros Iosifidis}

\address[first]{Faculty of Information Technology and Communication Sciences, Tampere University, Korkeakoulunkatu 1, FI-33720, Tampere, Finland}
\address[second]{Danmarks Nationalbank, Havnegade 5 1093 K\o benhavn K, Copenhagen, Denmark}
\address[third]{Department of Engineering, Electrical and Computer Engineering, Aarhus University, Finlandsgade 22, Hang{\o}vej 2, Denmark}

\begin{abstract}
Mid-price movement prediction based on limit order\blfootnote{\textbf{Disclaimer:} The views and conclusions expressed in this paper are solely those of the authors and do not necessarily reflect the views of Danmarks Nationalbank.} book data is a challenging task due to the complexity and dynamics of the limit order book. So far, there have been very limited attempts for extracting relevant features based on limit order book data. In this paper, we address this problem by designing a new set of handcrafted features and performing an extensive experimental evaluation on both liquid and illiquid stocks. More specifically, we present an extensive set of econometric features that capture statistical properties of the underlying securities for the task of mid-price prediction. The experimental evaluation consists of a head-to-head comparison with other handcrafted features from the literature and with features extracted from a long short-term memory autoencoder by means of a fully automated process. Moreover, we develop a new experimental protocol for online learning that treats the task above as a multi-objective optimization problem and predicts i) the direction of the next price movement and ii) the number of order book events that occur until the change takes place. In order to predict the mid-price movement, features are fed into nine different deep learning models based on multi-layer perceptrons, convolutional neural networks, and long short-term memory neural networks. The performance of the proposed method is then evaluated on liquid and illiquid stocks (i.e., TotalView-ITCH US and Nordic stocks). For some stocks, results suggest that the correct choice of a feature set and a model can lead to the successful prediction of how long it takes to have a stock price movement.
\end{abstract}



\end{frontmatter}


\section{Introduction}
\label{sec:introduction}
\noindent The automation of financial markets has increased the complexity of information analysis. This complexity can be effectively managed by the use of ordered trading universes like the limit order book (LOB). LOB is a formation that translates the daily unexecuted trading activity in price levels according to the type of orders (i.e., bid and ask side). The daily trading activity is a big data problem, since millions of trading events take place inside a trading session. Information extraction and digital signal (i.e., time series) analysis from every trading session provide the machine learning (ML) trader with useful instructions for orders, executions, and cancellations of trades. \\
\indent Traditional time series analysis methods have failed to capture the complexity of the contemporary trading markets adequately. For instance, the work in \cite{qian2017financial} and\cite{siami2018forecasting} suggest that classical machine learning and deep learning methods for financial metric predictions achieve better results compared to ARIMA and GARCH models. On the contrary, machine and deep learning methods have proved to be very effective mechanisms for time series analysis and prediction (e.g., \cite{chen2018artificial}, \cite{nousi2018machine}, \cite{sirignano2018universal}). The main advantage of these methods is their ability to capture non-linearities of the input data and filter them consecutively by creating new weighted features more relevant to the suggested problem.\\
\indent Despite their efficacy to predict time series, machine and deep learning methods are developed mainly through empirical testing. The majority of the literature (e.g.,\cite{velay2018stock}, \cite{dash2016hybrid}, \cite{gudelek2017deep}) that focuses on deep learning frameworks solely relies either on raw data or a limited number of features. So far, very little attention has been paid to the information a neural network should analyze for the mid-price prediction task. In this paper, we shed light on the information that the ML trader should consider utilizing in mid-price movement prediction. To this end, we employ an extensive list of econometric features\footnote{Econometrics features were used in the past for tasks such as identification of big changes in exchange rate volatility (i.e., \cite{wang2011pricing}), or bankruptcy prediction in \cite{zikeba2016ensemble}.} for mid-price prediction and make a head-to-head comparison with indicators derived from: i) technical and quantitative analysis (i.e., \cite{ntakaris2018mid}), ii) time-sensitive and time-insensitive features (i.e., \cite{doi:10.1080/14697688.2015.1032546} and  \cite{ntakaris2018benchmark}), and iii) features extracted through a fully automated process. This fully automated feature extraction process is conducted by a long short-term memory (LSTM) autoencoder (AE). \\
\indent We choose econometrics as motivation for our handcrafted features since it is the field of financial engineering that captures the empirical evidence of microstructure noise and causality of the data. Our data comes with variations in prices, known in the financial literature as volatility -- a measure that we incorporate into our handcrafted features.
Despite the general perception in academic literature that volatility itself is not a factor that affects stock returns, ample evidence exists to support the opposite. For instance, in \cite{guo2004limited} the author finds that volatility together with other proxies that are not directly observable in the data, like liquidity premium, affect stock returns. In the same direction, Lettau and Ludvigson \cite{lettau2001consumption} provide evidence that consumption-to-wealth ratio offers information for excess stock market returns, with volatility explaining a significant portion of these returns. Another example is the work by Chung and Chuwonganant \cite{chung2018market}, where authors find strong evidence that market volatility affects individual stock returns. Under this light, we believe that these are reliable indicators in considering econometrics as features for the task of mid-price movement prediction. \\
\indent We perform our analysis based on deep learning models which have recently been proposed for financial time series analysis. These models vary from multi-layer perceptrons (MLP) to convolutional neural networks (CNN) and recurrent neural networks (RNN) like LSTM. For our experiments, we use two TotalView-ITCH datasets from the US and the Nordic stock markets. We formulate these experiments based on two protocols: the first one (i.e., \enquote{Protocol I} in our experiments) is introduced here for the first time, and is based on online learning. The prediction of the mid-price movement takes place every next event and is treated as a multi-objective optimization problem, since it predicts when and in which direction the mid-price movement will happen. The second protocol (i.e., \enquote{Protocol II} in our experiments) is an existing protocol based on the work of Tsantekidis \emph{et al.} \cite{tsantekidis2018using}, ,according to which the mid-price movement prediction is treated as a three-class classification problem (i.e., up, down or stationary mid-price states) for every next $10^{th}$ event.\\
\indent The main contribution of our work lies on three pillars. The first pillar refers to the utilization of an extensive  econometric features list as input to deep learning models for mid-price movement prediction. The second pillar is related to an extensive evaluation of the newly introduced features with two other handcrafted feature sets and a feature set based on a fully automated process. We conduct a fair evaluation of these feature sets by using the same nine deep learning models for liquid and illiquid stocks, as well as unbalanced and balanced feature sets. Next, we test them not only on the newly introduced experimental protocol but also on a protocol suggested in the literature for the Nordic dataset (also utilized here). Our findings indicate that handcrafted features, which overperformed the fully automated feature extraction process (i.e., based on LSTM AE), transform the forecasting universe of high-frequency trading. More specifically, the present evaluation facilitates traders' task of selecting suitable features according to data, stock, and model availability. The third pillar, finally, refers to the development of a new experimental protocol that takes into consideration every trading event and is unaffected by time irregularities in high-frequency data. Our work suggests that feature extraction should be customized according to stock and model selection; similar findings can be in seen in \cite{goccken2016integrating}. The present research opens avenues for several other applications. For instance, the same sets of features can be tested for time series such as exchange rates or bitcoin price predictions. Furthermore, the newly introduced protocol can be the basis of every time series problem since it is event-driven and unaffected by time irregularities. Ultimately, there is no need for any type of data sampling, even for high-frequencies time resolution environments where datasets are massive.\\ 
\indent The remainder of the paper is organized as follows. We provide a comprehensive literature review in \hyperref[SS:LR]{Section II}. The problem statement is provided in \hyperref[SS:Pr]{Section III}. The list of handcrafted features follows in \hyperref[SS:FT]{Section IV}. In \hyperref[SS:Deep]{Section V}, we describe the various deep learning models adopted in our analysis, while in \hyperref[SS:Exper]{Section VI} we describe details of the datasets and the experimental protocol. In \hyperref[SS:Results]{Section VII} we provide the empirical results and \hyperref[SS:Con]{Section VIII} concludes the paper. A detailed description of the econometric features used in our experiments are provided in \hyperref[SS:Feat]{Appendix} together with results for Protocol II. 

\section{Literature Review}\label{SS:LR}
\noindent High-frequency LOB data analysis has captured the interest of the machine learning community. The complex and chaotic behavior of the data inflow gave space to the use of non-linear methods like the ones that we see in the machine and deep learning. For instance, Zhang \emph{et al.} \cite{zhang2019comparison} utilize neural networks for the prediction of Baltic Dry index and provide a head-to-head comparison with econometric models. The author in \cite{sirignano2016deep} develops a new type of deep neural network that captures the local behavior of a LOB for spatial distribution modeling. Dixon applies RNN  \cite{dixon2018sequence} on S\&P500 E-mini futures data for a metric prediction like price change forecasting. Minh \emph{et al.} \cite{minh2018deep} also propose RNN architecture for short-term stock predictions by utilizing successfully financial news and sentiment dictionary. In \cite{zhang2018deeplob}, authors apply a combined neural network model based on CNN and RNN for mid-price prediction. \\
\indent Metrics prediction, like mid-price, can be facilitated by the use of handcrafted features. Handcrafted features reveal hidden information as they are capable of translating time-series signals to meaningful trading instructions for the ML trader. Several authors worked towards this direction, like \cite{doi:10.1080/14697688.2015.1032546}, \cite{passalis2017time}, \cite{ntakaris2018benchmark}, \cite{tran2017tensor}, \cite{tran2018temporal}, \cite{zheng2012price} and \cite{sirignano2016deep}. These works present a limited set of features which varies from raw LOB data to change of price densities and imbalance volume metrics. Another work that provides a wider range of features is  presented by Ntakaris \emph{et al.} \cite{ntakaris2018mid}. The authors there extract handcrafted features based on the majority of the technical indicators and develop a new quantitative feature based on logistic regression, which outperformed the suggested feature list.  \\
\indent Handcrafted features represent only one part of the experimental protocol in the quest for mid-price movement prediction. Classification, via deep learning methods, is the continuation of a machine learning protocol. Many authors have used deep learning in financial literature for several problems. For example, Alberg and Lipton \cite{alberg2017improving} use MLPs and RNNs for companies' future fundamentals forecasting. Qian \cite{qian2017financial} utilizes machine and deep learning methods, like support vector machines (SVM), MLPs, denoising auto-encoder (DAE), and an assembled DAE-SVM model in order to predict future trends of stock's index prices. These machine and deep learning models outperformed traditional time series models like ARIMA and generalized autoregressive conditional heteroskedasticity (GARCH). Sezer \emph{et al.} \cite{sezer2017artificial} use MLPs and the three most commonly used technical indicators as inputs for stock price movement predictions.  \\
\indent Many authors utilize LOB data as input to their models. For instance,  Nousi \emph{et al.}\cite{nousi2018machine} examine the performance of several machine learning methods, like autoencoders (AE), bag-of-features algorithm, single hidden layer feedforward neural networks (SLFN), and MLPs for mid-price prediction. Han \emph{et al.} \cite{han2015machine} apply decision trees on LOB data and outperform support vector machines (SVM) for the problem of mid-price prediction. In the same direction, authors in \cite{kanagal2017market} apply similar methods on market order book data for market movement predictions. Doering \emph{et al.} \cite{doering2017convolutional} utilize event flow and limit order datasets for price-trend and price-volatility predictions based on a deep learning architecture.  Makinen \emph{et al.} \cite{makinen2018predicting} predict price jumps with the use of LSTM, where the input data is based on LOB data. A similar work, in terms of the neural model, is conducted in \cite{tsantekidis2017using} in order to forecast LOB's mid-price. \\
\indent To the best of our knowledge, this is the first time that an extensive list of econometric features based on high-frequency LOB data is proposed as input to several neural networks for mid-price prediction. We conduct a head-to-head comparison with state-of-the-art handcrafted features is conducted together with features based on a fully automated process; Finally, we report results extracted from two high-frequency datasets with two US and five Nordic stocks for both balanced and unbalanced sets.

\section{Problem Statement}\label{SS:Pr}
\noindent The problem under consideration is the mid-price movement prediction based on high-frequency LOB data. More specifically, we use message and limit order books as input for the suggested features. Message book (MB), as seen in \hyperref[tab:mbexample]{Table} \ref{tab:mbexample}, contains the flow of information which takes place at every event occurrence. The information displayed by every incoming event includes the timestamp of the order, execution or cancellation, the id of the trade, the price, the volume, the type of the event (i.e., order, execution or cancellation), and the side of the event (i.e., ask or bid). \\
\begin{table}[h!]
\centering
\captionsetup{width=.73\textwidth}
\scalebox{0.8}{
\begin{tabular}{cccccc}
\hline
Timestamp     & Id      & Price  & Quantity   & Event & Side \\
\hline
1275386347944 & 6505727 & 126200 & 400      & Cancellation & Ask  \\
1275386347981 & 6505741 & 126500 & 300      & Submission   & Ask  \\
1275386347981 & 6505741 & 126500 & 300      & Cancellation & Ask  \\
1275386348070 & 6511439 & 126100 & 17       & Execution    & Bid  \\
1275386348070 & 6511439 & 126100 & 17       & Submission   & Bid  \\
1275386348101 & 6511469 & 126600 & 300      & Cancellation & Ask  \\
\hline
\end{tabular}}
\caption[caption]{Message list example}
\label{tab:mbexample}
\end{table}
\indent LOB (\hyperref[tab:obexample]{Table} \ref{tab:obexample}) works under specific rules based on the operation of the trading system-exchange. The main advantage of an order book is that it accepts orders under limits (i.e., limit orders) and market orders. In the former case, the trader/broker is willing to sell or buy a financial instrument under a specific price. In the latter case, the action of buying or selling a stock at the current price takes place. LOBs accept orders by the liquidity providers who submit limit orders and the liquidity takers who submit market orders. These limit orders, which represent the unexecuted trading activity until a market order arrives or cancellation takes place, construct the LOB that is divided into levels. The best level consists of the highest bid and the lowest ask price orders, and their average price defines the so-called mid-price, whose movement we try to predict. \\
\indent We treat the mid-price movement prediction as a multi-objective optimization problem with two outputs -- one is related to classification and the other one to regression. The first part of our objective is to classify whether the mid-price will go up or down and the second part -- the regression part is to predict in how many events in the future this movement will happen. To further explain this, let us consider the following example: in order to extract the intraday labels, we measure starting from time $t_k$, in how many events the mid-price will change and in which direction (i.e., up or down). For instance, the mid-price will change in 10 events from now, and will go up. This means that our label at time $k$ is going to be \{1,10\}, where 1 is the direction of mid-price and 10 is the number of events that need to pass in order to see that movement taking place. \\
\indent We depart from this labeling system to answer the critical question of whether handcrafted features derived from econometrics can boost deep learning classification and regression performance. We conduct extensive experiments based on nine neural topologies (i.e., five MLPs, two CNNs, and two LSTMs) and two TotalView-ITCH datasets, and compare the performance of econometric features to three other feature sets. The first set is based on time-sensitive and time-insensitive features as presented in \cite{doi:10.1080/14697688.2015.1032546} and \cite{ntakaris2018benchmark}, the second feature set is based on technical and quantitative analysis, introduced in \cite{ntakaris2018mid}, and the third one is based on feature representations extracted automatically for the train of an LSTM AE with a description provided in \hyperref[sec:ae]{Section} \ref{sec:ae} .   

\begin{table}[ht]
\centering
\captionsetup{width=.99\textwidth}
\scalebox{0.76}{
\begin{tabular}{llllllll}
\hline
& & &\multicolumn{4}{c}{Level 1} & ...\\
 \cmidrule(l){4-7}  \cmidrule(l){8-7} 
& & &\multicolumn{2}{c}{Ask} & \multicolumn{2}{c}{Bid} &  \\
\cmidrule(l){1-3} \cmidrule(l){4-5} \cmidrule(l){6-7} 
Timestamp     & Mid-price & Spread & Price & Quantity & Price & Quantity & \\
\cmidrule(l){1-3}  \cmidrule(l){4-8}
1275386347944 & 126200     & 200    & 126300    & 300       & 126100    & 17        &  ...   \\
1275386347981 & 126200     & 200    & 126300    & 300       & 126100    & 17        &  ...   \\
1275386347981 & 126200     & 200    & 126300    & 300       & 126100    & 17        &  ...   \\
1275386348070 & 126050     & 100    & 126100    & 291       & 126000    & 2800      &  ...   \\
1275386348070 & 126050     & 100    & 126100    & 291       & 126000    & 2800      &  ...   \\
1275386348101 & 126050     & 100    & 126100    & 291       & 126000    & 2800      &  ...   \\
\hline
\end{tabular}}
\caption[caption]{Order book example}
\label{tab:obexample}
\end{table}

\section{Handcrafted Feature Pool}\label{SS:FT}
\noindent In this section we provide the nominal list (see \hyperref[tab:featuresets]{Table} \ref{tab:featuresets}) of the extensive econometric feature list together with the two other state-of-the-art handcrafted feature sets from the literature that are based on technical and quantitative analysis and time-insensitive and time-sensitive indicators. Description of the econometric features is seen in \hyperref[SS:Feat]{Appendix} while the description of technical and quantitative feature set and time-sensitive and time-insensitive set extracted from the LOB can be found in \cite{ntakaris2018mid}.\\
\begin{table}[h!]
\centering
\captionsetup{width=.99\textwidth}
\scalebox{0.7}{
\begin{tabular}{lll}
\toprule
\textbf{Econometric features} & \textbf{Tech \& Quant features} & \textbf{LOB features} \\
\midrule
&&\\
\underline{\textbf{Statistical Features}} & \underline{\textbf{Technical Indicators}} & \underline{\textbf{Basic}} \\
Mid-Price                            & Accumulation Distribution Line               & n LOB Levels   \\
Financial Duration                   & Awesome Oscillator                           & \\
Average Mid-Price Financial Duration & Accelerator Oscillator                       & \underline{\textbf{Time-Insensitive}}\\
Log-Returns                          & Average Directional Index                    & Spread \& Mid-Price\\
                                     & Average Directional Movement Index Rating    & Price Differences\\
\underline{\textbf{Volatility Measures}}      & Displaced Moving Average                     & Price\& Volume Means\\
Realized Volatility                  & Absolute Price Oscillator                    & Accumulated Differences\\
Realized Kernel                      & Aroon Indicator                              & \\
Realized Pre-Averaged Variance       & Aroon Oscillator                             & \underline{\textbf{Time-Sensitive}}\\
Realized Semi-Variance               & Average True Range                           & Price \& Volume Derivation\\
Realized Bipower Variation           & Bollinger Bands                              & Average Intensity per Type\\
Realized Bipower Variation (lag 2)   & Ichimoku Clouds                              & Relative Intensity Comparison\\
Realized Bipower Semi-Variance       & Chande Momentum Oscillator                   & Limit Activity Acceleration \\
Jump Variation                       & Chaikin Oscillator                           &\\
Spot Volatility                      & Chandelier Exit                              &\\
Average Spot Volatility              & Center of Gravity Oscillator                 &\\
                                     & Donchian Channels                            &\\
\underline{\textbf{Noise and Uncertainty Measures}} & Double Exponential Moving Average            &\\
Realized Quarticity                  & Detrended Price Oscillator                   &\\
Realized Quarticity Tripower         & Heikin-Ashi                                  &\\
Realized Quarticity Quadpower        & Highest High and Lowest Low                  &\\
Noise Variance \cite{Oomen2006}      & Hull MA                                      &\\
Noise Variance \cite{Zhang2005}      & Internal Bar Strength                        &\\
                                     & Keltner Channels                             &\\
\underline{\textbf{Price Discovery Features}} & Moving Average Convergence/Divergence Oscillator &\\
Weighted Mid-Price by Order Imbalance & Median Price &\\
Volume Imbalance                     & Momentum&\\
Bid-Ask Spread                       & Variable Moving Average&\\
Normalized Bid-Ask Spread            & Normalized Average True Range&\\
&Percentage Price Oscillator&\\
&Rate of Change&\\
&Relative Strength Index&\\
&Parabolic Stop and Reverse&\\
&Standard Deviation&\\
&Stochastic Relative Strength Index&\\
&T3-Triple Exponential Moving Average&\\
&Triple Exponential Moving Average&\\
&Triangular Moving Average&\\
&True Strength Index&\\
&Ultimate Oscillator&\\
&Weighted Close&\\
&Williams \%R&\\
&Zero-Lag Exponential Moving Average&\\
&Fractals&\\
&Linear Regression Line&\\
&Digital Filtering: Rational Transfer Function\\
&Digital Filtering: Savitzky-Golay Filter\\
&Digital Filtering: Zero-Phase Filter\\
&Remove Offset and Detrend\\
&Beta-like Calculation\\
& &  \\
&\underline{\textbf{Quantitative Indicators}}&\\
&Autocorrelation&\\
&Partial Correlation&\\
&Cointegration based on Engle-Granger test&\\
&Order Book Imbalance&\\
&Logistic Regression for Online Learning&\\
&&\\
\bottomrule
\end{tabular}}
\caption{Feature list for the three feature sets: Description for the newly introduced, based on Econometrics, handcrafted features can be found in \hyperref[SS:Feat]{Appendix}, where description for the Tech \& Quant and LOB feature sets can be found in \cite{ntakaris2018mid}}
\label{tab:featuresets}
\end{table}
\indent We extract our econometric features from both MB and LOB and divide them into four main categories:
Statistical features, volatility measures, noise measures, and price discovery features. The first category encompasses basic statistical features that are widely used in the literature (e.g., \cite{doi:10.1080/14697688.2015.1032546}, \cite{sirignano2016deep}). The logic behind the choice of the volatility measure features is the intimate relation between the volatility of the price process and the price movement itself. As such, we regard the volatility measures included in the present article to retain information useful to real-time price prediction. This is particularly true when the predicted objective is the next price movement.
Additionally, the econometric literature widely evidences the significant detrimental impact of the so-called microstructure noise in the measurement of fundamental quantities when working at the highest frequencies.
Furthermore, the noise process directly affects the underlying price process itself and as such contributes to the observed price movements. For these reasons we implement a number of estimates of the characteristics of the noise process, which we identify as the noise measures features set.\footnote{Most of the presented measures have been developed and are consistent estimators under broad assumptions on the underlying price process and contaminating noise process; we will not discuss these assumptions into details here as outside the scope of the article. Interested readers are referred to \cite{Andersen2010} and references within for an exhaustive review of the literature} The last group of features includes all those features related to the price discovery process; i.e., those that take into account the interaction of the two sides of the LOB. Several articles in the literature (e.g., \cite{ntakaris2018mid}, \cite{makinen2018predicting}) have focused and demonstrated the importance of accounting for the differences between the ask and bid side in order to improve the mid-price forecasting accuracy.\\
\indent Each of the features in \hyperref[tab:featuresets]{Table} \ref{tab:featuresets} operates under a different time duration. Time duration of the features plays an important role in capturing information about underline behavior of time series. More specifically, the feature extraction process consists of low frequency (e.g., technical indicators based on interpolation) and high-frequency features (e.g., adaptive logistic regression), which complement each other. Low frequency features identify long-term trends and structural data components, while high-frequency features capture discontinuities and rapid metric changes. This combination of features facilitates improves neural network perfromance (e.g., \cite{doi:10.1080/14697688.2015.1032546} and \cite{lahmiri2014wavelet}).

\section{Deep Learning}\label{SS:Deep}
\noindent The goal of this paper is to forecast the movement of the mid-price. The predicted output has dual information: the direction of the mid-price movement and the prediction of the number of events taking the mid-price to move up or down. An efficient way to do that is by using deep learning architectures. We consider three different neural networks types (i.e., MLPs, CNNs, and LSTMs) and run them seperately. We, then, examine their validity with respect to our optimization problem.

\subsection{MLP for Classification and Regression}\label{SS:MLP}
\noindent MLP (i.e., \cite{bishop1995neural}) is a type of neural network that shows a high degree of connectivity among its components/neurons (see \hyperref[fig:MLP]{Fig.} \ref{fig:MLP}). The strength of this connectivity is determined by the synaptic weights of the neural network. These synaptic weights are determined by a differentiable nonlinear activation function. These basic characteristics of the neural network complicate the analysis of MLPs' behavior. As a result, several MLP architectures have to be examined in order to see whether input data (i.e., handcrafted features) affect the outcome/prediction. The way that an MLP can be trained is based on a sequential data feeding process called batch learning. Batch learning is a process according to which the neural network adjusts the synaptic weights after the presentation of all the samples $\textbf{J} = \{ \textbf{x}(i), \textbf{d}(i)\}_{i = 1}^N$ in the training process, where $\textbf{x}(i)$ is the input multi-dimensional vector and $\textbf{d}(i)$ the response vector of the supervised problem at instance $i$, and the error function at instance $i$ is:

\begin{figure}[ht]
\centering
\includegraphics[scale=0.4]{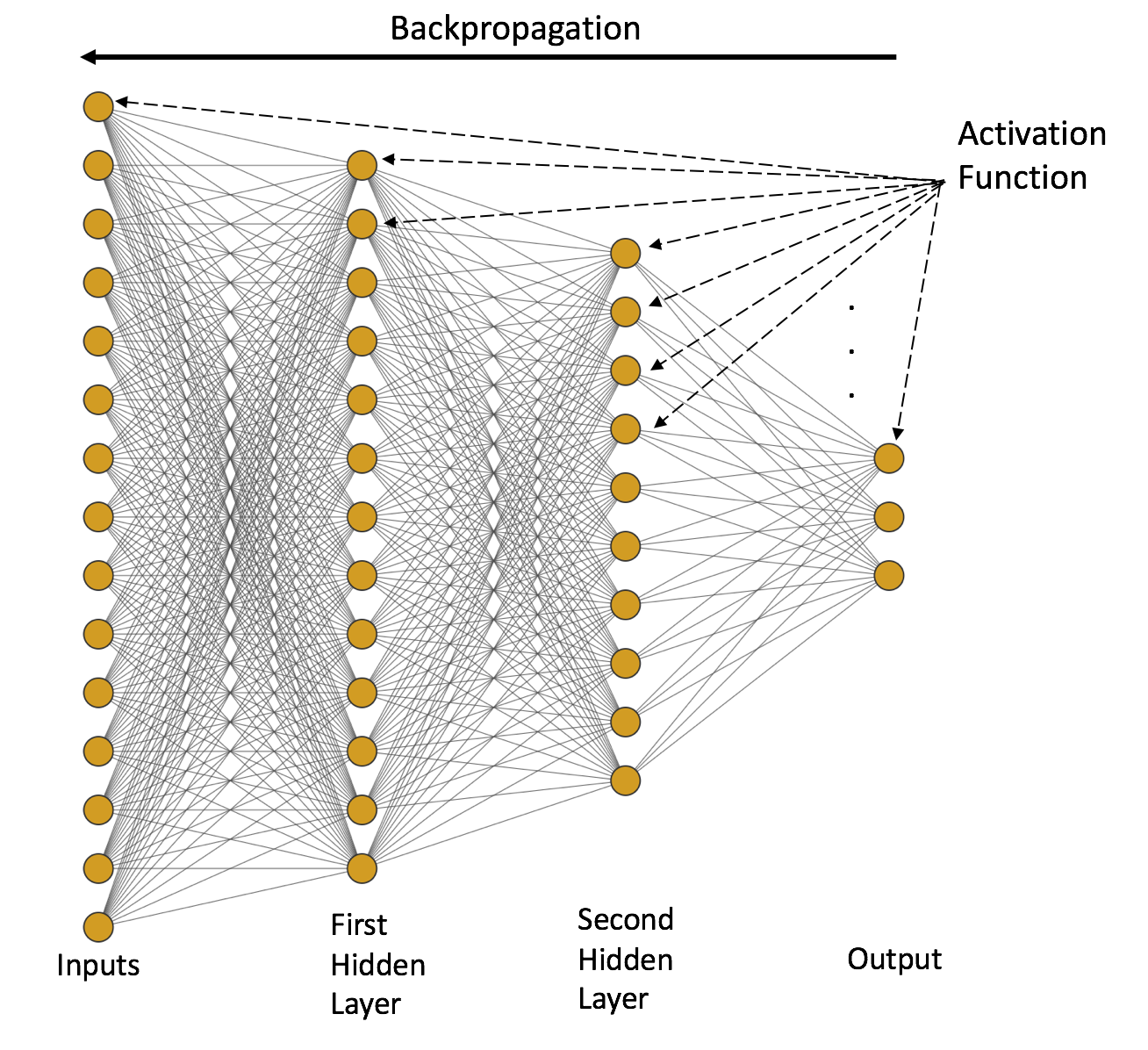}
\caption{Example of an MLP neural network with two hidden layers  and 4 units output. }
\label{fig:MLP}
\end{figure}

\begin{equation}
e^{(i)} = d^{(i)} - y^{(i)} 
\end{equation}

\noindent where $d^{(i)}$ is the $i^{th}$ element of the $\textbf{d}(i)$ and $y^{(i)}$ is the produced output term at instance $i$. The error function that we use for our experiments is bespoke to our supervised problem and its components are based on the binary cross entropy (for the classification task) and the mean squared error (for the regression task), as follows: 
\begin{equation}\label{eq:loss}
\mathcal{L}_{all} =\argmin_{\mathcal{L}_1, \mathcal{L}_2} \{\lambda\mathcal{L}_1 + (1-\lambda)\mathcal{L}_2\}
\end{equation}

\noindent where $\mathcal{L}_1$ $= -t$ $log$ $\hat{y}^{(i)}$ $-(1-t)$ $log(1-\hat{y}^{(i)})$, $t \in \{0,1\}$ and $\mathcal{L}_2 = \frac{1}{n}\sum\limits_{i=1}^{n}(y^{(i)} - \hat{y}^{(i)})^{2}$ with a free parameter $\lambda$, where $y^{(i)}$ and $\hat{y}^{(i)}$ be the ground truth and the predicted values of the $i^{th}$ training sample which belongs to $\mathbb{R}^N$, respectively. This customized function is part of the backpropagation algorithm that helps the neural network (e.g. MLP) to correct the synaptic weights in order to optimize \hyperref[eq:loss]{Eq.} \ref{eq:loss}. Backpropagation in our case follows the automatic differentiation (AD) reverse mode (i.e., \cite{baydin2015automatic}). Reverse AD facilitates the process of correcting the synaptic weights and it can be done as follows: 
Initially we define the input variables as $v_{i-n} = x_i, i = 1,...,n$, all the intermediate variables of the neural network as $v_i, i = 1, ..., k$ and $y_{m-i} = v_{k-i}$, $i = m-1, ... , 0$ be the output variables. Derivatives calculation is a two-step process. During the first phase the intermediate variables $v_i$ are populated and create the graph trace, whereas during the second phase derivatives are calculated based on the propagation of the adjoints $\bar{v}_i = \frac{\partial y_l}{\partial v_i}$. In general, the reverse AD performs the calculations from the output to the input starting from the output as seed: 
\begin{equation*}
\frac{\partial f}{\partial  y_{m-i}}  \leftarrow 1
\end{equation*} 
and moves to the inputs via the intermediate states based on the calculation:
\begin{equation*}
x_i \leftarrow \sum\limits_{j:i \in Pa(j)}^{} \frac{\partial f}{\partial x_j}\frac{\partial g_j}{\partial x_i}
\end{equation*}
where $Pa(j)$ denotes the parent formation of node $j$ and $g_j$ the intermediate functions of the graph. \\
\indent The next part of the MLP training is the learning process, which is defined as the method through which the loss function will reach the optimal solution via proper parameter updates. For this reason we choose the Nesterov accelerated gradient (NAG) method incorporated into the adaptive moment estimation (Adam) method named as Nadam by Dozat \cite{dozat2016incorporating}. Nadam applies the momentum step only once and takes into consideration the current momentum -- rather than the previous momentum -- vectors. This gives us the Nadam update parameters rule:
 
\begin{equation}
\theta_{t+1} := \theta_t - \frac{\eta}{\sqrt{\hat{v}_t} + \epsilon}(\beta_1\hat{m}_t + \frac{(1-\beta_1)\nabla_{\theta_t}\mathcal{L}_{all}(\theta_t)}{1-\beta^t_1})    
\end{equation}
where the first (i.e., mean) and second (i.e., variance) moment for the current momentum vector are, respectively:

\begin{equation}
\hat{m}_t = \frac{m_t}{1-\beta_1^t}, \,\,\,\,\,\, \hat{u}_t = \frac{v_t}{1-\beta_2^t}
\end{equation} 
with $m_t = \beta_1m_{t-1} + (1-\beta_1)\nabla_{\theta_t}\mathcal{L}_{all}(\theta_t)$, $v_t = \beta_2v_{t-1} + (1-\beta_1)\nabla_{\theta_t}^2\mathcal{L}_{all}(\theta_t)$ and learning rate $\eta  = 0.002$.

\subsection{CNN for Classification and Regression}
\begin{figure}[ht]
\centering
\subfloat[CNN based on the work presented in \cite{tsantekidis2017forecasting}. ]{%
  \includegraphics[width=0.8\columnwidth]{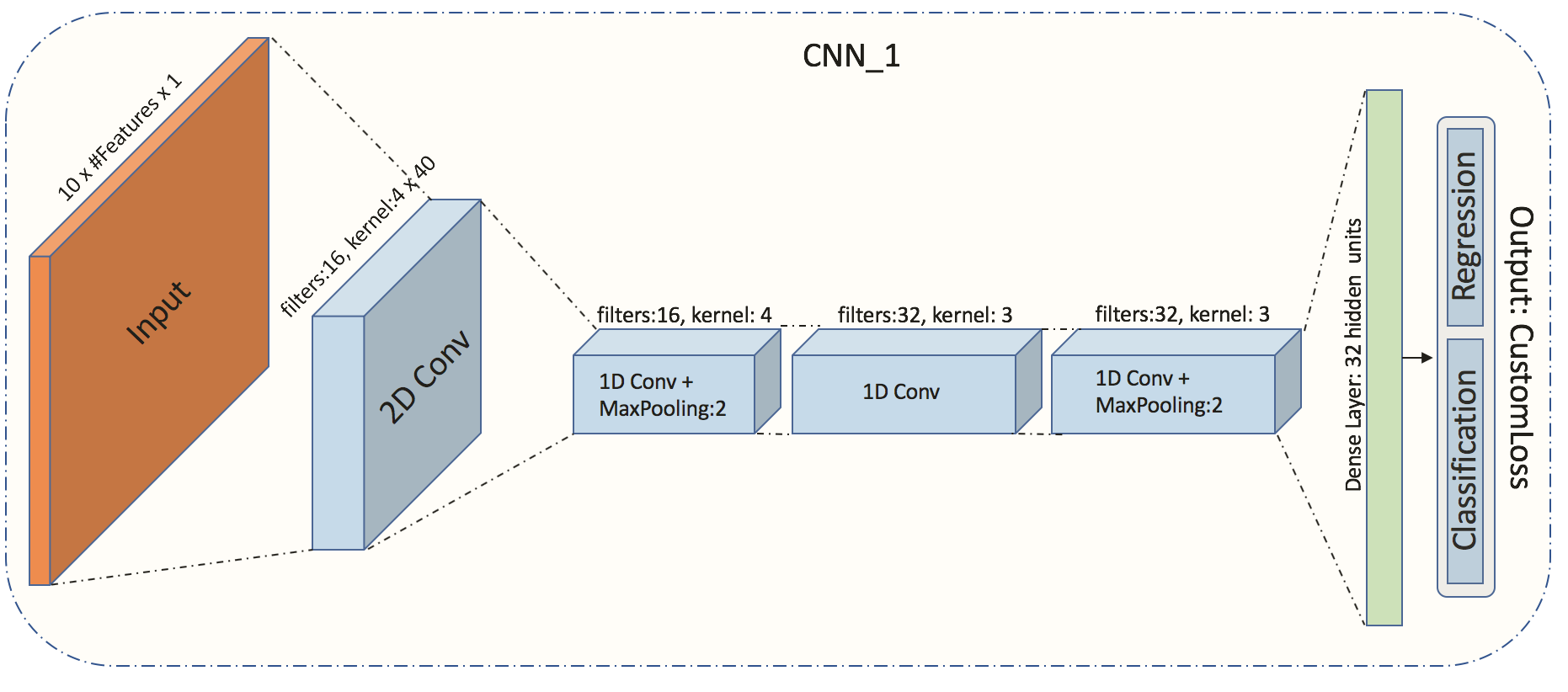}%
}\hspace{1cm}\vspace{0.3cm}
\subfloat[CNN with deeper topology that will be used later in the experimental protocols.]{%
  \includegraphics[width=0.8\columnwidth]{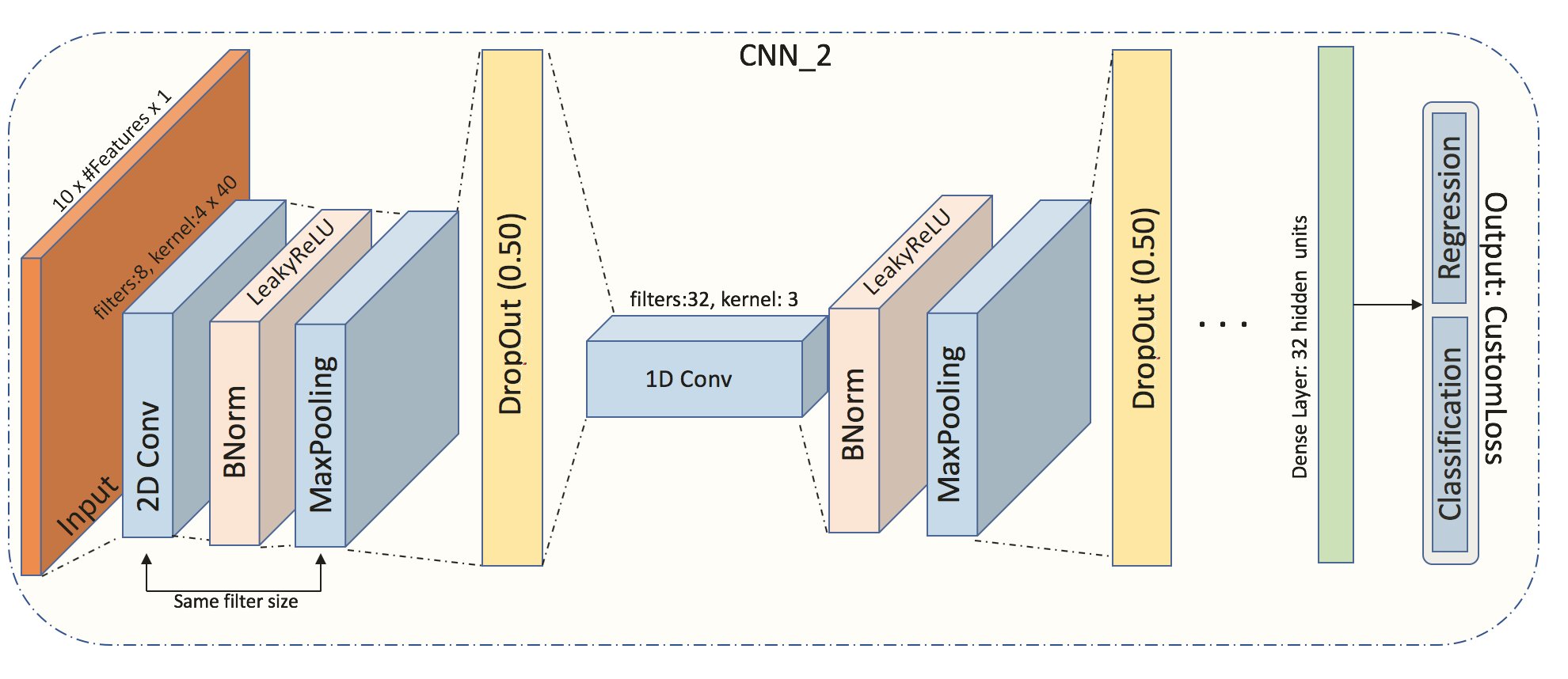}%
}
\caption{Two CNN examples that demonstrate their operation mechanisms. These two CNNs (i.e., CNN\_1 and CNN\_2) will later on utilized in the experiments. }
\label{fig:cnn}
\end{figure}

\noindent CNN, as described in \cite{Goodfellow-et-al-2016}, is a type of neural network that handles time series of multidimensional data for metric prediction. The main motivation for choosing this type of neural network is its capability for sparse connectivity between neural layers, for sharing the so-called tied weights and equivariant representation properties. More specifically, sparse connectivity can be achieved by using a kernel smaller than the sample input. This action reduces the amount of memory that is required for the training process. The second advantage of a CNN is the use of tied weights. Tied weights are shared among the inputs since the same amount of weights is applied to the inputs.\\
\indent CNN has three main parts: the convolution layer, the pooling layer, and the fully connected layer. The convolution layer extracts features from the input multi-dimensional signal expressed usually as a tensor or matrix. This process creates linear activations that run via a non-linear activation function such as the rectified linear activation function (ReLU) and the Leaky ReLU. Then the pooling layer will convert the local output based on a summary statistic related to the local outputs (e.g. max-pooling). The last step of the process is the connection to the fully connected layers (see examples in \hyperref[fig:cnn]{Fig.} \ref{fig:cnn}) that will perform the classification and regression tasks. These tasks are based on discrete time series events that formulate the (forward) convolution layer calculation as follows:

\begin{equation}\label{eq:conv}
y_{{i^{l+1}},{j^{l+1}}, d} = \sum\limits_{i = 0}^{H}\sum\limits_{j = 0}^{W}\sum\limits_{d = 0}^{D}f_{i,j,d} \times x_{{i^{l+1}+i},{j^{l+1}+j},d}^l
\end{equation} 
where $H, D,$and $D$ are the row, columns and depth dimension of the input tensor $\textbf{x}\in \mathbb{R}^{H^{l}\times W^{l}\times D^{l}}$ respectively, $\textbf{f} \in \mathbb{R}^{H^{l}\times W^{l}\times D^{l}}$ is the filter bank, and the indexing $(i^{l+1}+i, j^{l+1} + j,d)$ refers to the iterative local convolution of the filter bank  on the suggested input for the $l$-layer. Pooling is performed right after convolution; to conduct our experiments, we choose the formation of max pooling. The final step is the use of fully connected layers. The structure of these fully connected layers is the same as in \hyperref[SS:MLP]{Sec.} \ref{SS:MLP}. The process that we follow in order to train our CNN parameters (i.e., filter banks and synaptic/tied weights) is based on batch learning combined with reverse AD (i.e., backpropagation) as we did for the MLP case.   

\subsection{LSTM for Classification and Regression}
\noindent The ML trader has to consider the temporal behaviour of time series. The events that we have to deal with in the LOB universe are likewise formed in a sequential manner. Sequential systems, like RNNs, are based on computational graphs and are, thus, ideal for time series analysis. RNNs provide much flexibility in terms of architecture formation, which is described in \hyperref[eq:rnn]{Eq.} \ref{eq:rnn}:
\begin{equation}\label{eq:rnn}
\textbf{h}_t = f(\textbf{h}_{t-1},x_t;\theta)
\end{equation}  
where $\textbf{h}$ and $x$ are the state and the input at time $t$ and $\theta$ are the shared parameters for a transition function $f$ at time $t$. Since we use RNN for empirical calculations we choose to forecast mid-price by using gated RNNs (named LSTM) as presented in \cite{hochreiter1997long}. Motivation for choosing this type of gated RNN is its ability to create connections through time and account for the problem of vanishing (or exploding) gradients. Instead of applying just element-wise nonlinear input transformations, LSTM units (see LSTM's internal cell calculations in \hyperref[fig:cell]{Fig.} \ref{fig:cell}), contain processes which that take into consideration the sequential nature of time series. More specifically, an LSTM cell is equipped with gates that filter the information flow by applying weights internally. The first pass is the forget gate vector $f_i^{t}$:

\begin{figure}[ht]
\centering
\includegraphics[scale=0.34]{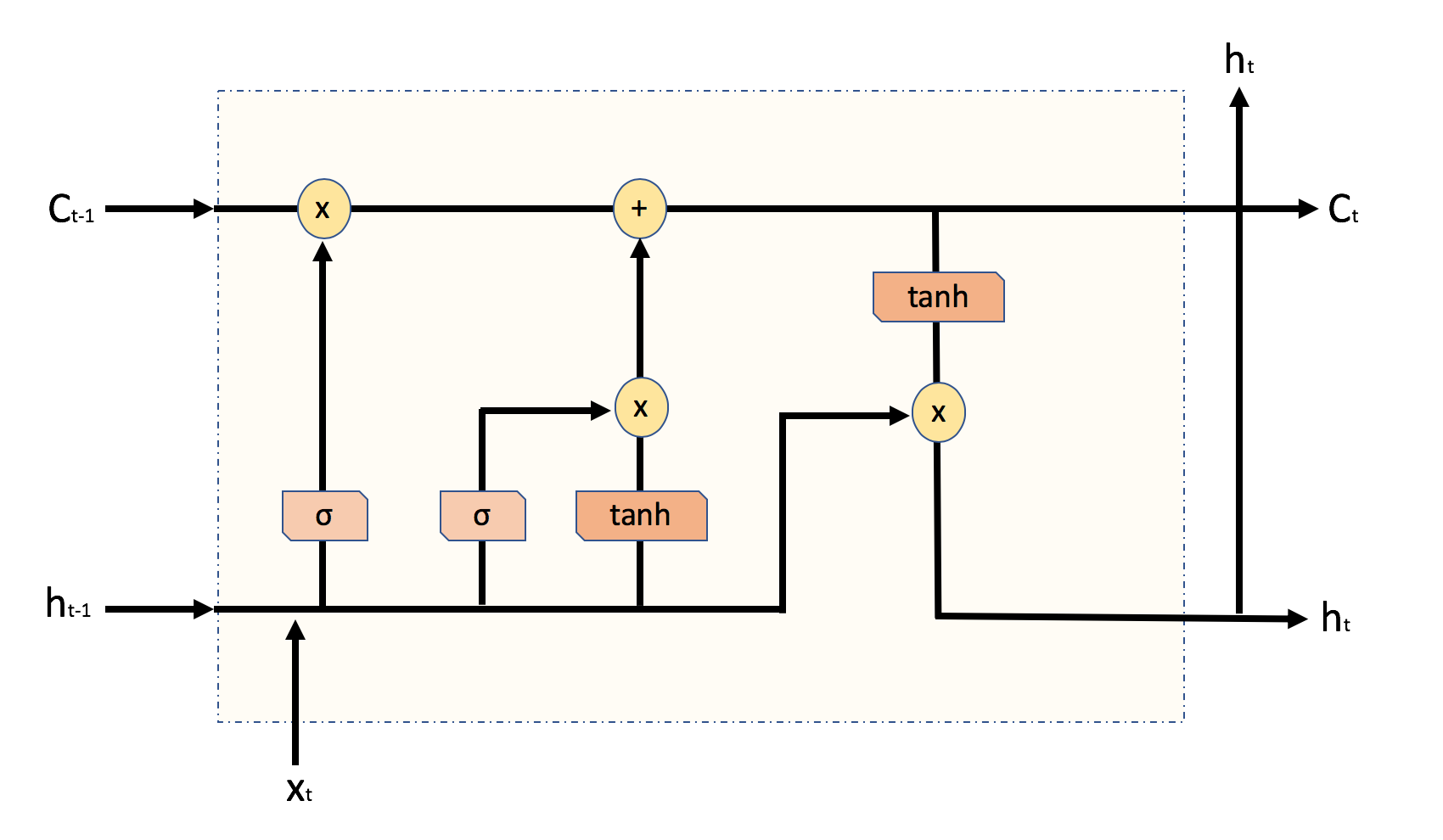}
\caption{Visual representation of LSTM's internal cell calculations.}
\label{fig:cell}
\end{figure}

\begin{equation}
f_i^{(t)} = \sigma\Big( \sum\limits_{j}W_{i,j}^f h_j^{(t-1)} + \sum\limits_{j}U_{i,j}^f x_j^{(t)}  + b_i^f\Big)
\end{equation}
where $\textbf{x}_i^{(t)}$ and $\textbf{h}_i^{(t)}$ are the current input and hidden state vectors of cell $i$ at time $t$, respectively. The attached weight matrices to these vectors are $W^f$ and $U^f$ for the forget gate vector with $b^f$ the bias term. The next pass is related to the information to be saved to the so-called "cell state". The cell state can be divided in two parts - the input vector and a $tanh$ layer as follows:

\begin{equation}
C_i^{(t)} = f_i^{(t)} C_i^{(t-1)} + g_i^{(t)}\sigma\Big( \sum\limits_{j}W_{i,j}^C h_j^{(t-1)} + \sum\limits_{j}U_{i,j}^C x_j^{(t)}  + b_i^C\Big)
\end{equation}
where $g^{(t)}$ is the input gate:

\begin{equation}
g_i^{(t)} = \sigma\Big( \sum\limits_{j}W_{i,j}^g h_j^{(t-1)} + \sum\limits_{j}U_{i,j}^g x_j^{(t)}  + b_i^g\Big)
\end{equation}
The last remaining part is the filtered output. More specifically, the LSTM output/hidden state will be formulated by the output gate vector $o_i^{(t)}$ which is calculated as follows:

\begin{equation}
o_i^{(t)} = \sigma\Big( \sum\limits_{j}W_{i,j}^o h_j^{(t-1)} + \sum\limits_{j}U_{i,j}^o x_j^{(t)}  + b_i^o\Big)
\end{equation}
and the final output $h_i^{(t)}$ is equal to:

\begin{equation}
h_i^{(t)} = o_i^{(t)} \ast tanh(C_i^{(t)}).
\end{equation}

\begin{figure}[ht]
\centering
\subfloat[LSTM based on the work in \cite{tsantekidis2018using}]{%
  \includegraphics[width=0.7\columnwidth]{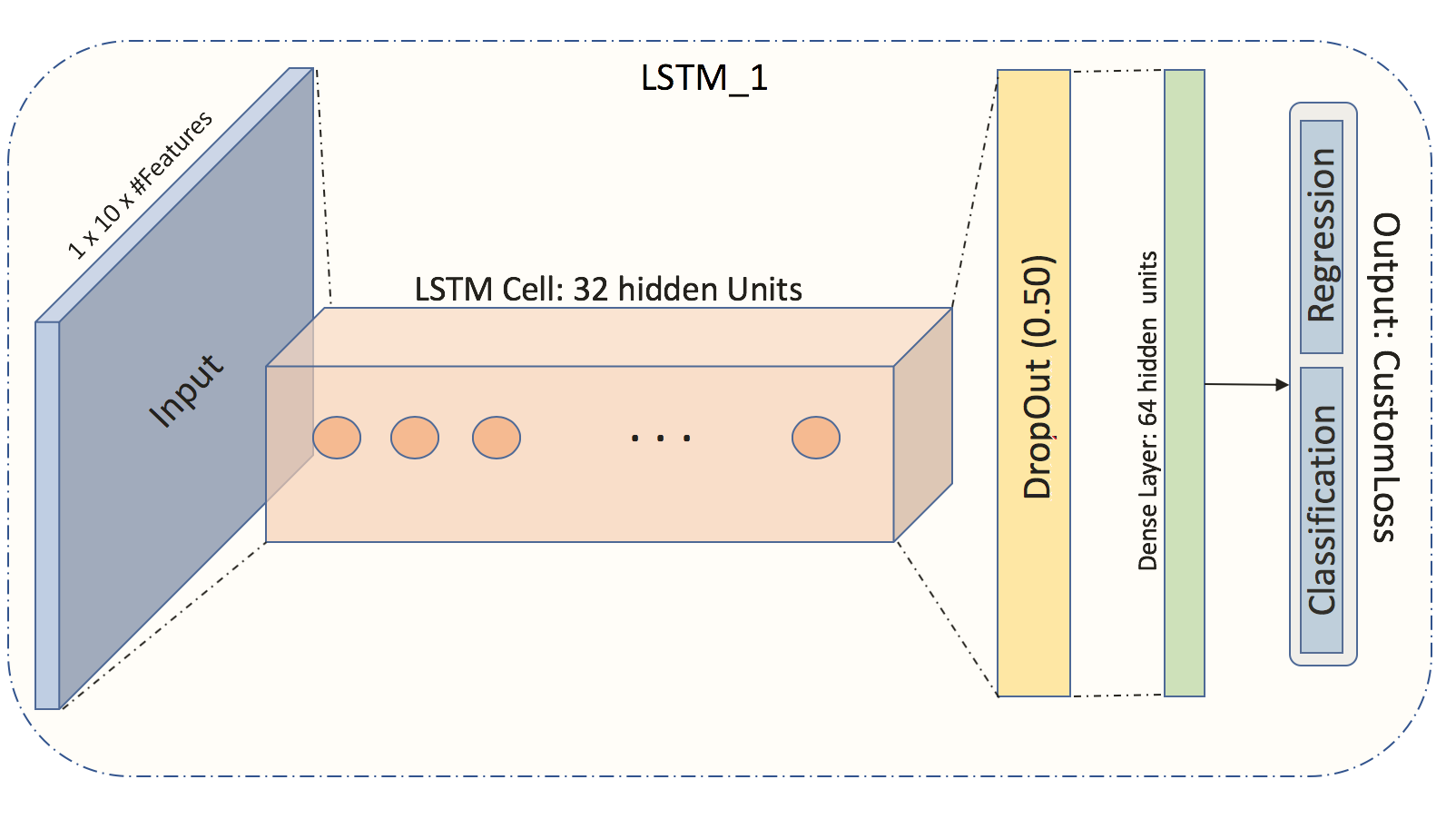}%
}\hspace{1cm}\vspace{0.3cm}
\subfloat[LSTM with attention layer]{%
  \includegraphics[width=0.7\columnwidth]{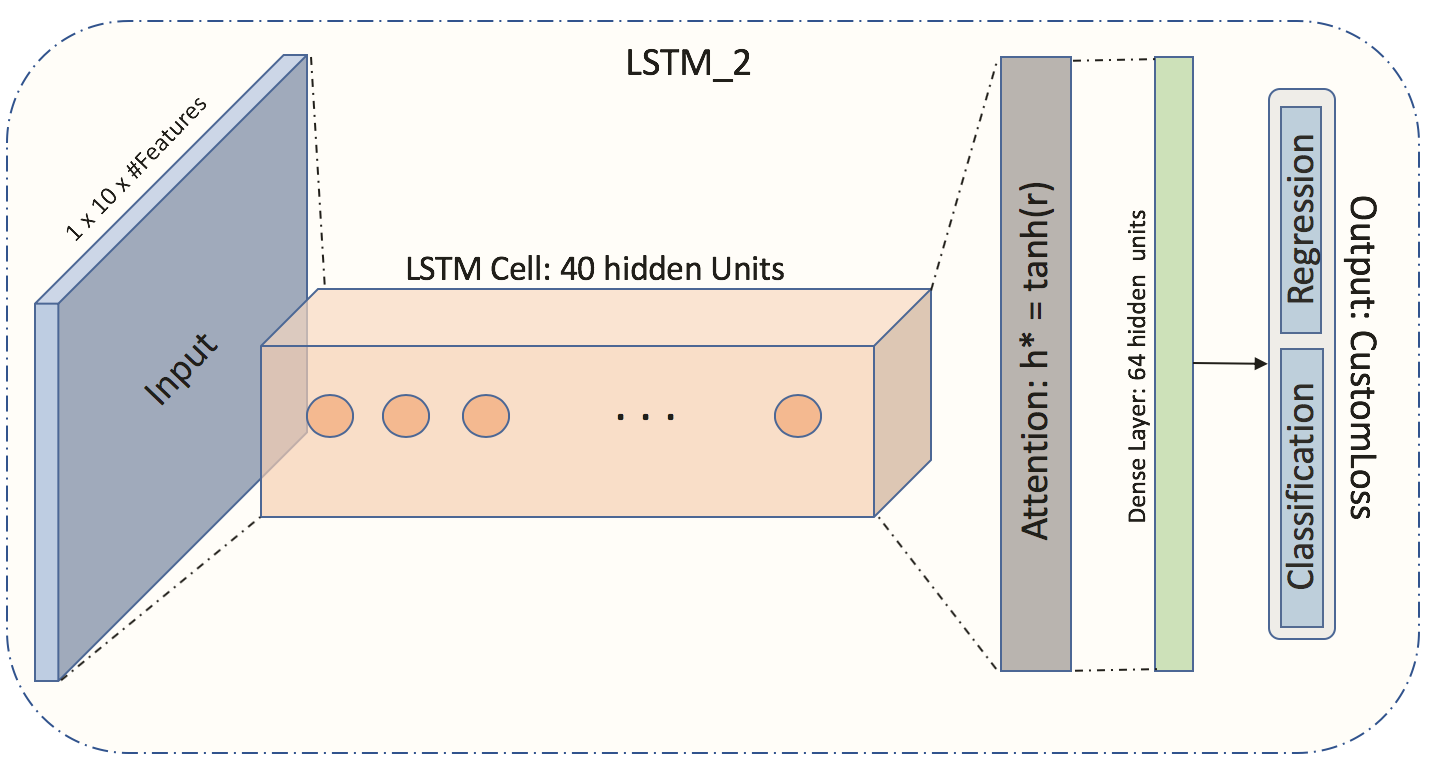}%
}
\caption{Two LSTM examples with one main LSTM block (orange colored box) with several hidden cell units (orange cycles). These two LSTMs (i.e., LSTM\_1 and LSTM\_2) will later on utilized in the experiments. }
\label{fig:lstm}
\end{figure}

The formation above refers to the case of a typical LSTM neural network, which we implement in \hyperref[SS:Exper]{Section} \ref{SS:Exper}. We also apply an attention mechanism to the LSTM architecture in order to weight/measure the significance of the input sequence. We follow the implementation in \cite{zhou2016attention} and \cite{makinen2018predicting} where the sequential LSTM outputs (i.e., hidden states $H^{(t)}$, $t \in \{1, ..., T\}$) are filtered via the following steps for every $K$-dimensional vector $w$:

\begin{equation}
M = tanh(H^{(t)})
\end{equation}

\begin{equation}
\alpha = \frac{e^{w_i^T \ast M}}{\sum\limits_{k=1}^{K}e^{w_k^T \ast M}}
\end{equation}

\begin{equation}
r = H^{(t)} \ast \alpha
\end{equation}
and the final LSTM with attention output is:

\begin{equation}
h^* = tanh(r).
\end{equation}
Here we use the same backpropagation mechanism as we did for MLPs. Examples of LSTM neural networks can be seen in \hyperref[fig:lstm]{Fig.} \ref{fig:lstm}

\subsection{Fully Automated Feature Extraction based on Autoencoders}\label{sec:ae}
\noindent

\begin{figure}[h!]
    \centering
    \includegraphics[scale=0.4]{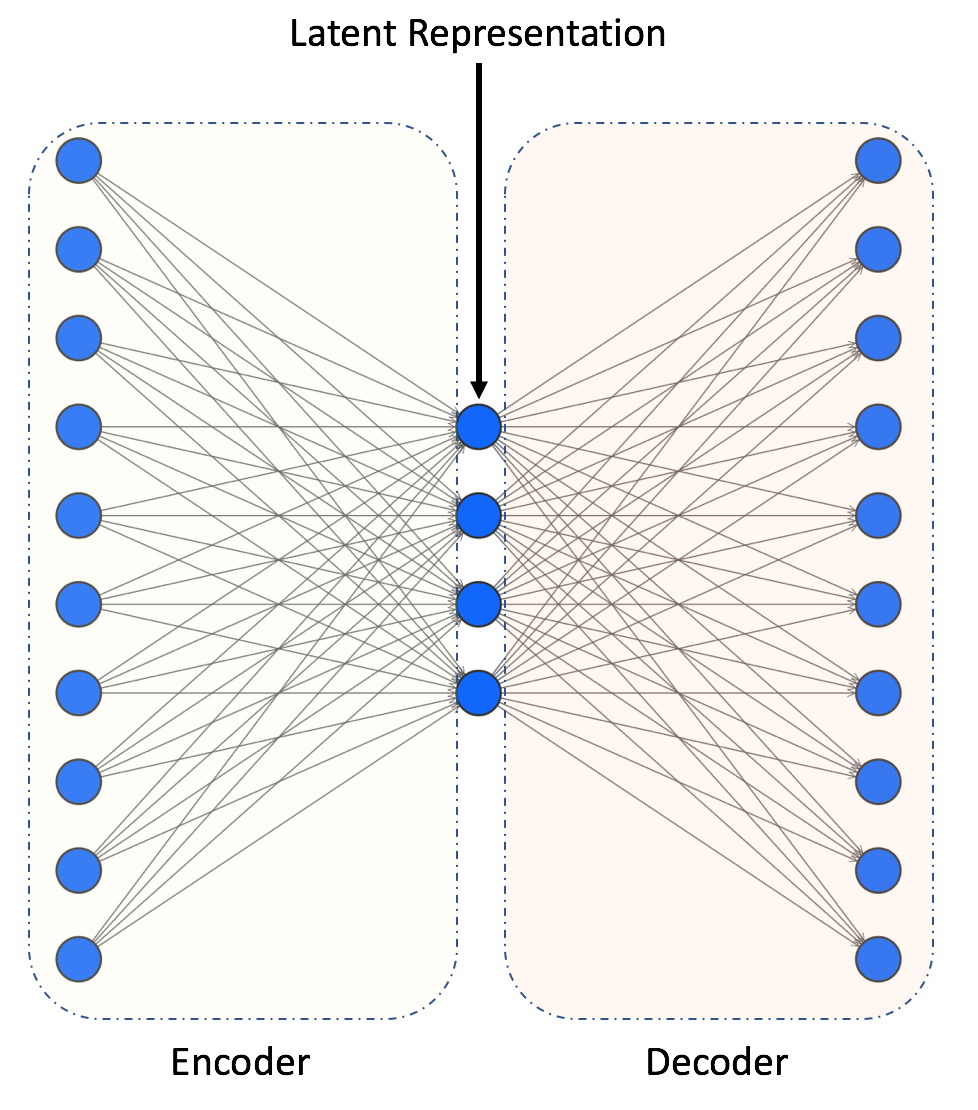}
    \caption{AE Example}
    \label{fig:AE}
\end{figure}

\noindent Autoencoders (AE) (i.e., \cite{liou2014autoencoder}, \cite{liou2008modeling}) are neural networks which operate on a self-feedback loop fashion. They do not require any labeling system since they depend on this semi-supervised protocol. This  type of neural network is devided in three main parts; the encoder, the latent representation, and the decoder (i.e. encoder and decoder). An example of AE can be seen in \hyperref[fig:AE]{Fig.} \ref{fig:AE}.

\noindent The basic structure of AE is defined as a mapping from encoder to decoder, the main objective being the following minimization problem:

\begin{equation}
f,g = \argmin_{f,g}|| X - (f \circ g)X ||^2
\end{equation}
where $f:X \rightarrow F$ and $g:F \rightarrow X$ with $X$ be the input raw LOB data in the present work.

The fully automated feature extraction process is based on the latent representation. This latent representation in the present work plays the role of the vector representation, which will, later on act as input to each of the suggetsed nine deep neural networks. In order to use this latent space as feature set we train an LSTM AE.\footnote{Details of the training are provided in \hyperref[SS:Results]{Section} \ref{SS:Results}.} LSTM AE has exactly the same structure as a simple AE with the difference that the filtering is based on LSTM layers for the encoding and deconding part. We choose LSTM AE since they take into consiration the temporal behaviour of our time series.

\section{Data Description and Experimental Protocols}\label{SS:Exper}
\noindent Our objective is to provide informative handcrafted features to ML traders and market makers for the task of mid-price movement prediction. Prediction of this movement requires in-depth analysis in terms of data selection (e.g., liquid or illiquid stocks) and experimental protocol development. For these reasons, our analysis consists of two TotalView-ITCH based on two US and five Nordic stocks and two experimental protocols. The first protocol, named Protocol I, is based on online prediction for every 10-block rolling events, and we introduce it here for the first time. The second protocol, named Protocol II, is derived from the literature (i.e., \cite{tsantekidis2018using}) and is based on mid-price movement prediction with 10-event lag. Both protocols are event driven, which means that there are no-missing values. However, Protocol II is based on independent 10-block events, which creates a lag of 10 events. Some of the suggested features can partially overcome this problem by finding averages or other types of transformations inside these blocks, but, still some information will be parsed. A possible solution to this problem comes from Protocol I where every single trading event is taken into consideration and, as a result, there are no missing values. We should also mention that LOB data is exposed to bid-ask\footnote{Bid-ask bounce is the rapid stock’s price bounce between bid and ask side.} bounce effect which may inject bias. We leave this topic for future research, where we plan to increase the rolling event block size in Protocol I since a wider block will, potentially, improve stability.  

\subsection{Data}
\begin{figure*}[h!]
\centering
\includegraphics[scale=0.48]{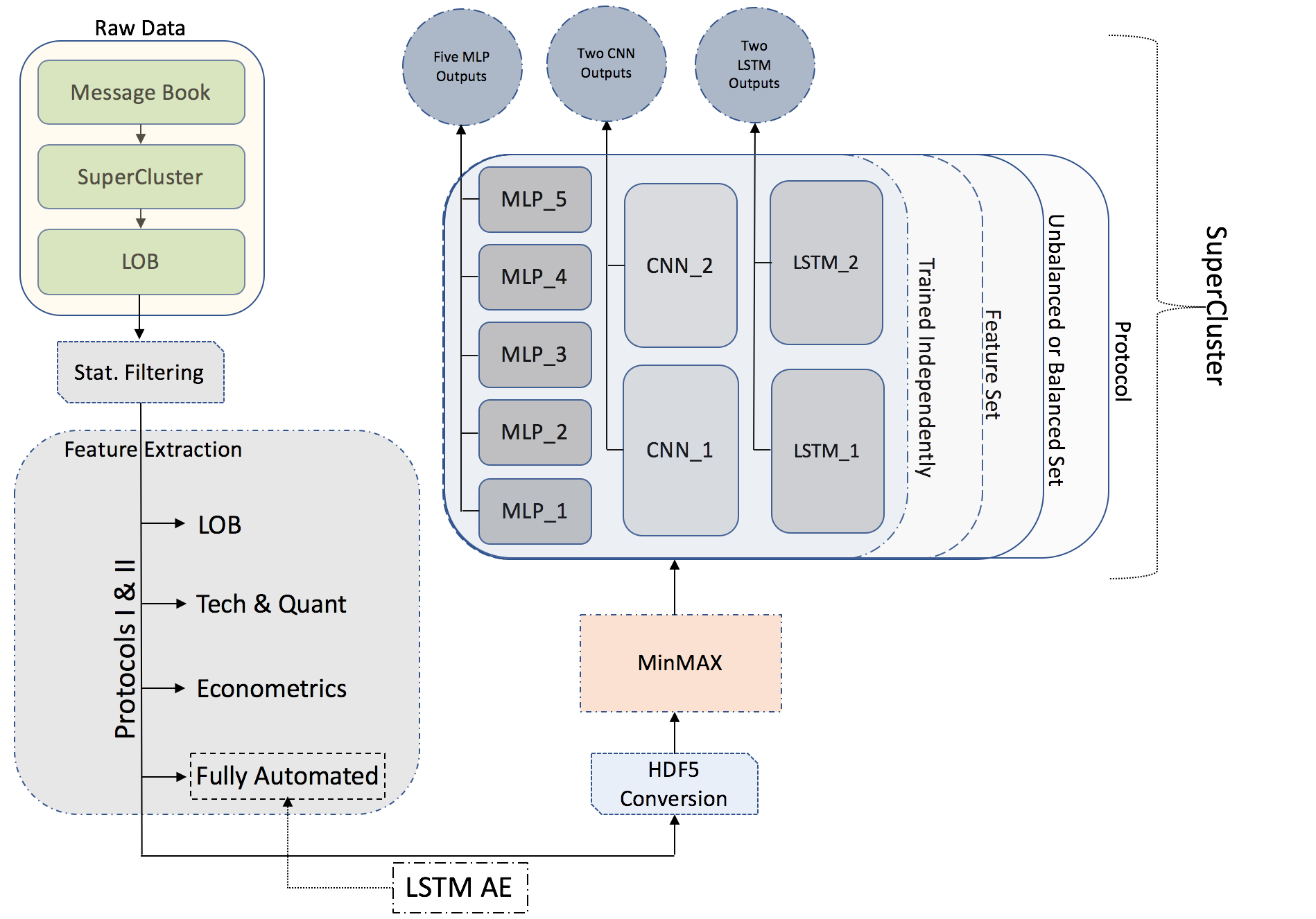}
\caption{This is a higher-level explanation of the steps that we follow for the present analysis. From left top to right bottom: The first step is to obtain the datasets for the US and Nordic stocks and send raw data (i.e., message books) to CSC superclusters and obtain the LOBs. The next step is to apply statistical filetring (description can be found in \hyperref[sec:Normal]{Section} \ref{sec:Normal}). What follows is the process of feature extraction for the four different feature sets for Protocols I \& II. An HDF5 conversion takes place right afterwards, and a MinMax normalization follows for every feature set case for both protocols. Next, each of the nine neural networks is trained independently for the four different feature lists based on unbalanced and balanced sets. The training process is based on python scripts, which are sent to CSC superclaster in order to obtain results for Protocol I \& II.}
\label{fig:General}
\end{figure*}

\noindent We utilize two TotalView-ITCH datasets based on two US and five Nordic stocks. The time resolution of the datasets is in milliseconds. For the US datasets, we use two stocks, Amazon and Google, whereas for the Nordic dataset we use Kesko Oyj, Outokumpu Oyj, Sampo Oyj, Rautaruukki, and Wartsila Oyj. We use ten business days for both datasets covering the periods: from 22.09.15 to 05.10.15 for the US dataset and from 01.06.10 to 14.06.10 for the Nordic dataset, respectively. The trading activity for these ten business days is 13,000,000 events for the US dataset and 4,000,000 events for the Nordic dataset. We use MBs in order to create relevant LOBs. We utilize super clustering computational power based on HP Apollo 6000 XL230a/SL230s supercluster to convert MBs to LOBs (i.e., LOBs are of depth 10 for both sides). We follow several pre-processing steps before we start training the deep learning models. A general description of the pre-processing process can be seen in \hyperref[fig:General]{Fig.} \ref{fig:General} 

\subsection{Protocol I}

\begin{figure}[ht]
\centering
\subfloat[Hypothesis test for stationarity check for the Nordic stock, Kesko Oyj. The plot represents a sample of 500 consecutive events.]{%
  \includegraphics[width=0.7\columnwidth]{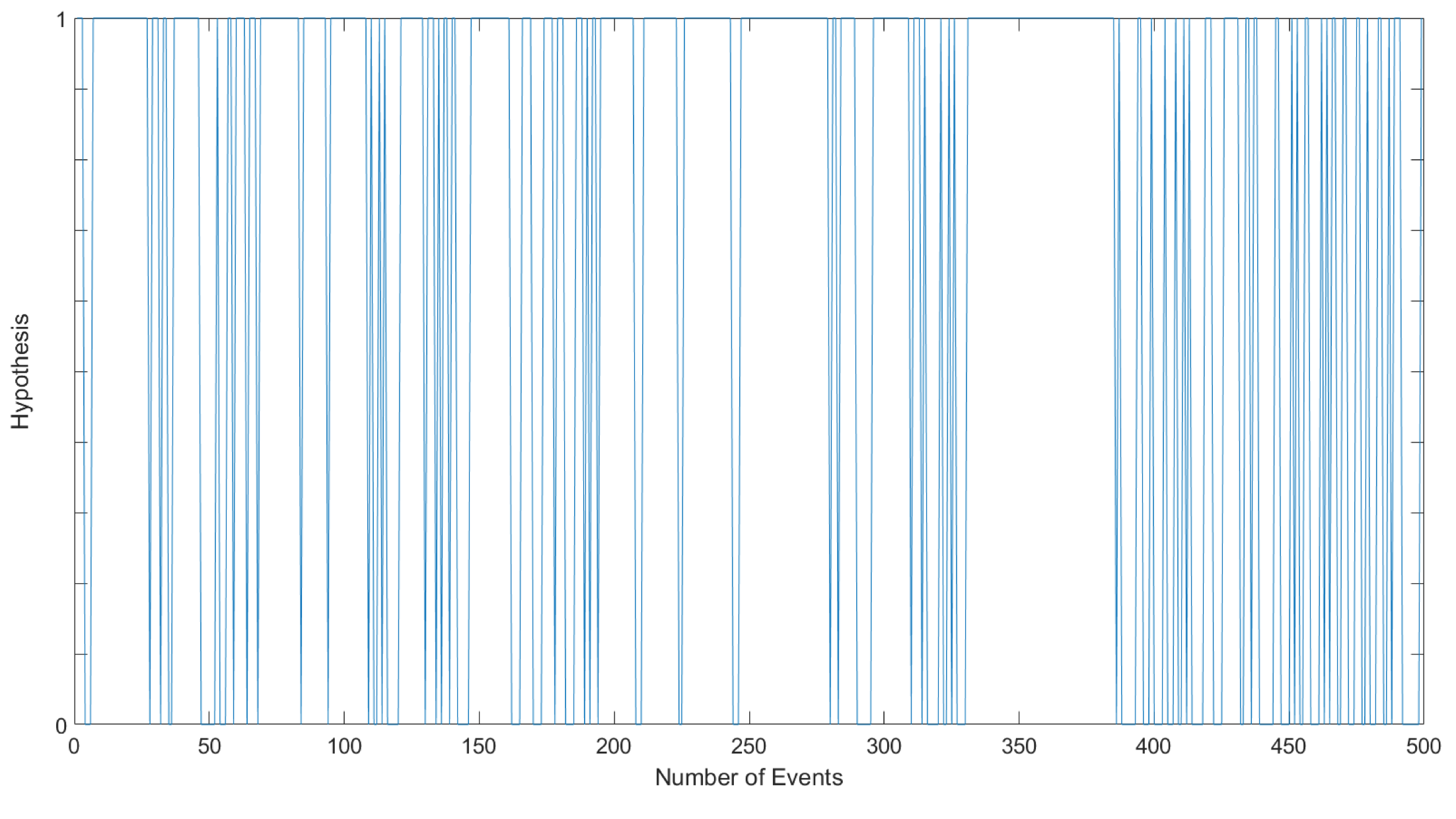}%
}\hspace{1cm}\vspace{0.3cm}
\subfloat[Hypothesis test for stationarity check for the US stock, Amazon. The plot represents a sample of 500 consecutive events.]{%
  \includegraphics[width=0.7\columnwidth]{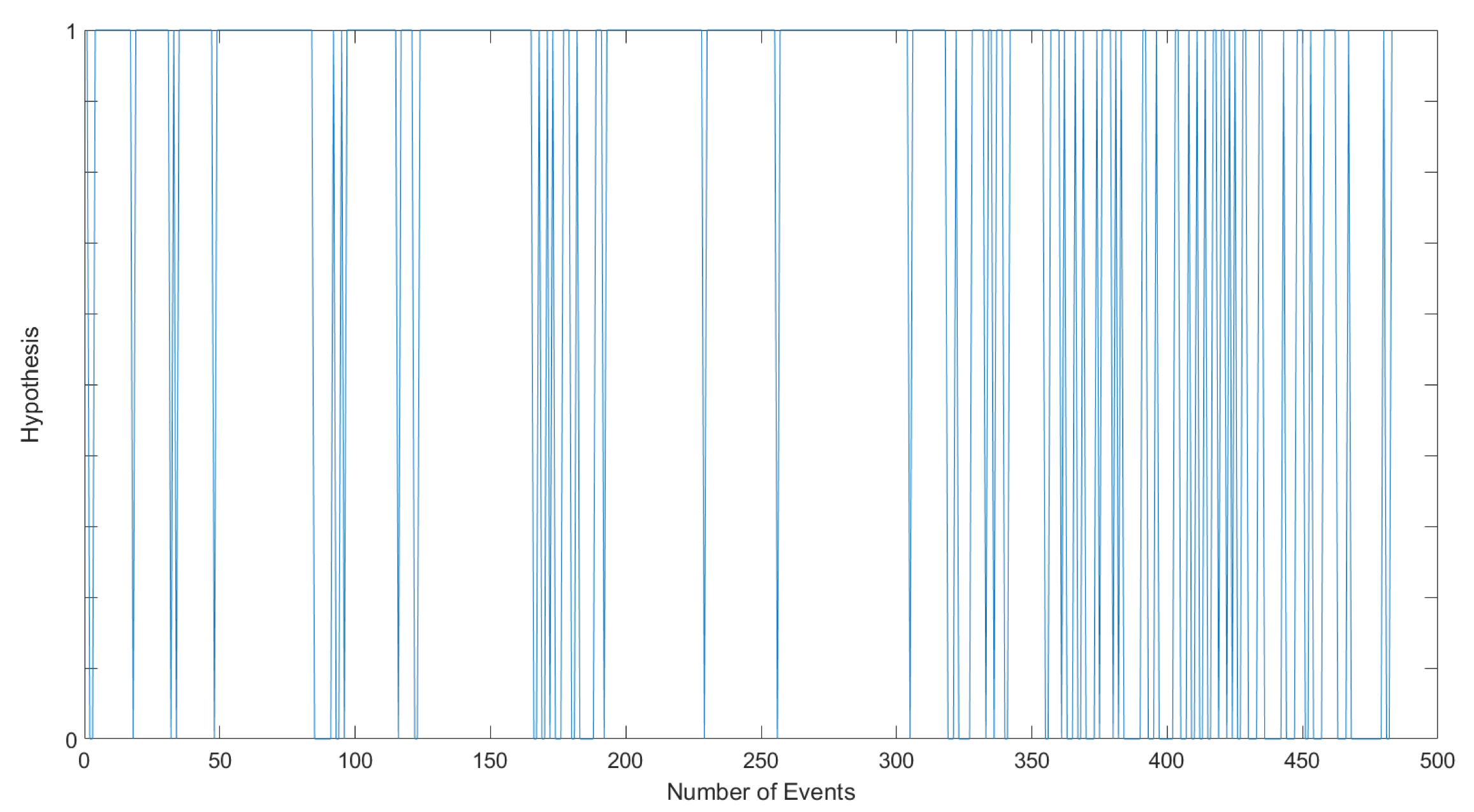}%
}
\caption{Hypothesis test for stationarity check, where constant transition from state 0 to state 1 is present.}
\label{fig:hyp}
\end{figure}

\noindent Both TotalView-ITCH datasets convey asynchronous information varying from events taking place at the same millisecond to events several minutes apart from each other. In order to address this issue, we develop Protocol I, which utilizes all the given events in an online manner. More specifically, our protocol extracts feature representation every ten events with an overlap of nine events for every next feature representation. 
We decided to use a 10-window block for our experiments due to the frequency \footnote{The average rate of change of the non-stationarity condition, for both TotalView-ITCH datasets, is changing in average every ten events.} of the stationarity present in both datasets. In order to identify whether our time series have unit roots, we perform an Engle-Granger cointegration test\footnote{Test implementation can be found in \cite{ntakaris2018mid}.}, with focus on the augmented Dickey-Fuller test, on the pair $Ask-Bid$ prices from LOBs level I. The hypothesis test shows that there is a constant alternation between its states (i.e. 1 for non-stationarity and 0 for stationarity of the suggested time series), which occurs several times during the day. This is indicative for both datasets as seen in \hyperref[fig:hyp]{Figure} \ref{fig:hyp}. These stationarity breaks supports the initial idea, as this presented by many authors (e.g., \cite{qi2008trend}, \cite{butler2011effects}, \cite{kim2004artificial}), that neural netwroks are capable of identifying underlying processes of a non-stationary time series. Neural networks are nonlinear and non-parametric adaptive-learning filters which operate with fewer assumptions compare to more traditional time series models like ARIMA and GARCH.

\begin{figure}[h!]
\centering
\includegraphics[scale=0.5]{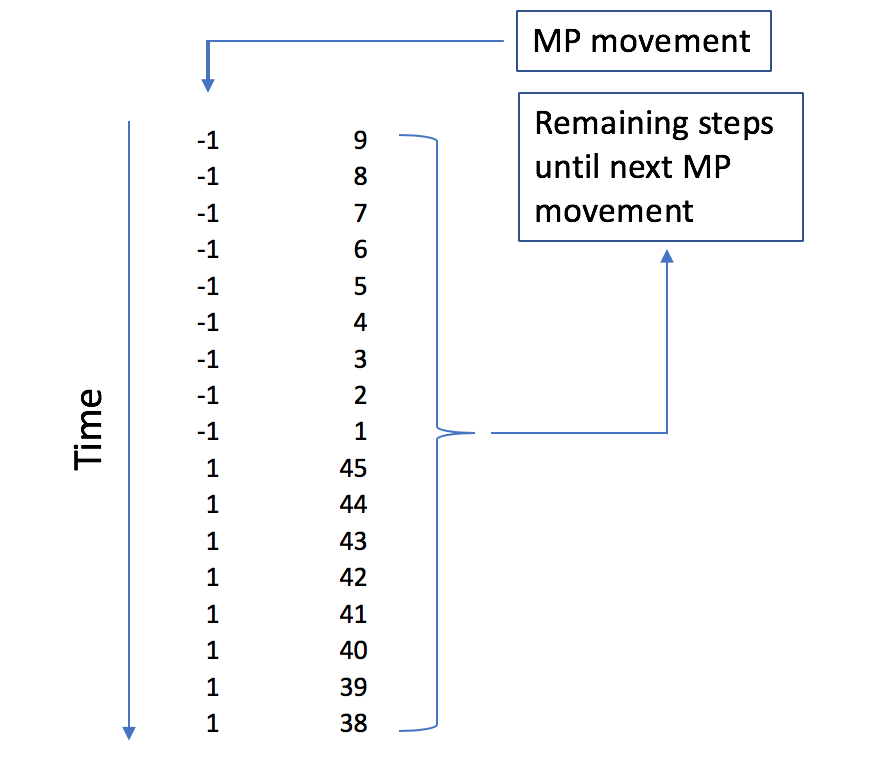}
\caption{Labeling sample for the dual prediction problem of our classification and regression objective. The left part represents the direction (i.e., up or down) of the mid-price (MP) movement while the right part represents the remaining steps until the next change in MP will take place.}
\label{fig:Labels}
\end{figure}

A visual description of our protocol can be seen in plot (a) in \hyperref[fig:featextra]{Fig.} \ref{fig:featextra}. The problem under consideration in Protocol I is to predict the movement of mid-price (i.e., classification: up or down) together with the number of events it takes for that movement to occur in the future (i.e., regression: number of events until next mid-price's movement change).
More specifically, in ordert to testing performance evaluation, we utilize f1 score for the classification task and RMSE (i.e., Root Mean Square Error) for the regression task. F1 score is defined as:

\begin{figure}[ht]
\centering
\subfloat[\textbf{Protocol I}: Feature extraction in an online manner with zero lag delay]{%
  \includegraphics[width=0.5\columnwidth]{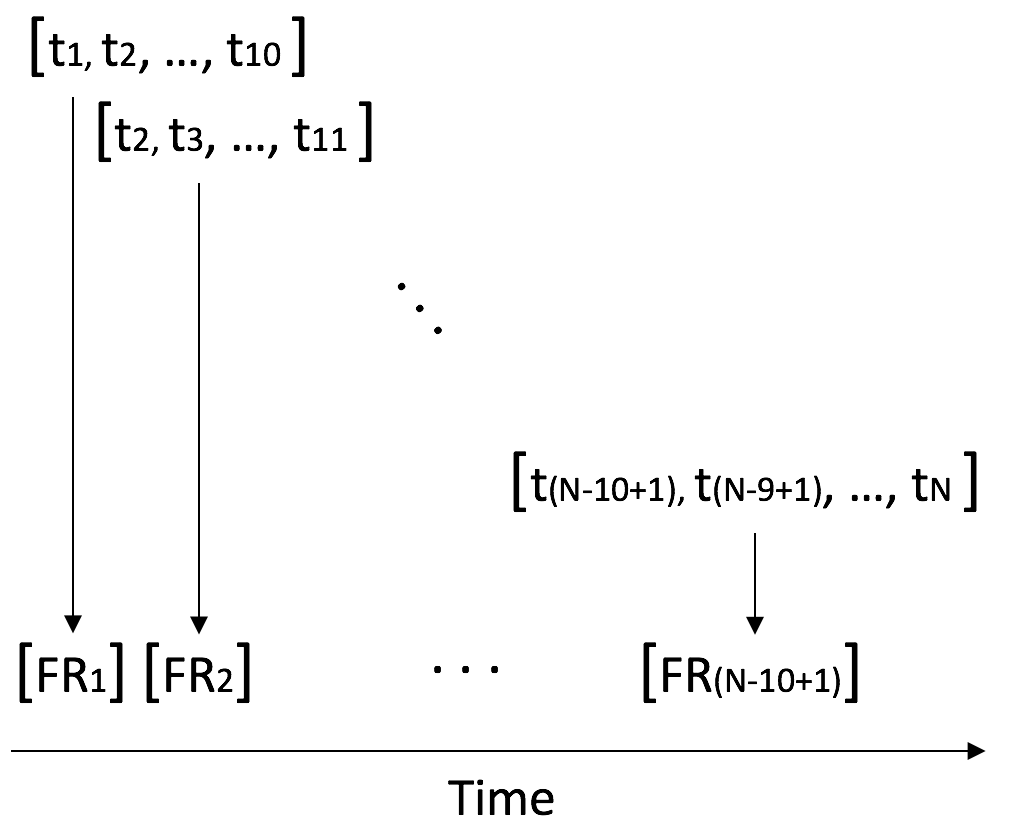}%
}\hspace{1cm}\vspace{0.3cm}
\subfloat[\textbf{Protocol II}: Feature extraction with 10 events lag]{%
  \includegraphics[width=0.6\columnwidth]{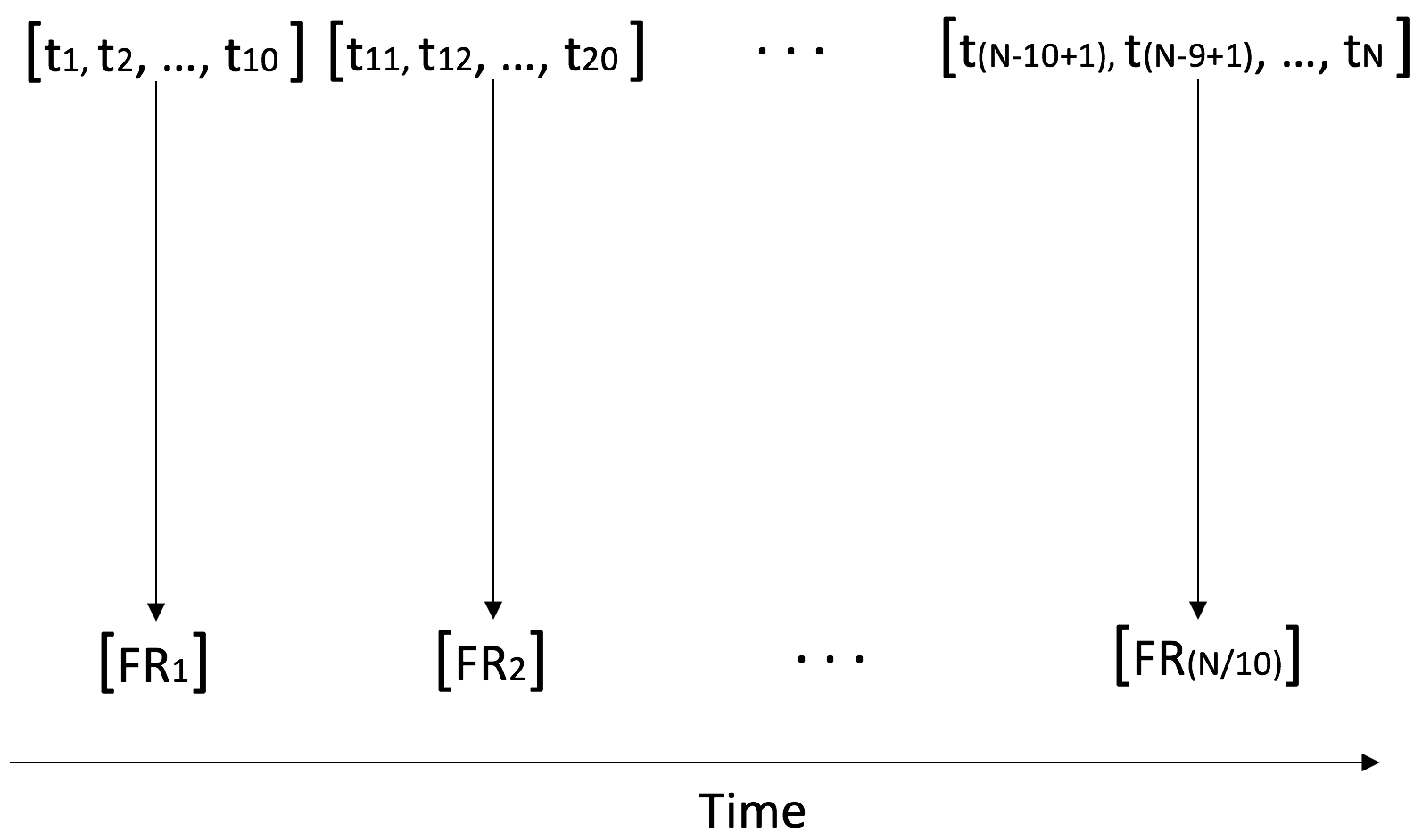}%
}
\caption{Feature extraction based on the two protocols for the task of mid-price movement prediction. For given time series ($t_1, t_2, ..., t_N$) there $N-10+1$ feature representations (FR) for Protocol I and $\frac{N}{10}$ FR for Protocol II. }
\label{fig:featextra}
\end{figure}

\begin{equation}
f1 = \frac{2 \times Recall \times Precision}{Recall + Precision},    
\end{equation}
\noindent with
\begin{equation}
Recall = \frac{TP}{TP + FN} 
\end{equation}
and 
\begin{equation}
Precision = \frac{TP}{TP + FP}
\end{equation}
\noindent where $TP$, $FN$, and $FP$ are the True-Positives, False-Negatives, and False-Positives, respectively, and RMSE is defined as:

\begin{equation}
RMSE = \sqrt{\frac{\sum_{i = 1}^{n}(P_i-O_i)^2}{n}},
\end{equation}
\noindent where $P_i$ and $O_i$ are the predicted and observed values of n samples, respectively.\\
\indent We have a labeling system that requires classification and regression. The first part of the dual labeling format contains the binary information 1 and -1 for the up and down mid-price movement, respectively. The second part of the labeling format represents the discretization of the numeric data expressed as the steps until the next mid-price change. A pictorial example of the above labeling system is in \hyperref[fig:Labels]{Fig.} \ref{fig:Labels}. The label extraction is described as follows:

\begin{enumerate}
\item $d(i) = \textbf{1}_{MP_{(i)}-MP_{(i-1)}>0}$ OR  $ - \textbf{1}_{MP_{(i)}-MP_{(i-1)}<0}$, where $i \in \mathbb{R}^{N-1}$, with $N$ be the number of the mid-prices (MP) samples,
\item $L(p) \leq d(i) < L(p+1)$, $1\leq p <Q$, where $L(p)$ is a vector that contains the bin limits in a monotonically increasing order and $Q$ is the number of bins equal to the total number of the non-zero elements in the vector of mid-price differences.    
\end{enumerate}

\subsection{Protocol II}
\noindent Protocol II is based on independent 10-event blocks for the creation of the feature representations as this can be seen in the plot (b) in  \hyperref[fig:featextra]{Fig.} \ref{fig:featextra}. More specifically, feature representations are based on the information that can be extracted from 10 events each time with these 10-event blocks independent from each other. Protocol II treats the problem of mid-price movement prediction as a three-class classification problem, with three states: up, down, and stationary condition for the mid-price movement. These changes in the mid-price are defined by means of the following calculations:

\[
l_t = 
\begin{cases}
        \hspace{0.24cm}1,  & \text{if} \frac{m_a(t)}{MP(t)}>1+\alpha \\[1em]
        -1, & \text{if} \frac{m_a(t)}{MP(t)}<1-\alpha \\[1em]
        \hspace{0.24cm}0,  & \text{otherwise}
\end{cases}
\]

\noindent where $MP(t)$ is the mid-price at time $t$, $m_a(t) = \frac{1}{r}\sum_{i = 1}^{r}MP(t+1)$ is the average of the future mid-price events with window size $r = 10$, and $\alpha$ determines the significance of the mid-price movement which is equal to $2 \times 10^{-5}$. 

\subsection{Data Normalization and Filtering}\label{sec:Normal}
\noindent The next step of the pre-processing step is data normalization. We perform a filtering and a normalization method during the feature extraction process and training. The first one is a statistical filtering method , while the second one is based on MinMax. More specifically, we perform the filtering methodology first and apply it directly on the raw MB data. The main idea of the methodology is to identify and eliminate any observation that does not reflect market activity. In the financial econometrics literature this is often referred to as data cleaning and its importance has been widely discussed in the literature (e.g., \cite{Dacorogna2001}, \cite{brownlees2006financial}, and \cite{Barndorff-Nielsen2009}).\footnote{While the advancement of technology has drastically reduced the number of outliers and misrecorded observations, their effect on the statistical analysis is still significant and the implementation of a cleaning procedure is, to this day, required to avoid biased results.} In more detail, to filter the raw data for outliers we follow a two-step procedure. We initially remove all transactions recorded outside official trading time and clearly misrecorded transactions.\footnote{These can be, for example, observations with a price equal to zero.} We then proceed by implementing a more elaborate filtering algorithm, which takes into account the statistical properties of the series to assess the validity of each observation according to its likelihood of being an outlier.\footnote{The methodology follows closely \cite{brownlees2006financial}}
More specifically, for a $k$ size window, we identify a set of (centered) neighbouring observations for each data point. To avoid including prices too distant in time, the window size $k$ should be chosen according to the trading intensity of the series. We then compute the trimmed mean of the neighboring set and mark as an outlier the considered observation if it falls more than $\alpha+\gamma$ standard deviations away from the neighbors' mean. Where $\gamma$ is a granularity parameter, which should be chosen as a multiple of the tick size. The idea behind $\gamma$ is to create a lower positive bound for the price variation. This is particularly important for the cleaning procedure as it is not uncommon to observe a sequence of equal mid prices in the LOB, which would lead to a zero variance and a consequent rejection of every price different from the mean value.
Technically, be $X_i$ the $i^{th}$ element of a time series of observations $X$, we check:
\begin{equation}\label{eq:filter}
(|X_i-\bar{X}_i(k)|<\alpha*s_i(k)+\gamma)
\end{equation}
where $s_i (k)$ and $\bar{X}_i (k)$ are respectively the sample standard deviation and the trimmed mean computed over a neighborhood of $k$ observations around $X_i$.
Hence, we identify and remove observation $X_i$ if \eqref{eq:filter} is true and keep it otherwise.
The normalization procedure is based on MinMax for the handcrafted features, as follows:
\begin{equation}
MM = \frac{X_{(i)} - X_{min}}{X_{max} - X_{min}}, i \in \mathbb{R}^N,
\end{equation}
where $N$ is the total sample size for every feature vector $\textbf{X}$ and $X_{(i)}$ is the $i^{th}$ element of $X$.

\section{Results \& Discussion}\label{SS:Results}

\begin{figure}[h!]
\centering
\subfloat[This is Protocol I, where we test the four sets of features (i.e., Econ, Tech-Quant, LOB, and fully automated), via nine deep learning models (i.e., five MLPs, two CNNs, and two LSTMs) for mid-price prediction. The mid-price prediction in this protocol is a combined prediction of when the next mid-price movement will happen and in which direction. This protocol is based on online learning architecture. We test this protocol for both US and Nordic stocks.]{%
  \includegraphics[width=0.7\columnwidth]{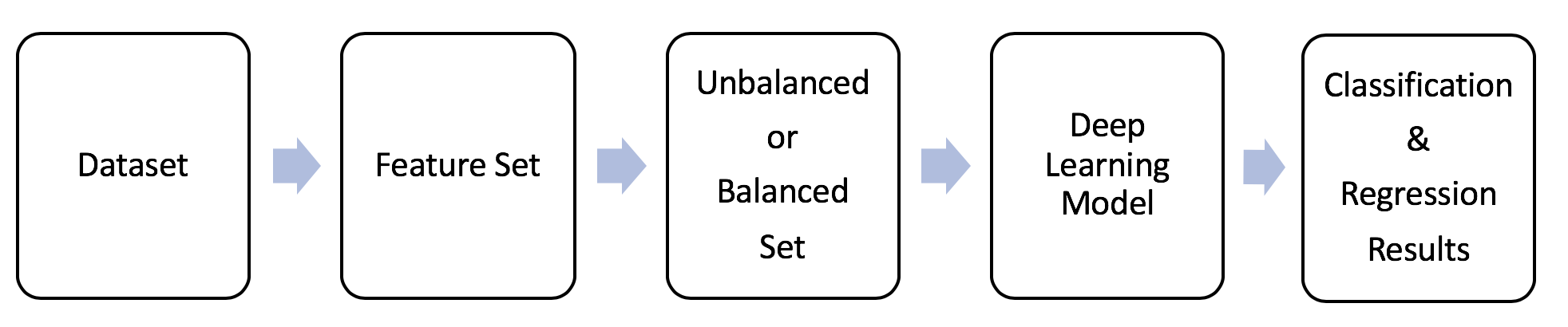}%
}\\
\subfloat[This is Protocol II, where we test the four sets of features (i.e., Econ, Tech-Quant, LOB, and fully automated), via nine deep learning models (i.e., five MLPs, two CNNs, and two LSTMs) for mid-price prediction. The mid-price prediction in this protocol is a three-class problem with states for up, down, and stationary mid-price movement. Protocol I predicts every 10th event from the current mid-price state. We test this protocol for both US and Nordic stocks.]{%
  \includegraphics[width=0.7\columnwidth]{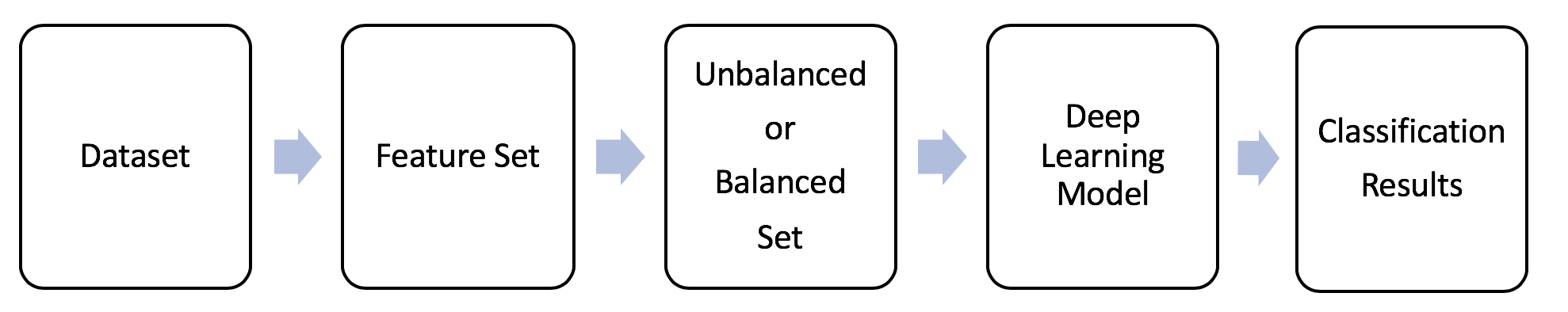}%
}\vspace{0.3cm}
\caption{Plots (a) and (b) show the process for predicting the mid-price movement based on Protocol I and Protocol II, respectively. In both protocols, the first step is the choice of dataset. The ML trader has to choose the US or Nordic stock(s) (e.g., there is the option of choosing a stock or the 'Joint' case where all the stocks from the US or Nordic markets used for training). The second step is to choose the feature set. The ML trader has to choose one of the four suggested feature sets, which are: the newly introduced econometric set, the one that is based on technical and quantitative indicators, another one based on time-sensitive and time-insensitive LOB features, and the last one based on fully automated features. The third step is whether the prediction should be based on a balanced or unbalanced set. The fourth step is the choice of one of the suggested nine deep learning models. The final step is the one that differs in Protocol I and Protocol II. The difference lies in the fact that Protocol I is a combined classification and regression optimization problem with zero event lag and Protocol II is a three-class classification problem based on a 10-event lag.}
\label{fig:Protocols}
\end{figure}

\noindent In this section, we provide results of the experiments we conducted, based on two massive LOB datasets from the US (i.e., two stocks: Amazon and Google) and Nordic (i.e., five stocks: Kesko Oyj, Outokumpu Oyj, Sampo Oyj, Rautaruukki, Wartsila Oyj) stock markets. We also discuss the perfromance of the handcrafted feature extraction universe for mid-price movement prediction and test its efficacy against a fully automated process. What is more, we make a head-to-head comparison of the three handcrafted feature sets, namely: i) \enquote{Limit Order Book (LOB):L}, based on the works of \cite{doi:10.1080/14697688.2015.1032546} and \cite{ntakaris2018benchmark}, ii) \enquote{Tech-Quant:T-Q}, based on \cite{ntakaris2018mid}, and iii) \enquote{Econ:E}, which uses econometric features. Finally, we compare these three sets of handcrafted features with features extracted based on an LSTM autoencoder. \\
\indent Latent representations are extracted after training an LSTM AE.  This training employs an extensive grid search, in which the best perfromance is reported. The grid search is based on symmetrical, assymetrical, shallow, deep, overcomplete, and undercomplete LSTM AE. The provided options vary from: i) the encoder with maximum depth up to four hidden LSTM layers with different numbers of filters varing according to the list \{128, 64, 18, 9\}, ii) the decoder with maximum depth up to four hidden LSTM layers with different numbers of filters varing according to the list \{128, 64, 18, 9\}, and iii) the latent representation with different options varying  according to the list \{5, 10, 20, 50, and 130\}. The best performance reported is based on a symmetrical and undercomplete LSTM AE of four hidden LSTM layers with 128, 64, 18, and 9 filters respectively, and 10 for the latent representation vector size. The list of the suggested grid search is limited; however, we believe it provides a wide range of combinations in order to make a fair comparison of a fully automated feature extraction process against advanced handcrafted features. We should also mention that, despite the extensive grid search on the LSTM AE, we limited our search to up to four hidden units for the encoding and deconding parts with four different filter options. Further analysis on the topic is required.\\
\indent In order to scrutinize the efficacy of the handcrafted and fully automated features, we use two experimental protocols and nine deep learning models, and present results based on unbalanced and balanced inputs. In particular, we test the four feature sets according to two protocols: the newly introduced experimental protocol (i.e., Protocol I) for online learning, as we explain in \hyperref[SS:Exper]{Section} \ref{SS:Exper}, and Protocol II, that follows \cite{tsantekidis2018using}. Protocol I is suitable for online learning, whose main objective is to predict when a change in the mid-price will happen (i.e., regression problem) and in which direction, for instance, up or down (i.e., two-class classification problem). Protocol II predicts the mid-price movement direction for every next $10^{th}$ event, where feature representations are based on independent 10-event blocks. Authors in \cite{tsantekidis2018using} used a joint training set of the five Nordic stocks for seven trading days and the next three days as testing for mid-price movement prediction (i.e., up, down, and stationary movement). We incorporate the same idea here, under the name \enquote{Joint}, and we also use the same 7-3 training and testing proportion for each stock individually for both US and Nordic datasets. A general idea for both protocols can be seen in \hyperref[fig:Protocols]{Fig.} \ref{fig:Protocols}.\\
\indent Protocol I and Protocol II use three types of deep neural networks as classifiers and regressors. In particular, we utilize five different MLPs, two CNNs, and two LSTMs. Motivation for choosing MLPs is the fact that such a simple neural network can perform extremely well when descriptive handcrafted features are used as input. The next type of neural network that we use is CNN. The first CNN, named \enquote{CNN\_1} is based on \cite{tsantekidis2017forecasting}, whereas the second one, named \enquote{CNN\_2} is based on the grid search that we describe below. The last type of neural network that we utilize is LSTM. We use two different architectures: the first one, named \enquote{LSTM\_1}, is based on \cite{tsantekidis2018using}, and the second one, named \enquote{LSTM\_2} is based on LSTM with attention mechanism. In total, we train independently nine deep neural networks for each of the two experimental protocols separately. Details of these nine topologies can be found in \hyperref[tab:DeepModels]{Table} \ref{tab:DeepModels}. \\
\indent We report results for nine different neural networks, two of which are based on existing works as shown above. For the remaining seven neural networks we conduct the following grid search:
\begin{itemize}
\item For MLPs we set a limit up to three hidden layers, where for the number of nodes we set the options \{ 4, 9, 18, 64, 128, 256, and 512\} nodes per layer and for dropout 20\% and 50\%. We report results based on five MLPs since these neural networks achieved good results for several cases (see \hyperref[sec:Results]{Section} \ref{sec:Results} for results discussion). 
\item For CNN we conduct an extensive grid search limited to up to three convolutional layers (with the option of 1-dimensional and 2-dimensional convolutional layer types) with 8, 16, and 32 filters and kernels size options \{4$\times$10, 4$\times$20, 4$\times$30, and 4$\times$40 for the 2-dimensional case and 3 and 4 for the 1-dimensional case\}. Dropout options are restricted to 20\% and 50\%. We report only one CNN since we noticed that shallower CNN architectures had very poor performance and no significant difference for the deeper ones.
\item For LSTM we follow the same approach with up to three hidden layers and five options for hidden LSTM units \{9, 18, 32, 64, 128\} and the option of attention layer. We report only one LSTM performance since all other topologies perfromed worse for our task.
\end{itemize}
\begin{table}[h!]
\centering
\captionsetup{width=.99\textwidth}
\scalebox{0.6}{
\begin{tabular}{cl}\hline
\toprule
\multirow{1}{*}{\textbf{Model}}&\multirow{1}{*}{\textbf{Topology}}\\
\midrule
\multirowcell{2}{MLP\_1}& $\bullet$ Dense layer with 4 units with Tanh activation  \\
                        & $\bullet$ Output \\ 
\midrule
\multirowcell{4}{MLP\_2}& $\bullet$ Dense layer with 512 units and Tanh activation  \\
                        & $\bullet$ 20\% Dropout \\ 
                        & $\bullet$ Dense layer with 256 units and Tanh activation\\ 
                        & $\bullet$ Output\\ 
\midrule
\multirowcell{4}{MLP\_3}& $\bullet$ Dense layer with 256 units and Tanh activation  \\
                        & $\bullet$ 20\% Dropout \\ 
                        & $\bullet$ Dense layer with 256 units and Tanh activation \\ 
                        & $\bullet$ Output\\ 
\midrule
\multirowcell{6}{MLP\_4}& $\bullet$ Dense layer with 256 units and Tanh activation  \\
                        & $\bullet$ 20\% Dropout \\ 
                        & $\bullet$ Dense layer with 256 units and Tanh activation\\ 
                        & $\bullet$ 20\% Dropout\\ 
                        & $\bullet$ Dense layer with 256 units and Tanh activation\\
                        & $\bullet$ Output\\ 
\midrule
\multirowcell{6}{MLP\_5}& $\bullet$ Dense layer with 128 units and Tanh activation  \\
                        & $\bullet$ 20\% Dropout \\ 
                        & $\bullet$ Dense layer with 128 units and Tanh activation \\
                        & $\bullet$ 20\% Dropout \\
                        & $\bullet$ Dense layer with 128 units and Tanh activation \\
                        & $\bullet$ Output \\ 
\midrule
\midrule
\multirowcell{1}{CNN\_1}& $\bullet$ 2D Convolution layer with 16 filters, 4 x 40 kernel size \\
                        & $\bullet$ 1D Convolution layer with 16 filters, 4 as kernel size \\
                        & $\bullet$ Maxpooling size of 2 \\
                        & $\bullet$ 1D Convolution layer with 32 filters, 3 as kernel size \\
                        & $\bullet$ 1D Convolution layer with 32 filters, 3 as kernel size \\
                        & $\bullet$ Maxpooling size of 2 \\
                        & $\bullet$ Dense layer with 32 neurons \\
                        & $\bullet$ Output \\
\midrule
\multirowcell{16}{CNN\_2}& $\bullet$ 2D Convolution layer with 8 filters, 4 x 40 kernel size,                            2 x 2 stride size and same output size \\
                        & $\bullet$ BatchNormalization \\
                        & $\bullet$ LeakyReLU with 0.1 slope \\
                        & $\bullet$ 2 x 2 MaxPooling with the same output size \\
                        & $\bullet$ 50\% Dropout \\
                        & $\bullet$ 1D Convolution layer with 32 filters, 3 as kernel size, 5 as stride and same output size \\
                        & $\bullet$ BatchNormalization \\
                        & $\bullet$ LeakyReLU with 0.1 slope \\
                        & $\bullet$ Maxpooling size of 2 with the same output size \\
                        & $\bullet$ 50\% Dropout \\
                        & $\bullet$ 1D Convolution layer with 32 filters, 3 as kernel size, 5 as stride and same output size \\
                        & $\bullet$  LeakyReLU with 0.1 slope \\
                        & $\bullet$  Maxpooling size of 2 with the same output size \\
                        & $\bullet$  50\% Dropout \\
                        & $\bullet$  Dense layer with 8 units \\
                        & $\bullet$  Output \\
\midrule
\midrule
\multirowcell{1}{LSTM\_1}& $\bullet$ LSTM layer with 32 units \\
                         & $\bullet$ Dropout \\
                         & $\bullet$ PReLU \\
                         & $\bullet$ Dense layer with 64 units \\
                         & $\bullet$ Output \\
\midrule
\multirowcell{1}{LSTM\_2}& $\bullet$ LSTM layer with 40 units \\
                         & $\bullet$ PReLU \\
                         & $\bullet$ Attention layer \\
                         & $\bullet$ Dense layer with 40 units \\
                         & $\bullet$ Output \\
\bottomrule 
\end{tabular}}
\caption[caption]{List of the nine deep learning models that are used for the two experimental protocols. Output, in the neural networks above, means that for Protocol I the output is a dense layer with 1 unit and linear activation function for the regression task and a dense layer with two units and softmax activation function. for Protocol II, the output is a dense layer with three units and softmax activation function.}
\label{tab:DeepModels}
\end{table}
\indent The training of these nine neural networks takes place at CSC super-cluster where we use Pascal P100 and K80 GPUs. We use multi-GPUs, under Keras (i.e., \cite{chollet2015keras}) framework, in order to reduce the training time. The models, apart from CNN\_1 and LSTM\_1, use the Nesterov-Adam optimizer with a learning rate of 0.002, with mean squared error and binary cross-entropy for the dual output of Protocol I where this dual output is weighted by 0.01 and 0.99, respectively, and categorical cross-entropy as loss function for Protocol II. Additionally, we use 250 epochs to train our models with data shuffling and validation ratio of 0.2. Finally, in order to control overfitting we utilize Dropout to the majority of the suggested neural networks. By dropping out some nodes (i.e., a dropped out node have a zero output) from neural network topologies we control node dependencies and we achieve more robust results.

\subsection{Results}\label{sec:Results}
\begin{figure}[h!]
  \centering
  \subfloat[F1 scores based on the unbalanced (top) and balanced (bottom) sets]{\includegraphics[scale=0.2]{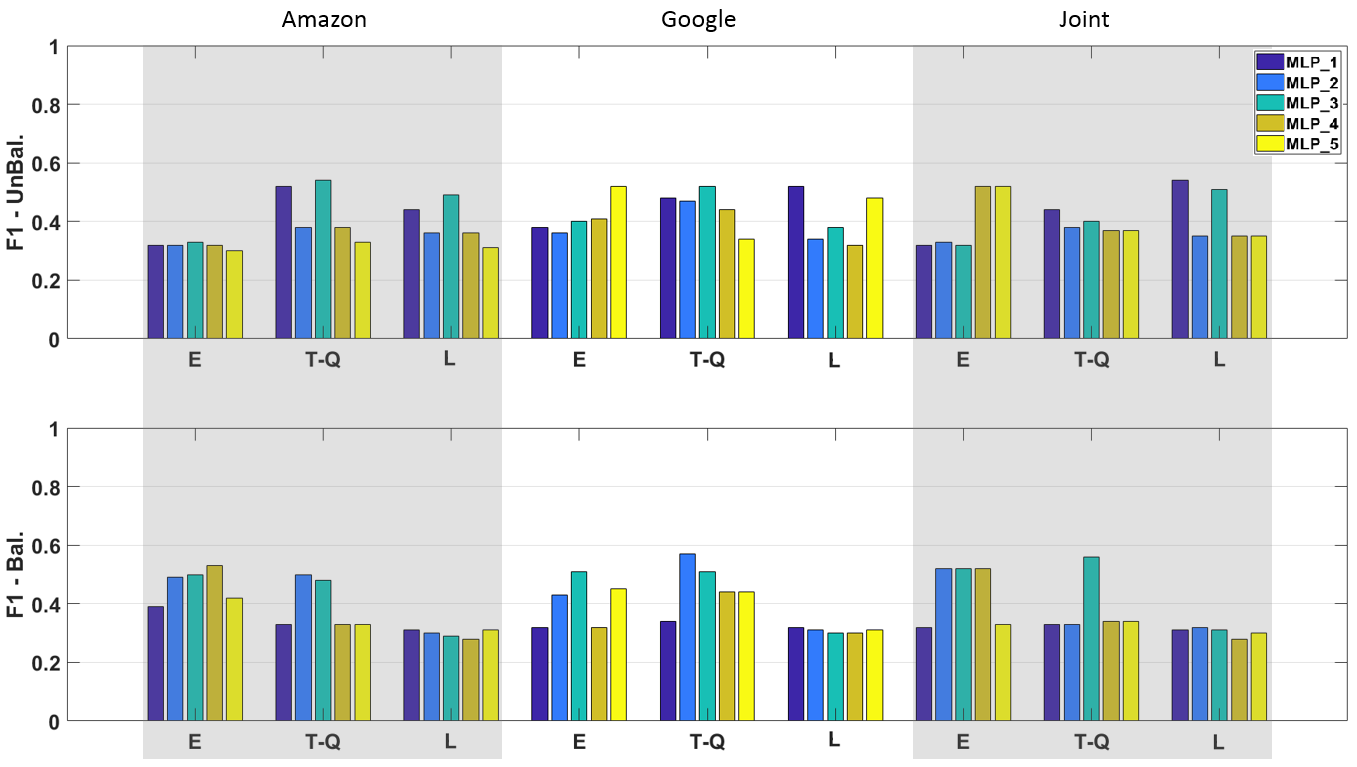}}\quad
  \subfloat[RMSE scores based on the unbalanced (top) and balanced (bottom) sets]{\includegraphics[scale=0.20]{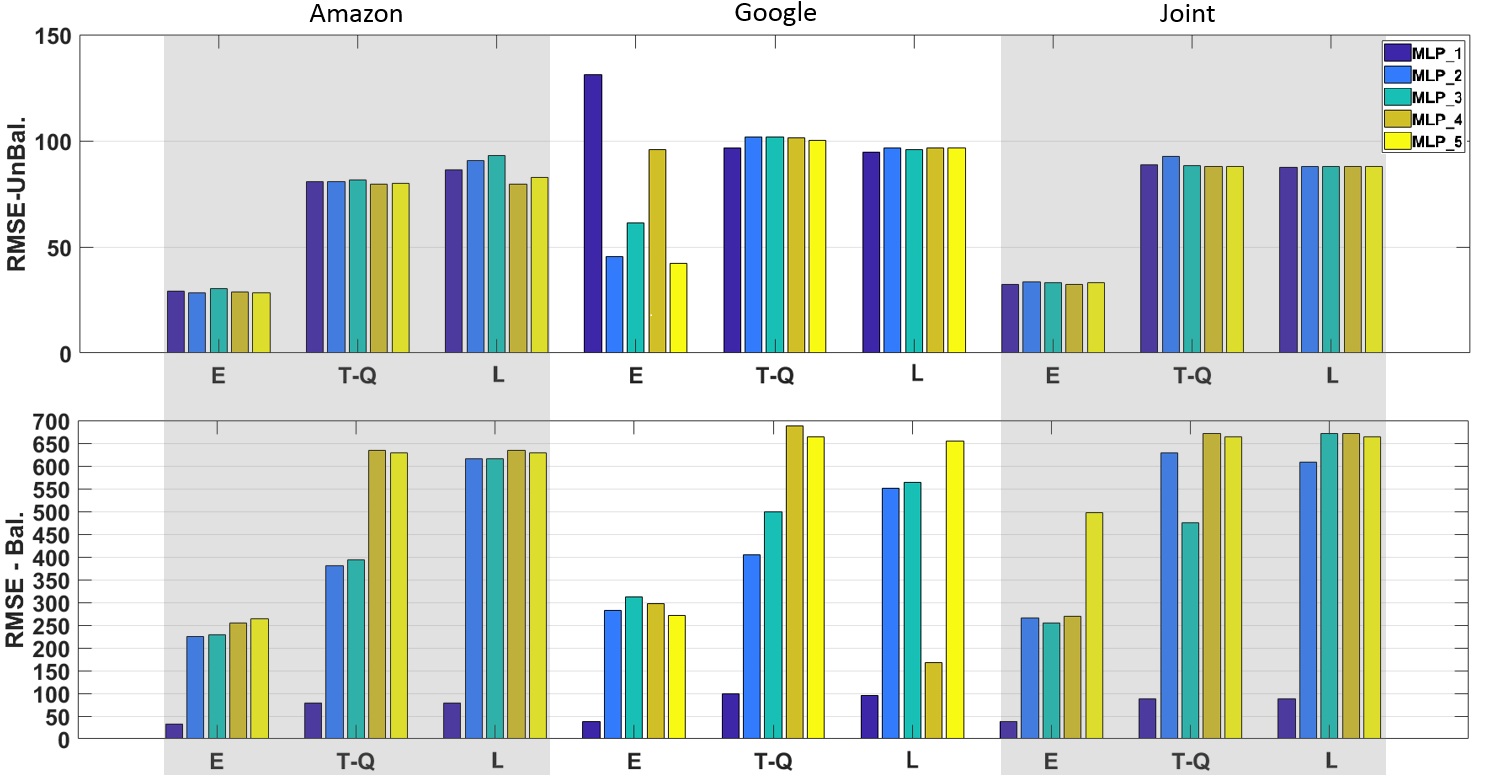}}\\
\subfloat[F1 scores based on the unbalanced (top) and balanced (bottom) sets]{\includegraphics[scale=0.20]{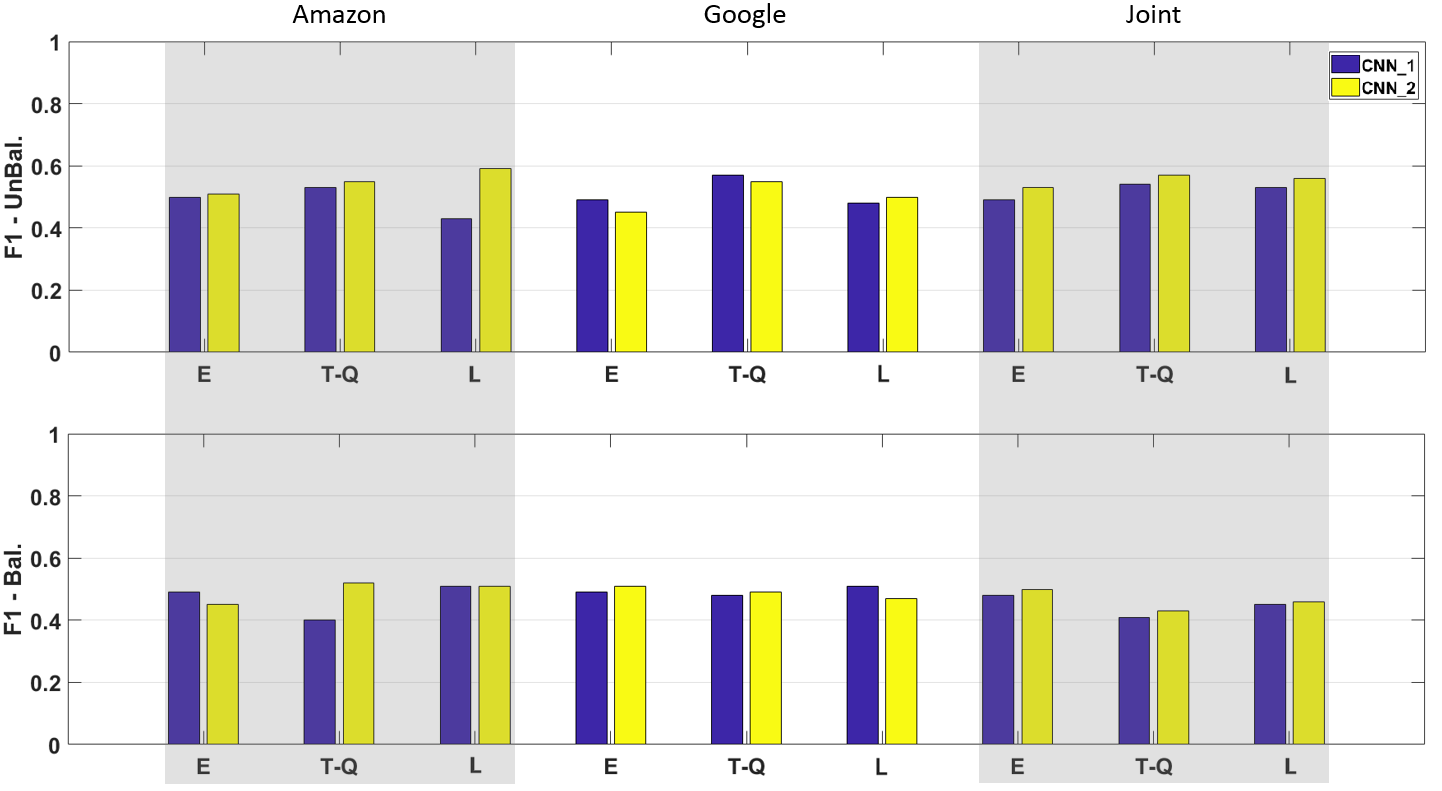}}\quad
  \subfloat[RMSE scores based on the unbalanced (top) and balanced (bottom) sets]{\includegraphics[scale=0.20]{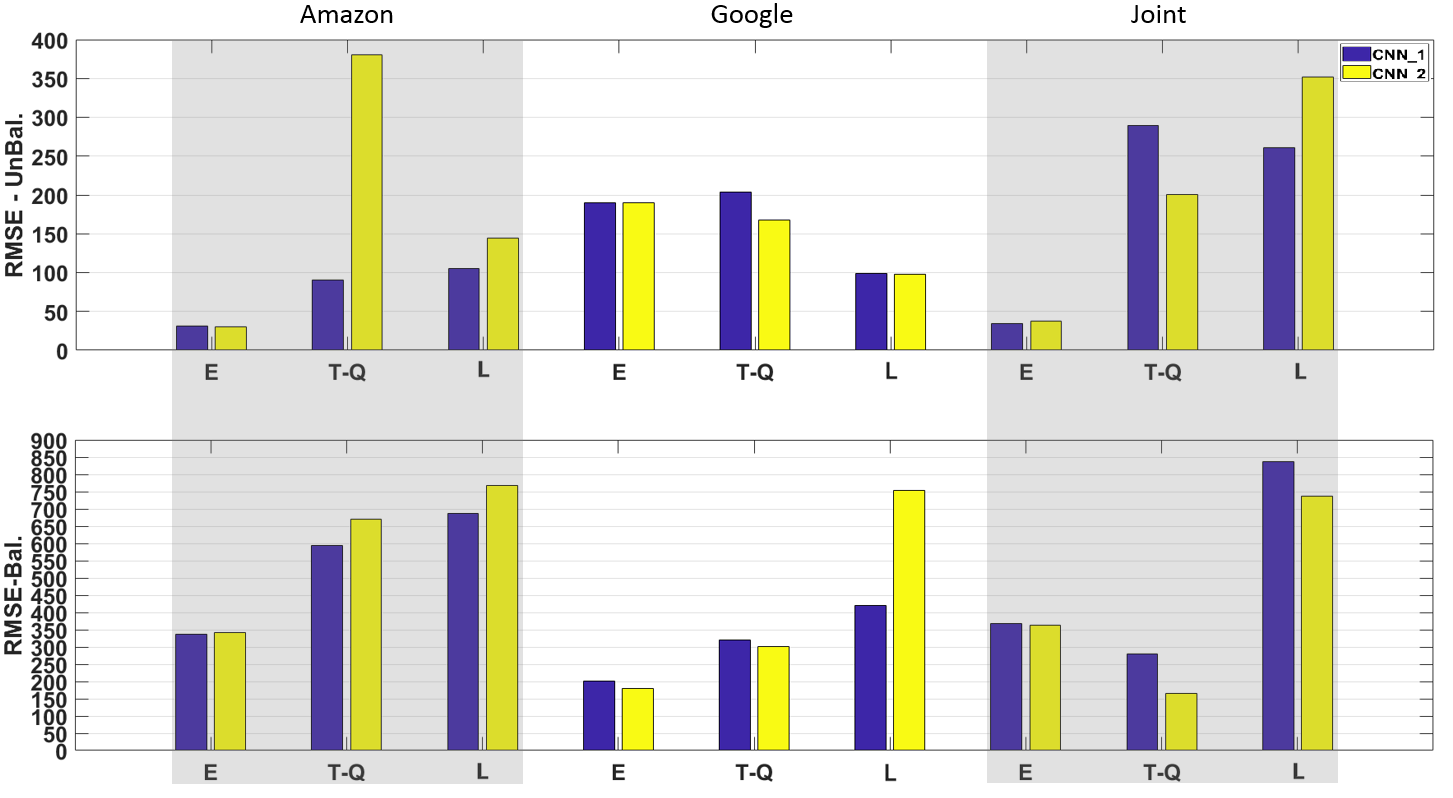}}\\
\subfloat[F1 scores based on the unbalanced (top) and balanced (bottom) sets]{\includegraphics[scale=0.20]{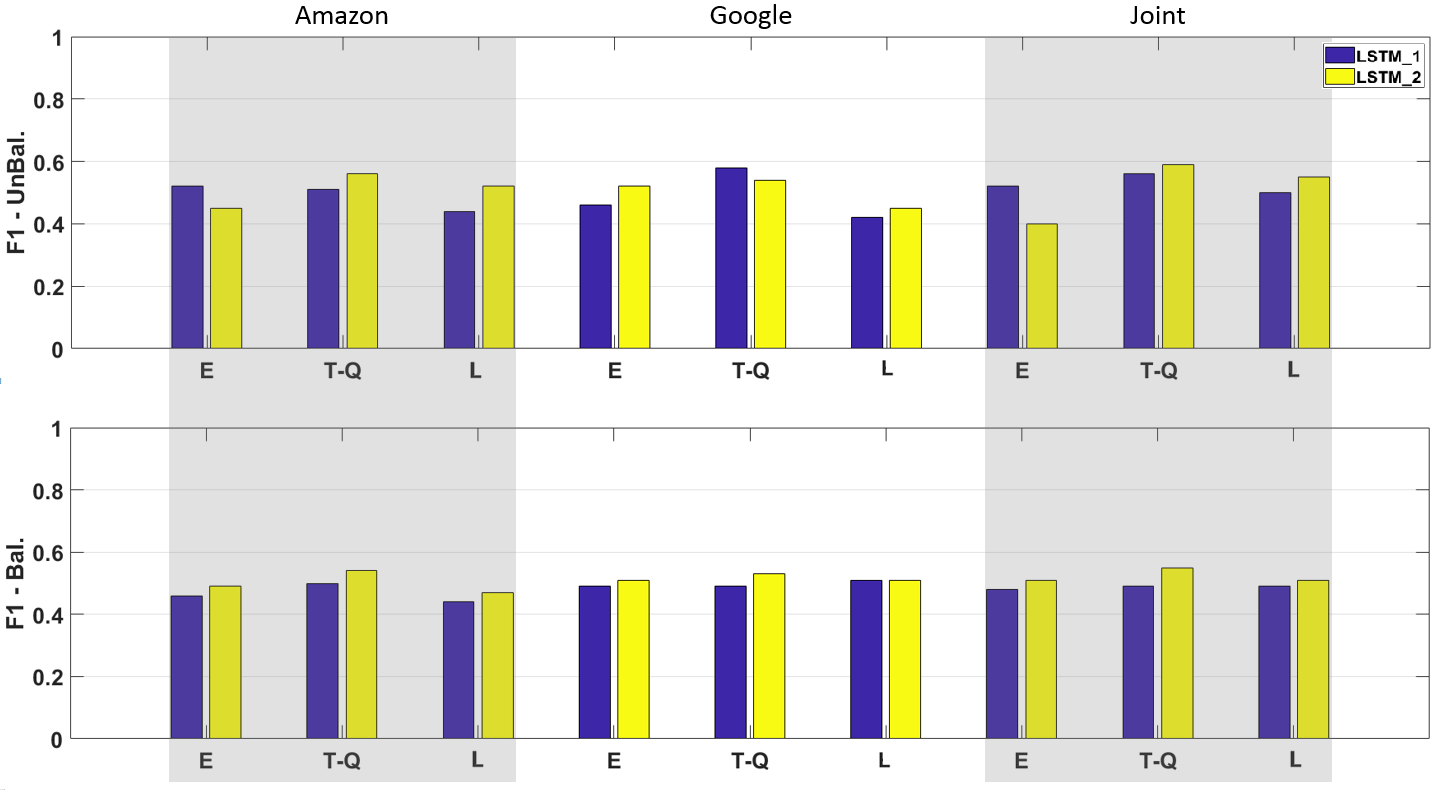}}\quad
  \subfloat[RMSE scores based on the unbalanced (top) and balanced (bottom) sets]{\includegraphics[scale=0.20]{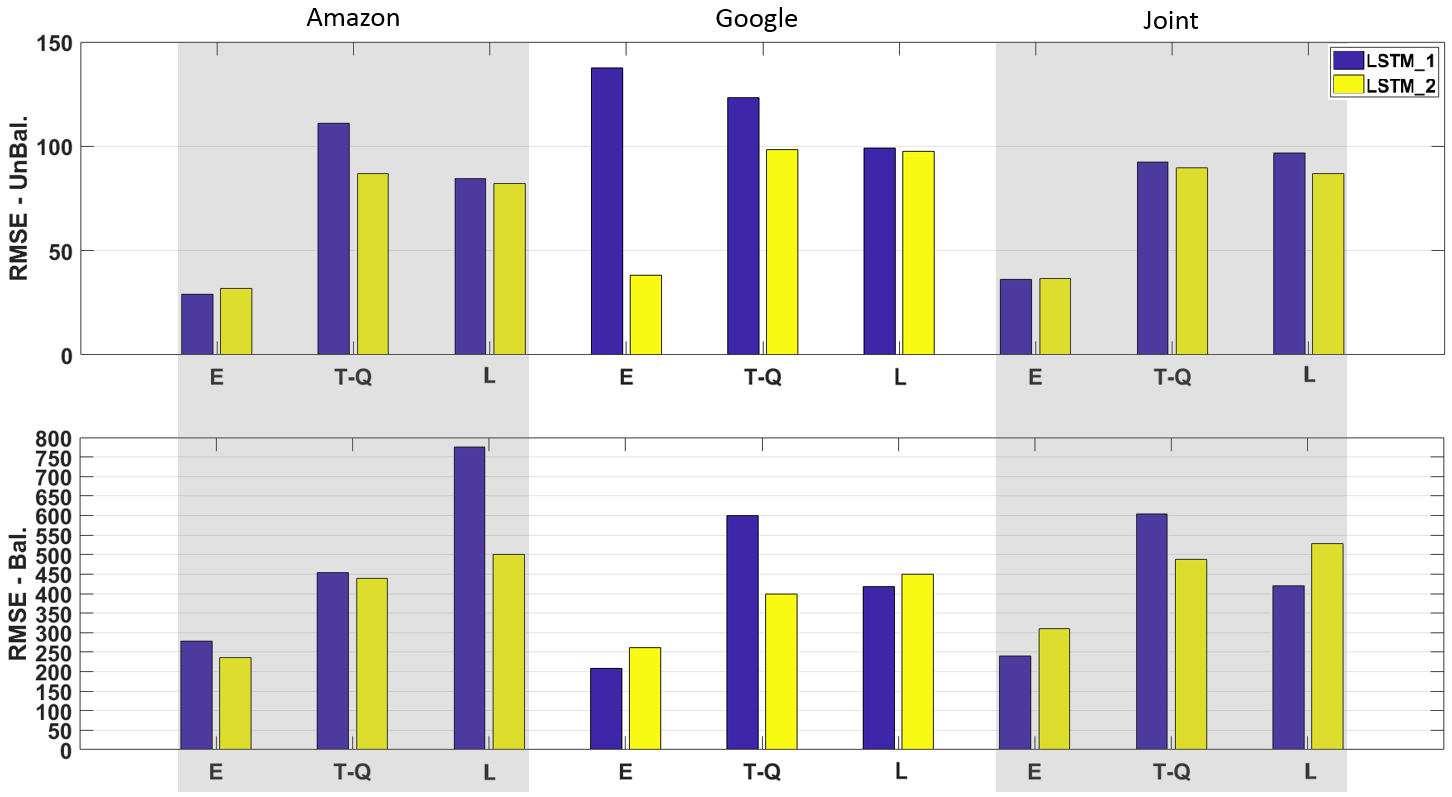}}
\caption{F1 (left column plots) and RMSE (right column plots) scores for the nine deep learning models based on the US data}
\label{fig:US}
\end{figure}

\begin{figure}[h!]
  \centering
  \subfloat[F1 scores based on the unbalanced (top) and balanced (bottom) sets]{\includegraphics[scale=0.2030]{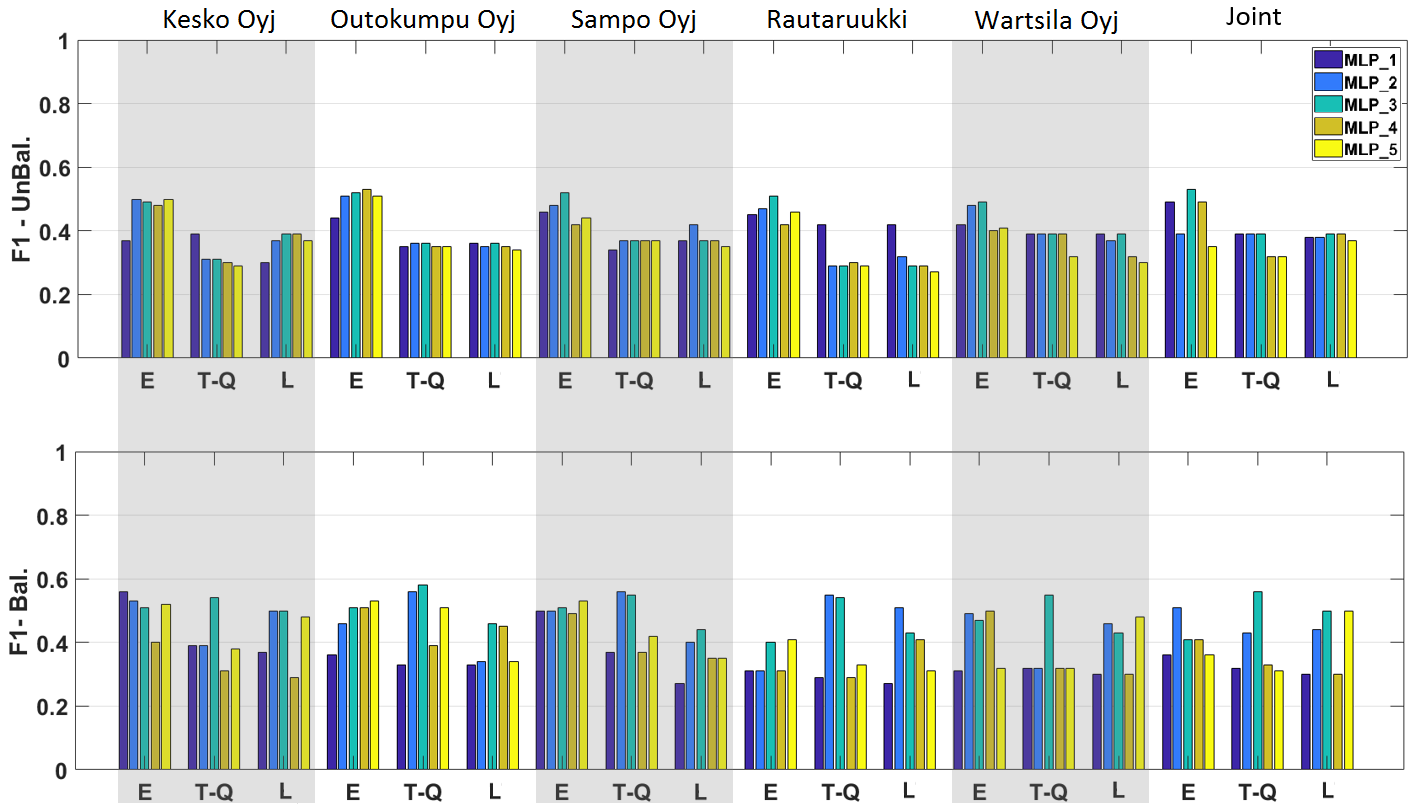}}\quad
  \subfloat[RMSE scores based on the unbalanced (top) and balanced (bottom) sets]{\includegraphics[scale=0.2045]{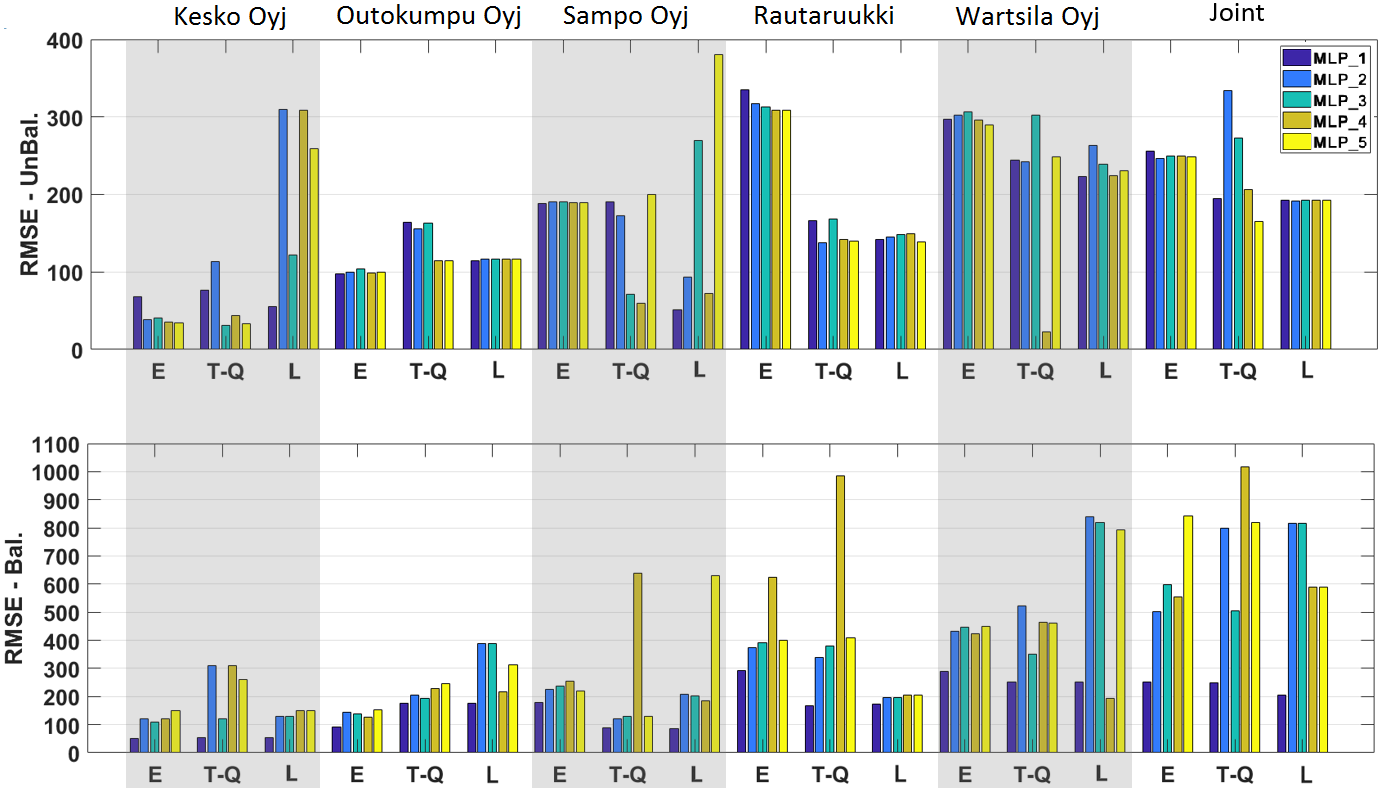}}\\
\subfloat[F1 scores based on the unbalanced (top) and balanced (bottom) sets]{\includegraphics[scale=0.208]{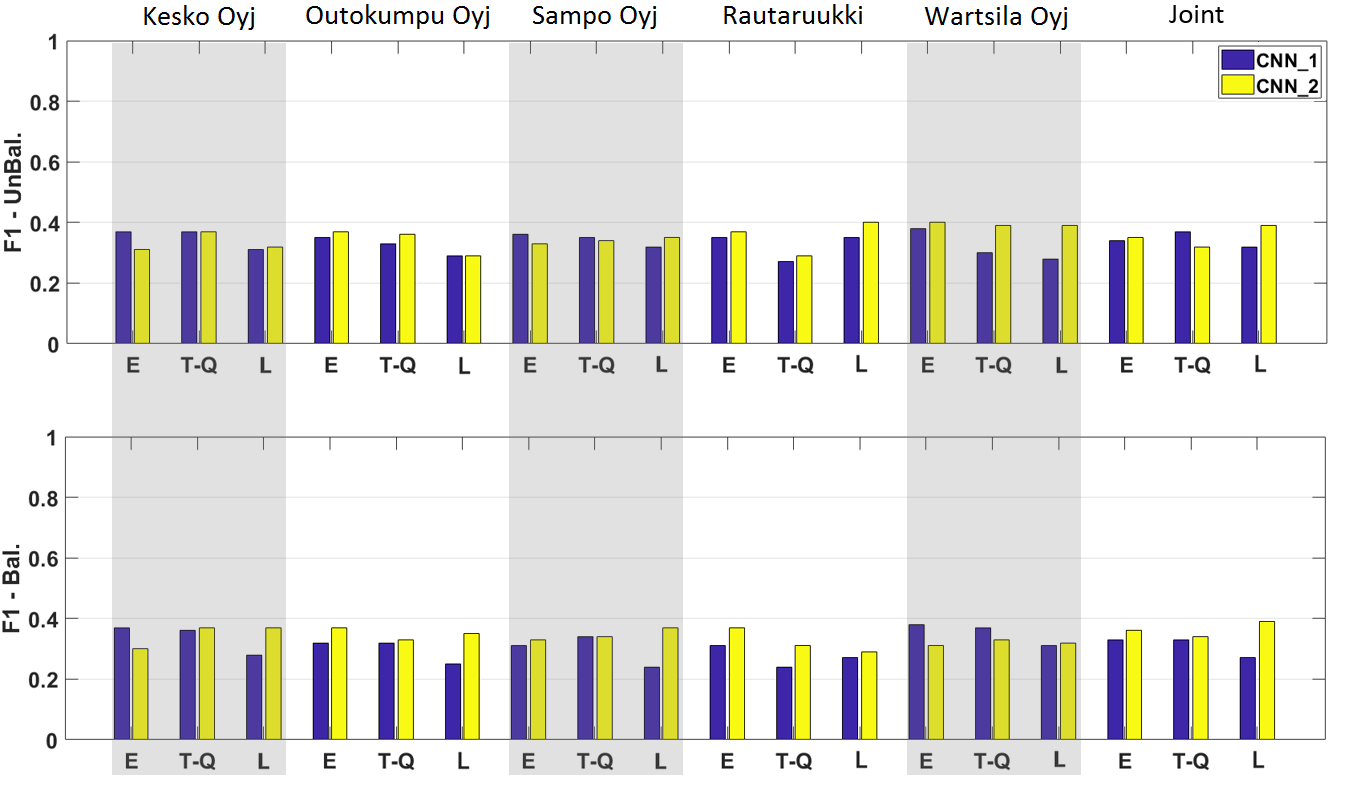}}\quad
  \subfloat[RMSE scores based on the unbalanced (top) and balanced (bottom) sets]{\includegraphics[scale=0.2045]{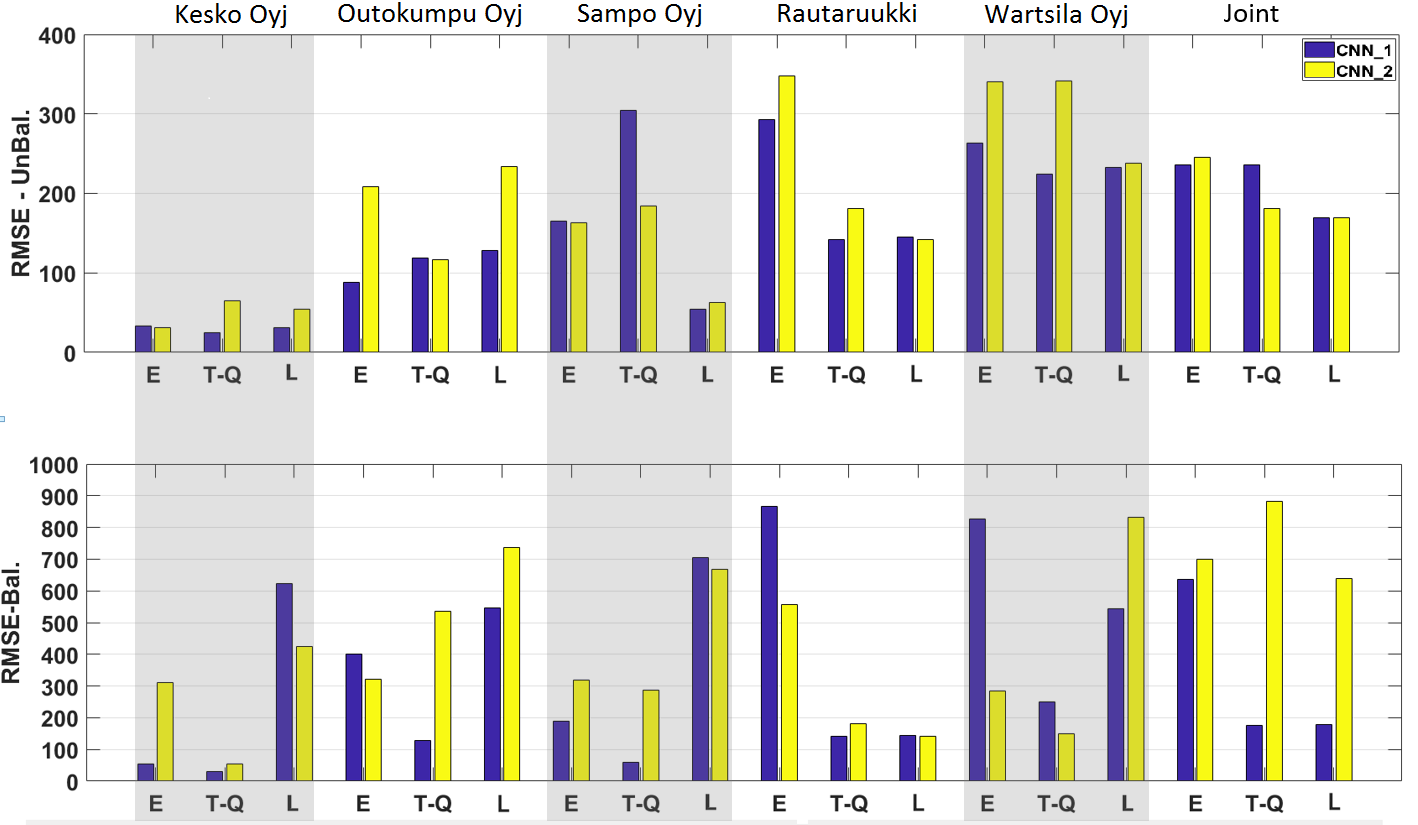}}\\
 \subfloat[F1 scores based on the unbalanced (top) and balanced (bottom) sets]{\includegraphics[scale=0.2045]{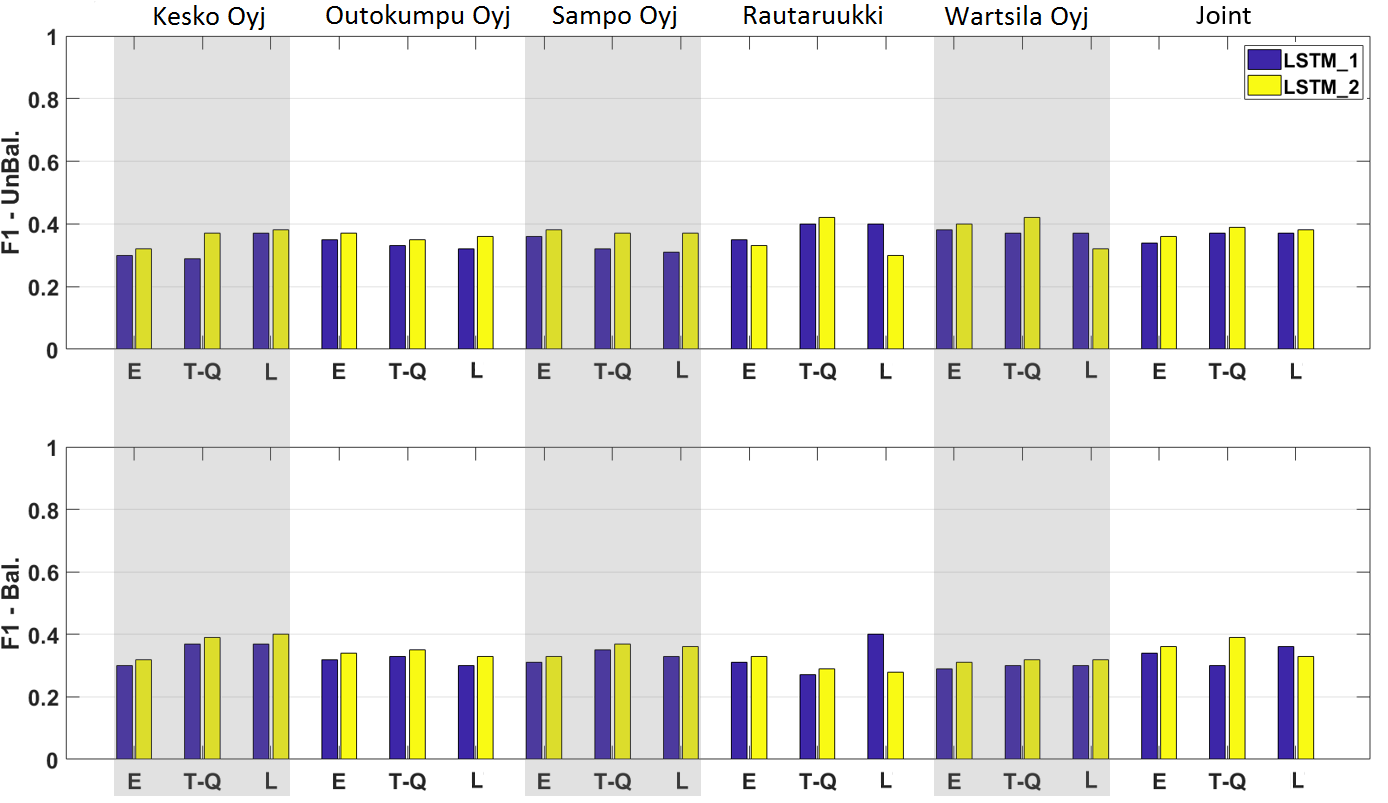}}\quad
  \subfloat[RMSE scores based on the unbalanced (top) and balanced (bottom) sets]{\includegraphics[scale=0.2045]{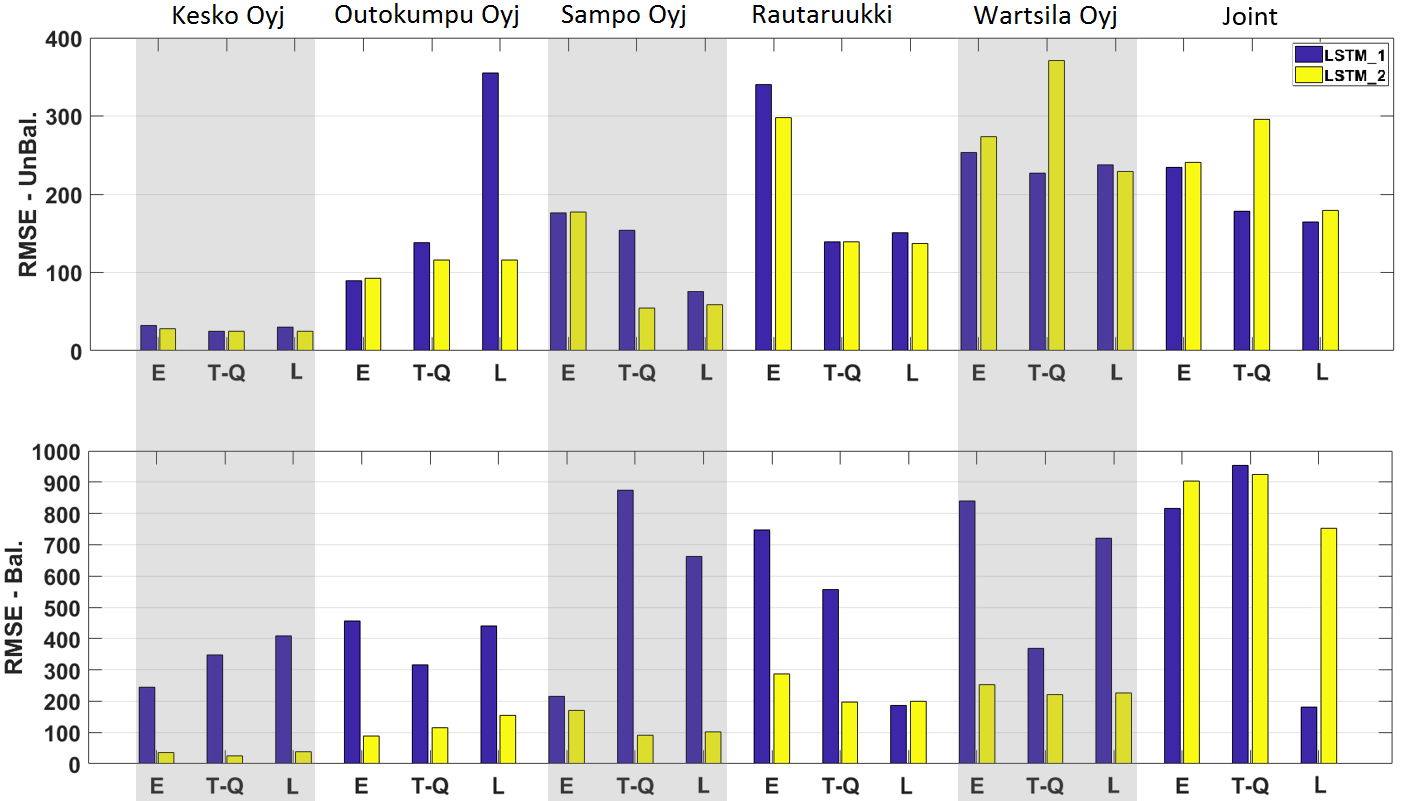}}
\caption{F1 (left column plots) and RMSE (right column plots) scores for the nine deep learning models based on the Nordic data}
\label{fig:Nordic}
\end{figure}

\noindent We present our results in separate tables for Protocol I (see \hyperref[tab:USF1]{Table} \ref{tab:USF1} - \hyperref[tab:NORDICRMSEAE]{Table} \ref{tab:NORDICRMSEAE}) and Protocol II (see \hyperref[SS:ProI]{Appendix} \ref{SS:ProI}). For each protocol, we split the results (i.e., f1 score and RMSE for Protocol I and f1 scores for Protocol II) for both US and Nordic datasets. We would like to mention that results derived from the LSTM AE, for both f1 and RMSE scores, are presented in seperate tables (see \hyperref[tab:USF1AE]{Tables} \ref{tab:USF1AE} \&\hyperref[tab:NORDICRMSEAE]{} \ref{tab:NORDICRMSEAE} for Protocol I and \hyperref[tab:US2]{Tables} \ref{tab:US2} \&\hyperref[tab:NORDIC2]{} \ref{tab:NORDIC2} for Protocol II). Since handcrafted feature results overperformed the fully automated feature set we empasize more on their perfromance by providing tables together with bar plots (see \hyperref[fig:US]{Fig.} \ref{fig:US} \&\hyperref[fig:Nordic]{} \ref{fig:Nordic}).  
Each of the tables contains the full head-to-head comparison for the three handcrafted features sets for each of the nine different deep learning models separately. For instance, \hyperref[tab:NordicF1]{Table} \ref{tab:NordicF1} contains f1 scores for the Nordic stocks based on Protocol I. The table has five main columns (i.e., Model, Stock, Econ, Tech-Quant, and LOB) and six subcolumns divided into three pairs (i.e., UnBal. and Bal.). The first main column contains the nine deep neural networks; the second main column contains the five independent and different Nordic stocks, in which the sixth row for every model is the joint training set based on these five stocks; and the third, fourth and fifth main columns represent the three handcrafted feature sets. Moreover, for every feature set, we present results for unbalanced and balanced cases, whereas for the balanced cases we use random undersampling for the majority class. Even though balanced datasets do not project a realistic trading scenario (i.e., trading fees are not applicable), it is important to give an equal opportunity to the minority class, which can be an ML trader's trading position. More specifically, for Protocol I and the classification task, the Nordic dataset has 45\% for the downward movement and 55\% for the upward, while for the US dataset is 47\% for the downward movement and 53\% for the upward. The undersampling offers an 85\% data reduction for the Nordic set and 90\% for the US set. For better interpretation of Protocol I we provide bar plots which show the reaction of every deep learning model and dataset for the unbalanced and balanced cases (see \hyperref[fig:US]{Fig.} \ref{fig:US} and \hyperref[fig:Nordic]{Fig.} \ref{fig:Nordic}). Protocol II and the Nordic dataset exhibits a 75\% for the stationary condition, with the remaining 25\% being equally divided to the upward and downward mid-price movement before undersampling. For the US dataset 73\% belongs to the stationary condition, 20\% to the upward movement and the remaining 7\% to the downward movement. The undersampling offers a 30\% data reduction for the Nordic dataset and 10\% data reduction for the US dataset.

\begin{table*}[htb!]
\begin{minipage}{0.5\linewidth}
\centering
\scalebox{0.7}{
        \begin{tabular}{crcccccc}\hline
        \multirow{4}{*}{Model}&\multirow{4}{*}{Stock}& \multicolumn{2}{c}{Econ} & \multicolumn{2}{c}{Tech-Quant}  & \multicolumn{2}{c}{LOB}\\
        \cmidrule(lr){3-4} \cmidrule(lr){5-6} \cmidrule(lr){7-8}
         & & UnBal. & Bal. & UnBal. & Bal.& UnBal. & Bal. \\
        \midrule
\multirowcell{3}{MLP\_1}&{Amazon} & 0.32& 0.39 &0.52 &0.33 &0.44 &0.31 \\
                        &{Google}& 0.38& 0.32 &0.48 &0.34 &0.52 &0.32 \\  
                        &{Joint}& 0.32&0.32 &0.44 &0.33 &0.54 &0.31\\ 
\midrule
\multirowcell{3}{MLP\_2}&{Amazon} & 0.32&0.49 &0.38 &0.50 &0.36 &0.30 \\
                        &{Google}& 0.36&0.43 &0.47 &\underline{\textbf{0.57}} &0.34 &0.31 \\ 
                        &{Joint}& 0.33&0.52 &0.38 &0.33 &0.35 &0.32\\ 
\midrule
\multirowcell{3}{MLP\_3}&{Amazon} & 0.33&0.50 &0.54 &0.48 &0.49 &0.29 \\
                        &{Google}& 0.40& 0.51 &0.52 &0.51 &0.38 &0.30 \\ 
                        &{Joint}& 0.32& 0.52 &0.40 &\underline{\textbf{0.56}} &0.51 &0.31\\ 
\midrule
\multirowcell{3}{MLP\_4}&{Amazon} & 0.32 &0.53 &0.38 &0.33 &0.36&0.28 \\
                        &{Google}& 0.51 &0.32 &0.44 &0.44 &0.32 &0.30 \\ 
                        &{Joint}& 0.52&0.52 &0.37 &0.34 &0.35 &0.28\\ 
\midrule
\multirowcell{3}{MLP\_5}&{Amazon} & 0.30&0.42 &0.33 &0.33 &0.31 &0.31 \\
                        &{Google}& 0.43&0.45 &0.34 &0.44 &0.48 &0.31 \\  
                        &{Joint}& 0.52&0.33 &0.37 &0.34 &0.35&0.30\\ 
\midrule
\midrule
\multirowcell{3}{CNN\_1}&{Amazon} & 0.50&0.49 &0.53 &0.40 &0.43 &0.51 \\
                        &{Google}& 0.49&0.49 &0.57 &0.48 &0.48 &0.51 \\ 
                        &{Joint}& 0.49& 0.48 &0.54 &0.41 &0.53 &0.45\\ 
\midrule
\multirowcell{3}{CNN\_2}&{Amazon} & 0.51 &0.45 &0.55 &0.52 &0.56 &0.51 \\
                        &{Google}& 0.45 &0.51 &0.55 &0.49 &0.50 &0.47 \\ 
                        &{Joint}& 0.53 & 0.50 & 0.57 &0.43 &0.56 &0.46\\ 
\midrule
\midrule
\multirowcell{3}{LSTM\_1}&{Amazon} & 0.52&0.46 &0.51 &0.50 &0.44 &0.44 \\
                         &{Google}& 0.46&0.49 &\underline{\textbf{0.58}} &0.49 &0.42 &0.51 \\ 
                         &{Joint}& 0.52&0.48 &0.56 &0.49 &0.50 &0.49\\ 
\midrule
\multirowcell{3}{LSTM\_2}&{Amazon} & 0.45&0.49 &0.56 &0.54 &0.52 &0.47 \\ 
                         &{Google}& 0.52&0.51 &0.54 &0.53 &0.45 & 0.51 \\ 
                         &{Joint}& 0.40&0.51 &\underline{\textbf{0.59}} &0.55 &0.55 &0.51\\
\bottomrule  
        \end{tabular}}
        \caption{Protocol I: f1 scores for the US stocks. \\{\tiny \textit{Note: Highlighted text shows the best f1 performance for: \\1) Joint/Unbalanced, 2) Joint/Balanced, \\ 3) Stock-Specific/Unbalanced, and \\4) Stock-Specific/Balanced cases\\}}}
        \label{tab:USF1}
\end{minipage}
\hspace{0.001cm}
\begin{minipage}{0.5\linewidth}
\centering
\scalebox{0.7}{
        \begin{tabular}{crcccccc}\hline
        \multirow{4}{*}{Model}&\multirow{4}{*}{Stock}& \multicolumn{2}{c}{Econ} & \multicolumn{2}{c}{Tech-Quant}  & \multicolumn{2}{c}{LOB}\\
        \cmidrule(lr){3-4} \cmidrule(lr){5-6} \cmidrule(lr){7-8}
         & & UnBal. & Bal. & UnBal. & Bal.& UnBal. & Bal. \\
        \midrule
\multirowcell{3}{MLP\_1}&{Amazon} & 28.99&\underline{\textbf{33.02}} &80.64 &79.63 &86.22 &79.61 \\
                        &{Google}& 131.18&37.53 &96.75 &98.25 &94.59 &96.11 \\  
                        &{Joint}& 32.32&38.43 &88.88 &87.70 &87.49 &87.72\\ 
\midrule
\multirowcell{3}{MLP\_2}&{Amazon} & \underline{\textbf{28.39}}&224.91 &80.67 &381.22 &90.65 &603.53 \\
                        &{Google}& 45.57&282.40 &101.91 &404.47 &96.51 &550.23\\ 
                        &{Joint}& 33.34&266.32 &92.79 &628.33 &87.86 &608.31\\ 
\midrule
\multirowcell{3}{MLP\_3}&{Amazon} & 30.44&227.90 &81.37 &393.43 &793.08&615.51 \\
                        &{Google}& 61.28&312.42 &101.74 &498.35 &95.90 &563.60 \\ 
                        &{Joint}& 33.31&255.48 &88.30 &474.82 &87.78 &671.49\\ 
\midrule
\multirowcell{3}{MLP\_4}&{Amazon} & 28.61&253.89 &79.60 &634.33 &79.60 &634.31 \\
                        &{Google}& 95.73&296.81 &101.54 &688.81 &96.43 &167.81 \\ 
                        &{Joint}& \underline{\textbf{32.29}}&269.86 &87.71 &671.49 &\underline{\textbf{87.70}} &671.67\\ 
\midrule
\multirowcell{3}{MLP\_5}&{Amazon} & 28.40&264.38 &79.78 &628.99 &82.89 &629.01 \\
                        &{Google}& 42.07&271.69 &100.01 &664.08 &96.47 & 654.19\\  
                        &{Joint}& 33.22&496.80 &87.73 &663.93 &87.70 &663.89\\
\midrule                                                
\midrule
\multirowcell{3}{CNN\_1}&{Amazon} & 31.21&337.89 &90.35 &592.95 &104.90 &686.93 \\
                        &{Google}& 190.19&201.45 &203.65 &319.34 &99.39 &421.14 \\ 
                        &{Joint}& 34.85&367.70 &289.76 &278.90 &260.46 &836.35\\ 
\midrule
\multirowcell{3}{CNN\_2}&{Amazon} & 30.47&342.41 &380.90 &671.19 &144.22 &767.75 \\
                        &{Google}& 189.87&178.98 &167.89 &302.33 &97.63&753.54 \\ 
                        &{Joint}& 37.46&362.68 &200.67 &165.23 &352.20 &737.46\\
\midrule                                                
\midrule
\multirowcell{3}{LSTM\_1}&{Amazon} & 29.08&277.94 &110.86 &454.85 &84.64 &774.62 \\
                        &{Google}& 137.65&207.48 &123.36 &599.94 &99.13 &418.62\\ 
                        &{Joint}& 36.05&240.14 &92.58 &604.54 &96.58 &421.04\\ 
\midrule
\multirowcell{3}{LSTM\_2}&{Amazon} & 32.03&235.14 &86.81 &440.10 &82.03 &500.47 \\ 
                         &{Google}& 38.37&262.28 &98.38 &398.17 &97.45 &449.66 \\ 
                         &{Joint}& 36.65&309.80&89.69 &487.73 &86.95 &527.83\\ 
\bottomrule  
        \end{tabular}}
        \caption{Protocol I: RMSE scores for the US stocks. \\{\tiny \textit{Note: Highlighted text shows the best RMSE performance for: \\1) Joint/Unbalanced, 2) Joint/Balanced, \\ 3) Stock-Specific/Unbalanced, and \\4) Stock-Specific/Balanced cases\\}}}
        \label{tab:USRMSE}
\end{minipage}
\end{table*}

\begin{table}[htb!]
\begin{minipage}{0.5\linewidth}
\centering
\scalebox{0.6}{
        \begin{tabular}{crcccccc}\hline
        \multirow{4}{*}{Model}&\multirow{4}{*}{Stock}& \multicolumn{2}{c}{Econ} & \multicolumn{2}{c}{Tech-Quant}  & \multicolumn{2}{c}{LOB}\\
        \cmidrule(lr){3-4} \cmidrule(lr){5-6} \cmidrule(lr){7-8}
         & & UnBal. & Bal. & UnBal. & Bal.& UnBal. & Bal. \\
        \midrule
        \multirowcell{6}{MLP\_1}&{Kesko Oyj} &0.37& 0.56 &0.39 &0.39 &0.39 &0.37 \\
                                &{Outokumpu Oyj}& 0.44&0.36 &0.35 &0.33 &0.36 &0.33 \\ 
                                &{Sampo Oyj}& 0.46&0.50 &0.34 &0.37 &0.37 &0.35\\ 
                                &{Rautaruukki}& 0.45&0.31 &0.42 &0.29 &0.42 &0.27\\  
                                &{Wartsila Oyj}& 0.42&0.31 &0.39 &0.32 &0.39 &0.30\\ 
                                &{Joint}& 0.49&0.36 &0.39&0.32 &0.38 &0.30\\ 
        \midrule
        \multirowcell{6}{MLP\_2}&{Kesko Oyj} & 0.50&0.53 &0.31 &0.39 &0.37 &0.50 \\
                                &{Outokumpu Oyj}& 0.51&0.46 &0.36&0.56 &0.35&0.34 \\ 
                                &{Sampo Oyj}& 0.48&0.50 &0.37 &0.56 &0.42 &0.40\\ 
                                &{Rautaruukki}& 0.47&0.31 &0.29 &0.55 &0.32&0.51\\ 
                                &{Wartsila Oyj}& 0.48&0.49 &0.39 &0.32 &0.37 &0.46\\ 
                                &{Joint}& 0.39&0.51 &0.39 &0.43 &0.38 &0.44\\ 
        \midrule
        \multirowcell{6}{MLP\_3}&{Kesko Oyj} & 0.49&0.51 &0.31 &0.54 & 0.39 &0.50 \\
                                &{Outokumpu Oyj}& 0.52&0.51 &0.36 &\underline{\textbf{0.58}} &0.36 &0.46 \\ 
                                &{Sampo Oyj}& 0.52&0.51 &0.37 &0.55 &0.37&0.44\\ 
                                &{Rautaruukki}& 0.51&0.40 &0.29 & 0.54 &0.29&0.43\\ 
                                &{Wartsila Oyj}& 0.49&0.47&0.39 &0.55 &0.39 &0.43\\ 
                                &{Joint}&  \underline{\textbf{0.53}}&0.41 &0.39 & \underline{\textbf{0.56}} &0.39 &0.50\\ 
        \midrule
        \multirowcell{6}{MLP\_4}&{Kesko Oyj} & 0.48&0.40&0.30 &0.31 &0.39 &0.29 \\
                                &{Outokumpu Oyj}& \underline{\textbf{0.53}}&0.51 &0.35 &0.39 &0.35 &0.45 \\ 
                                &{Sampo Oyj}& 0.42&0.49 &0.37 &0.37 &0.37 &0.35\\ 
                                &{Rautaruukki}& 0.42&0.31 &0.30 &0.29 &0.29 &0.41\\ 
                                &{Wartsila Oyj}& 0.40&0.50 &0.39 &0.32 &0.32 &0.30\\ 
                                &{Joint}& 0.49&0.41 &0.32 &0.33 &0.39&0.30\\ 
        \midrule
        \multirowcell{6}{MLP\_5}&{Kesko Oyj} & 0.50 &0.52 &0.29 &0.38 &0.37 &0.48 \\
                                &{Outokumpu Oyj}& 0.51&0.53 &0.35 &0.51 &0.34 &0.34 \\ 
                                &{Sampo Oyj}& 0.44&0.53 &0.37 &0.42 &0.35 &0.35\\ 
                                &{Rautaruukki}& 0.46&0.41 &0.29 &0.33 &0.27 &0.31\\ 
                                &{Wartsila Oyj}& 0.41&0.32 &0.32 &0.32 &0.30 &0.48\\ 
                                &{Joint}& 0.35&0.36 &0.32 &0.31 &0.37 &0.50\\ 
        \midrule
        \midrule
        \multirowcell{6}{CNN\_1}&{Kesko Oyj} & 0.37&0.37 &0.37 &0.36 &0.31 &0.28 \\
                                &{Outokumpu Oyj}& 0.35&0.32 &0.33 &0.32 &0.29 &0.25\\ 
                                &{Sampo Oyj}& 0.36&0.31 &0.35 &0.34 &0.32 &0.24\\ 
                                &{Rautaruukki}& 0.35&0.31 &0.27 &0.24 &0.35 &0.27\\ 
                                &{Wartsila Oyj}& 0.38&0.38 &0.30 &0.37 &0.28 &0.31\\ 
                                &{Joint}& 0.34&0.33& 0.37&0.33&0.32 &0.27\\ 
        \midrule
        \multirowcell{6}{CNN\_2}&{Kesko Oyj} & 0.31&0.32&0.37 &0.31 &0.32 &0.39 \\
                                &{Outokumpu Oyj}& 0.37&0.37 &0.36 &0.33 &0.29 &0.35 \\
                                &{Sampo Oyj}& 0.33&0.33 &0.34 &0.34&0.35 &0.37\\ 
                                &{Rautaruukki}& 0.37&0.37 &0.29 &0.31 &0.40 &0.29\\ 
                                &{Wartsila Oyj}& 0.40&0.31 &0.39 &0.33 &0.39 &0.32\\ 
                                &{Joint}& 0.35&0.36 &0.32&0.34&0.39 &0.39\\
        \midrule
        \midrule
        \multirowcell{6}{LSTM\_1}&{Kesko Oyj} & 0.30&0.30 &0.29 &0.37 &0.37 &0.37 \\
                                 &{Outokumpu Oyj}& 0.35&0.32 &0.33 &0.33 &0.32 &0.30 \\ 
                                 &{Sampo Oyj}& 0.36&0.31 &0.32&0.35&0.31 &0.33\\ 
                                 &{Rautaruukki}& 0.35&0.31 &0.40 &0.27 &0.40 &0.40\\  
                                 &{Wartsila Oyj}& 0.38&0.29 &0.37 &0.30 &0.37 &0.30\\ 
                                 &{Joint}& 0.34&0.34 &0.37 &0.30&0.37 &0.36\\ 
        \midrule
        \multirowcell{6}{LSTM\_2}&{Kesko Oyj} & 0.32&0.32 &0.37 &0.39 &0.38 &0.40 \\
                                 &{Outokumpu Oyj}& 0.37&0.34 &0.35 &0.35 &0.36 &0.33 \\ 
                                 &{Sampo Oyj}& 0.38&0.33&0.37 &0.37 &0.37 &0.36\\ 
                                 &{Rautaruukki}& 0.33&0.33 &0.42 &0.29 &0.30 &0.28\\   
                                 &{Wartsila Oyj}& 0.40&0.31 &0.42 &0.32 &0.32 &0.32\\ 
                                 &{Joint}& 0.36&0.36& 0.39&0.39 &0.38 &0.33\\         
        \bottomrule 
        \end{tabular}}
        \caption{Protocol I: f1 scores for the Nordic stocks. \\{\tiny \textit{Note: Highlighted text shows the best f1 performance for: \\1) Joint/Unbalanced, 2) Joint/Balanced, \\ 3) Stock-Specific/Unbalanced, and \\4) Stock-Specific/Balanced cases\\}}}
        \label{tab:NordicF1}
\end{minipage}
\hspace{0.0001cm}
\begin{minipage}{0.5\linewidth}
\centering
\scalebox{0.60}{
        \begin{tabular}{crcccccc}\hline
        \multirow{4}{*}{Model}&\multirow{4}{*}{Stock}& \multicolumn{2}{c}{Econ} & \multicolumn{2}{c}{Tech-Quant}  & \multicolumn{2}{c}{LOB}\\
        \cmidrule(lr){3-4} \cmidrule(lr){5-6} \cmidrule(lr){7-8}
         & & UnBal. & Bal. & UnBal. & Bal.& UnBal. & Bal. \\
        \midrule
        \multirowcell{6}{MLP\_1}&{Kesko Oyj} &68.62 & 50.97&76.99 &55.23 &173.92 &54.94 \\
                                &{Outokumpu Oyj}& 97.76 &91.29 &164.00 &176.92 &114.58 &176.93\\
                                &{Sampo Oyj}& 187.98 &178.04 &190.84 &87.32 &51.71 &86.69\\ 
                                &{Rautaruukki}& 334.61 &292.36 &166.61 &167.00 &142.40 &173.41\\  
                                &{Wartsila Oyj}& 297.42 &289.06 &244.18 &250.69 &223.17 &250.74\\ 
                                &{Joint}& 255.92 &252.13 &195.15&250.02 &192.19 &205.41\\ 
        \midrule
        \multirowcell{6}{MLP\_2}&{Kesko Oyj} &38.65& 119.89&113.20 &309.49 &539.66 &128.88 \\
                                &{Outokumpu Oyj}& 99.89&143.86&155.78 &203.97 &116.83 &388.18 \\ 
                                &{Sampo Oyj}& 190.41&226.04 &172.69 &119.33 &93.07 &206.68\\ 
                                &{Rautaruukki}& 317.66&374.41 &137.48 &340.20 &144.91 &196.54\\  
                                &{Wartsila Oyj}& 302.42&433.15 &241.99 &522.49 &262.87 &839.49\\ 
                                &{Joint}& 246.80&502.29 &334.22&799.74 &191.54 &815.36\\ 
        \midrule
        \multirowcell{6}{MLP\_3}&{Kesko Oyj} &40.20& 108.74&31.13 &122.02 &447.78 &128.45 \\
                                &{Outokumpu Oyj}& 103.84&136.95 &163.30 &194.48 &117.14 &388.12\\ 
                                &{Sampo Oyj}& 190.81&236.69&71.01 &128.70 &269.54 &201.89\\ 
                                &{Rautaruukki}& 312.41&390.14 &168.69 &378.60 &149.69 &196.54\\  
                                &{Wartsila Oyj}& 306.61&446.21 &302.83 &349.45 &224.24 &820.50\\ 
                                &{Joint}& 249.72&599.45 &272.35&504.21 &192.46 &815.36\\ 
        \midrule
        \multirowcell{6}{MLP\_4}&{Kesko Oyj} &102.52& 121.94&43.46 &308.61 &105.17&148.69 \\
                                &{Outokumpu Oyj}& 98.44&126.24 &114.98 &229.33 &116.83 &216.99 \\ 
                                &{Sampo Oyj}& 189.59&254.34 &59.40 &639.52 &72.42 &185.65\\ 
                                &{Rautaruukki}& 309.03&625.19 &142.40 &985.50 &149.69 &204.93\\  
                                &{Wartsila Oyj}& 295.85&423.50 &22.84 &464.79 &224.24 &716.44\\ 
                                &{Joint}& 250.11&555.76 &206.55 &1015.87&192.46 &589.94\\ 
        \midrule
        \multirowcell{6}{MLP\_5}&{Kesko Oyj} &99.79& 148.71&33.19 &259.19 &650.56 &148.43 \\
                                &{Outokumpu Oyj}& 99.79&151.41&114.83 &246.73 &117.14 &312.73 \\ 
                                &{Sampo Oyj}& 189.36&220.94&199.50 &129.55 &379.95 &629.82\\ 
                                &{Rautaruukki}& 308.72&399.87 &140.08 &409.47 &138.74 &205.36\\  
                                &{Wartsila Oyj}& 290.03&450.53 &248.25 &460.97 &230.90 &793.99\\ 
                                &{Joint}& 248.34&841.92 &\underline{\textbf{165.29}}&818.41 &193.10 &589.15\\ 
        \midrule
        \midrule
        \multirowcell{6}{CNN\_1}&{Kesko Oyj} &32.73& 55.43&25.32 &30.37 &30.98 &623.58 \\
                                &{Outokumpu Oyj}& 88.04&401.17 &118.89 &128.44 &128.44 &545.72 \\ 
                                &{Sampo Oyj}& 164.81&189.73 &304.42 &59.40 &54.07 &704.06\\ 
                                &{Rautaruukki}& 292.60&865.16 &141.77 &150.66 &145.31 &261.21\\  
                                &{Wartsila Oyj}& 262.92&827.61 &224.24 &248.79 &232.48 &544.67\\ 
                                &{Joint}& 235.99&636.80 &236.17&\underline{\textbf{176.25}}& 169.12 &178.67\\ 
        \midrule
        \multirowcell{6}{CNN\_2}&{Kesko Oyj} &30.75& 310.08&64.88 &54.35 &54.85 &424.82 \\
                                &{Outokumpu Oyj}& 208.71&321.86 &116.56 &535.32 &233.86 &737.63\\ 
                                &{Sampo Oyj}& 163.52&317.88 &184.45 &288.18 &62.67 &669.31\\ 
                                &{Rautaruukki}& 347.81&556.37 &181.09 &168.62 &142.08 &304.91\\  
                                &{Wartsila Oyj}& 340.68&284.84 &340.92 &149.73 &238.14 &832.23\\ 
                                &{Joint}& 245.35&700.03 &180.72&880.98 &169.03 &639.19\\
        \midrule
        \midrule
        \multirowcell{6}{LSTM\_1}&{Kesko Oyj} &32.50& 245.49&25.24 &346.99 &30.37 &409.85 \\
                                &{Outokumpu Oyj}& 89.83&455.90 &138.15 &315.56 &354.52 &440.01\\ 
                                &{Sampo Oyj}& 176.26&216.65 &154.00 &875.29 &75.96 &662.33\\ 
                                &{Rautaruukki}& 339.78&746.22 &138.94 &556.63&150.66 &187.29\\  
                                &{Wartsila Oyj}& 253.11&839.99 & 226.97 &369.42 &237.22 &719.71\\ 
                                &{Joint}& 234.65&816.23 &178.73&952.69 &166.10 &180.74\\ 
        \midrule
        \multirowcell{6}{LSTM\_2}&{Kesko Oyj} &28.22& 34.63&\underline{\textbf{24.81}} &\underline{\textbf{24.66}} &24.88 &39.29 \\
                                 &{Outokumpu Oyj}& 92.34&89.71 &116.09 &114.20 &115.78 &153.77\\ 
                                 &{Sampo Oyj}& 177.53&171.03 &54.07 &90.62 &59.23 &100.59\\ 
                                 &{Rautaruukki}& 297.96&285.85 &139.15 &198.07 &136.83 &200.85\\  
                                 &{Wartsila Oyj}& 273.83&253.96 &370.96 &220.45 &229.40 &227.30\\ 
                                 &{Joint}& 240.97&903.34 &295.32&925.90 &179.12 &752.62 \\
        \bottomrule 
        \end{tabular}}
        \caption{Protocol I: RMSE scores based on Nordic stocks for the handcrafted features. \\{\tiny \textit{Note: Highlighted text shows the best RMSE performance for: \\1) Joint/Unbalanced, 2) Joint/Balanced, \\ 3) Stock-Specific/Unbalanced, and \\4) Stock-Specific/Balanced cases\\}}}
        \label{tab:NordicRMSE}
\end{minipage}
\end{table}

\begin{table}[htb!]
\centering
\scalebox{0.7}{
        \begin{tabular}{crcccc}\hline
        \multirow{4}{*}{Model}&\multirow{4}{*}{Stock}& \multicolumn{2}{c}{LSTM AE - f1} & \multicolumn{2}{c}{LSTM AE - RMSE} \\
        \cmidrule(lr){3-4}  \cmidrule(lr){5-6}
         & & UnBal. & Bal. & UnBal. & Bal \\
        \midrule
\multirowcell{3}{MLP\_1} &{Amazon} & 0.37 & 0.36  & 79.81	&90.79\\
                        		       &{Google}  & 0.34  & 0.34 &109.80	& 110.98\\  
                                           &{Joint}      & 0.31 & 0.35  & 121.49	&100.78\\ 
\midrule
\multirowcell{3}{MLP\_2}   &{Amazon} & 0.31 & 0.36 &88.37&116.29 \\
                        			&{Google}   & 0.34 & 0.34 &96.51&  136.51\\ 
                                             &{Joint}      & 0.35 & 0.35 &101.44&125.44\\ 
\midrule
\multirowcell{3}{MLP\_3} &{Amazon}  & 0.31 & 0.36 &83.73&116.89 \\
                                           &{Google}    & 0.33 & 0.34 &96.05& 134.96\\ 
                                           &{Joint}        & 0.31 & 0.35&121.54&126.52\\ 
\midrule
\multirowcell{3}{MLP\_4} &{Amazon} & 0.36 & 0.36  &\underline{\textbf{79.50}}&117.54\\
                                           &{Google}   & 0.32 & 0.34 &103.76&134.17 \\ 
                                           &{Joint}       & 0.31& 0.35&133.24& 126.10\\ 
\midrule
\multirowcell{3}{MLP\_5} &{Amazon} & 0.31 & \underline{\textbf{0.37}} &81.86&105.38 \\
                                           &{Google}& 0.32 &0.34  &98.29&124.58\\  
                                           &{Joint}& 0.35 & 0.35 &88.80&114.09\\ 
\midrule
\midrule
\multirowcell{3}{CNN\_1} &{Amazon} & 0.36&0.36 &187.06& 90.92\\
                                           &{Google}& 0.34&0.34 &95.92&109.60 \\ 
                                           &{Joint}& 0.35& 0.35 &363.45&110.31\\ 
\midrule
\multirowcell{3}{CNN\_2} &{Amazon} & 0.36 &0.31 &82.66& 90.18\\
                                            &{Google}& 0.34 &0.32 &113.39& 109.24\\ 
                                            &{Joint}& 0.36 & 0.35 &216.62&101.36\\ 
\midrule
\midrule
\multirowcell{3}{LSTM\_1} &{Amazon} & 0.31&0.31 &79.89& \underline{\textbf{87.70}}\\
                                             &{Google}& 0.34&0.32 &98.02&109.24 \\ 
                                             &{Joint}& 0.31&0.35 &87.70&100.52\\ 
\midrule
\multirowcell{3}{LSTM\_2} &{Amazon} & 0.46&0.31 &80.03&90.17 \\ 
                                              &{Google}& \underline{\textbf{0.53}}&0.32 &97.13 & 109.06\\ 
                                              &{Joint}& 0.31&0.35&90.17&87.75 \\
\bottomrule  
\end{tabular}
}
\caption{Protocol I: f1 and RMSE scores based on US stocks for the fully-automated features. \\{\tiny \textit{Note: Highlighted text shows the best f1 performance for: \\1) Joint/Unbalanced, 2) Joint/Balanced, \\ 3) Stock-Specific/Unbalanced, and \\4) Stock-Specific/Balanced cases\\}}}
\label{tab:USF1AE}
\end{table}

\begin{table}[htb!]
\centering
\scalebox{0.7}{
        \begin{tabular}{crcccc}\hline
        \multirow{4}{*}{Model}&\multirow{4}{*}{Stock}& \multicolumn{2}{c}{LSTM AE - f1} & \multicolumn{2}{c}{LSTM AE - RMSE} \\
        \cmidrule(lr){3-4}  \cmidrule(lr){5-6}
         & & UnBal. & Bal. & UnBal. & Bal.  \\
        \midrule
\multirowcell{6}{MLP\_1} &{Kesko Oyj} &0.29 & 0.37&39.11 &48.90  \\
                                          &{Outokumpu Oyj}&0.33 &0.33 &186.68& 197.53\\
                                          &{Sampo Oyj}& 0.35 &0.35&63.78&79.56 \\ 
                                          &{Rautaruukki}& 0.27 &0.27 &182.43 &201.86 \\  
                                          &{Wartsila Oyj}& 0.37 &0.34 &282.11 &309.47 \\ 
                                          &{Joint}& 0.37 &0.37 &207.90&190.24 \\ 
        \midrule
        \multirowcell{6}{MLP\_2}&{Kesko Oyj} &0.37& 0.29&51.70 &47.35 \\
                                		&{Outokumpu Oyj}& 0.34&0.34&202.91 &199.90 \\ 
                                		&{Sampo Oyj}& 0.35&0.32&87.96 &146.54 \\ 
                                		&{Rautaruukki}& 0.27&0.40 &210.69 &212.94 \\  
                                		&{Wartsila Oyj}& 0.43&0.36 &299.77 &395.76 \\ 
                                		&{Joint}& 0.30&0.37 &236.10&207.82 \\ 
        \midrule
        \multirowcell{6}{MLP\_3}&{Kesko Oyj} &0.29& 0.37&47.12&52.91 \\
                                		&{Outokumpu Oyj}& 0.34&0.32&204.05 &183.82 \\ 
                                		&{Sampo Oyj}& 0.32&0.32&92.06 &67.94 \\ 
                                		&{Rautaruukki}& 0.40&0.40 &207.93 &174.24\\  
                                		&{Wartsila Oyj}& 0.37&0.36 &303.98 &396.01 \\ 
                                		&{Joint}& 0.30&0.37 &232.29&209.93 \\ 
        \midrule
        \multirowcell{6}{MLP\_4}&{Kesko Oyj} &0.29& 0.37&55.23 &52.55 \\
                                		&{Outokumpu Oyj}& 0.34&0.32 &202.99 &200.38  \\ 
                                		&{Sampo Oyj}& 0.35&0.33 &90.66&58.94 \\ 
                                		&{Rautaruukki}& 0.40&0.40 &209.65 &202.04 \\  
                                		&{Wartsila Oyj}& 0.37&0.36 &299.49 &327.36 \\ 
                                		&{Joint}& 0.30&0.37 &233.67 &204.09\\ 
        \midrule
        \multirowcell{6}{MLP\_5}&{Kesko Oyj} &0.29& 0.37&45.40&52.50  \\
                                		&{Outokumpu Oyj}& 0.33&0.32&195.46 &202.28 \\ 
                                		&{Sampo Oyj}& 0.35&0.33&79.83 &111.58 \\ 
                                		&{Rautaruukki}& 0.40&0.40 &196.63 &245.04 \\  
                                		&{Wartsila Oyj}& 0.37&0.37&292.99 &344.89 \\ 
                                		&{Joint}& 0.30&0.37 &222.40&202.40\\ 
        \midrule
        \midrule
        \multirowcell{6}{CNN\_1}&{Kesko Oyj} &0.33& 0.38&44.30 &288.28  \\
                                		&{Outokumpu Oyj}& 0.33&0.37&186.07 &299.16  \\ 
                                		&{Sampo Oyj}& 0.35&0.32 &61.76 &655.34 \\ 
                                		&{Rautaruukki}& 0.40&0.27 &173.71 &510.65 \\  
                                		&{Wartsila Oyj}& 0.37&0.30 &279.80 &300.04 \\ 
                                		&{Joint}& 0.40&0.37 &206.89&305.67\\ 
        \midrule
        \multirowcell{6}{CNN\_2}&{Kesko Oyj} &0.37& 0.37&40.20 &70.80 \\
                                		&{Outokumpu Oyj}& 0.33&0.37&185.48 & 181.14\\ 
                                		&{Sampo Oyj}& 0.35&0.35 &62.34 &429.33\\ 
                                		&{Rautaruukki}& 0.27&0.27 &176.98 &350.67 \\  
                                		&{Wartsila Oyj}& 0.43&0.30 &280.23 &432.86 \\ 
                                		&{Joint}& \underline{\textbf{0.49}}&0.37 &\underline{\textbf{204.61}}&350.43 \\
        \midrule
        \midrule
        \multirowcell{6}{LSTM\_1}&{Kesko Oyj} &0.37 & 0.38&39.10 &42.10  \\
                                		&{Outokumpu Oyj}&0.35&0.33 &185.28 &179.89 \\ 
                                		&{Sampo Oyj}& 0.35&0.35 &61.65 &67.34 \\ 
                                		&{Rautaruukki}& 0.35&0.43&181.27 &223.70\\ 
                                		&{Wartsila Oyj}& 0.35&0.30 & 279.46 &256.78 \\ 
                                		&{Joint}& 0.37&0.37 &234.78&110.54 \\ 
        \midrule
        \multirowcell{6}{LSTM\_2}&{Kesko Oyj} &0.42& 0.37&\underline{\textbf{39.03}} &40.19 \\
                                 		&{Outokumpu Oyj}& \underline{\textbf{0.47}}&0.45 &185.24 &178.67\\ 
                                 		&{Sampo Oyj}& 0.36&0.35 &61.69&69.42 \\ 
                                 		&{Rautaruukki}& 0.35&0.30 &180.49 &167.89 \\  
                                 		&{Wartsila Oyj}& 0.42&0.30 &279.13 &243.32 \\ 
                                 		&{Joint}& 0.38&0.37 &212.89&89.90 \\
\bottomrule  
        \end{tabular}}
       \caption{Protocol I: f1 and RMSE scores based on Nordic stocks for the fully-automated features. \\{\tiny \textit{Note: Highlighted text shows the best RMSE performance for: \\1) Joint/Unbalanced, 2) Joint/Balanced, \\ 3) Stock-Specific/Unbalanced, and \\4) Stock-Specific/Balanced cases\\}}}
        \label{tab:NORDICRMSEAE}
\end{table}


\subsection{Discussion}
\noindent The conducted experiments reveal some interesting results for both experimental protocols and datasets selection. Both protocols forecast the mid-price movement, with Protocol I forecasting the mid-price movement every next event and Protocol I with a lag of 10 events. Protocol I provides more information regarding the high-frequency activity since it takes into consideration every trading event. We cannot directly compare the two protocols since both tackle the problem of mid-price forecasting from a different angle. However, by focusing on each protocol seperately, we can see that: for Protocol I, the best classification score comes from US dataset and best regression score from Nordic dataset, while, for Protocol II, the best classification score comes again from the US dataset. \\
\indent Each one of the nine neural networks has to perform a dual task, regression and classification simultaneously. To begin with, the Joint (i.e., the full range of stocks is used for training) reports for the Nordic dataset the best f1 performance that comes from MLP\_3, for both unbalanced and balanced datasets under the Econ feature set with 53\% and  56\% for the Tech-Quant set. This MLP did not perform well for the regression task where the RMSE was above 165.29. For the stock specific case: we achieve the best classification performance of 53\% f1 score for Outokumpu Oyj under MLP\_4 and the Econ feature set with RMSE of 98.44. Thisvstock-specific performance of the MLP\_4 is the best trade-off between classification and regression for the Nordic dataset. If we want to focus on the regression task only, we can choose the more advanced model, LSTM\_2, with RMSE of approximately 24 for both unbalanced and balanced Tech-Quant feature sets for Kesko Oyj. \\
\indent For the US dataset, the new protocol presents more interesting results. For the Joint case, where both Amazon and Google used for training, the LSTM\_2 achieves 59\% f1 score and RMSE of 89.69, whereas, for the stock specific case, LSTM\_1 under the Tech-Quant feature set achieves 58\% f1 score and high RMSE of 123.36 for Google and the unbalanced case. If we focus only on the regression part, we can choose the entire MLP universe and the Econ feature set for Amazon and the Joint case. The newly introduced Econ feature set performed very well for the regression task also for LSTM\_2 across the entire protocol for the unbalanced dataset. One more interesting observation is that the Econ feature set together with the shallower MLP\_1 and the balanced set reports very low RMSE for Amazon, Google, and the Joint cases, respectively. That means that the Econ feature set, for the Amazon and Joint case, were able to predict that the mid-price will change its direction in a millisecond duration. Here, it is vital to report that the daily trading activity, for the US and Nordic stocks, contains several trades with the same timestamp/millisecond. Approximately 30\% of the trades, in the US dataset, occur in a millisecond, whereas this percentage for the Nordic dataset is 36\%. \\
\indent For Protocol II and the Joint case we achieve the best forecasting performance of 51\% f1 for the Nordic dataset based on MLP\_4 (which is one of our deeper MLP architectures) under the Tech-Quant feature set and the unbalanced case. For the Joint case in the US dataset, we achieve the best f1 performance of 65\% based on MLP\_4 under the Tech-Quant feature set and the balanced case. In terms of individual stock performance for the Nordic case we achieve 63\% f1 score for Kesko Oyj, and our shallower MLP (i.e., MLP\_1) under the Tech-Quant set, while for the US dataset we achieve an f1 performance of 65\% for Google based on MLP\_4 for the balanced case. We can see that MLPs for Protocol II were able to retain the information that the Tech-Quant feature set carries. The majority of the Tech-Quant features was derived from technical analysis, a type of analysis which is based on geometrical pattern identification of agglutinated times series like ours. What is more, the data size affected the performance of models and feature sets. For instance, Kesko Oyj, which scored the highest f1 score, is the stock with the least daily trading activity compared to the rest of the Nordic stocks and of course compared to the massive US dataset. Finally, we would like to point out that we limited the experiments to two US and five Nordic stocks; we leave the extension of the present evaluation on wider LOB datasets for future reseach that will help us to identify similarities among stock categories and time periods.

\section{Conclusion}\label{SS:Con}
\noindent In this paper, we extracted handcrafted features based on the econometric literature for mid-price prediction using deep learning techniques. Our work is the first of its kind since we do not only utilize an extensive feature set list, based on econometrics for the mid-price prediction task, but we also provide a fair comparison with two other existing state-of-the-art handcrafted and fully automated feature sets . Our extensive experimental setup, based on liquid and illiquid stocks (i.e., two US and five Nordic stocks) showed superiority of the suggested handcrafted feature sets against the fully automated process derived from an LSTM AE. What is more, our research sheds light on the area of deep learning and feature engineering by providing information based on online mid-price predictions. Our findings suggest that extensive analysis of the input signal leads to high forecasting performance even with simpler neural network architects like shallow MLPs, particularly when advanced features capture the relevant information edge. More specifically, econometric features and deep learning predicted that the mid-price would change direction in a millisecond duration for Amazon and the Joint (i.e., training on both Amazon and Google) cases. Although these results are promising, our study here also suggests that selection of features and models should be differentiated for liquid and illiquid stocks. 

\clearpage

\section*{Appendices}

\addcontentsline{toc}{section}{Appendices}
\renewcommand{\thesubsection}{\Alph{subsection}}

\renewcommand{\thesection}{\Alph{section}}

\subsection{Feature Pool}\label{SS:Feat}


\setcounter{equation}{0}
\setcounter{table}{0}
\setcounter{figure}{0}
\renewcommand{\thesection}{A.\arabic{section}}
\renewcommand{\theequation}{A.\arabic{equation}}
\renewcommand\thetable{A.\arabic{table}} 
\renewcommand\thefigure{A.\arabic{figure}}

\subsubsection{\textbf{Statistical features}} 
    \begin{itemize}
        \item Mid price is defined as:
	\begin{equation}
	MP = \frac{Ask_{best} + Bid_{best}}{2}
	\end{equation}
        	\item Financial duration is defined as:
	\begin{equation}
	FD = T_t - T_{t-1},
	\end{equation}
	where $T$ denotes the time instance at time $t$.
         \item Average mid-price financial duration is defined as:
	\begin{equation}
	AMPD_{l} = \frac{\Big\{ \mathcal{T}_1, \mathcal{T}_2, ..., \mathcal{T}_N \Big\}_{i=1}^{N}}{\Big\{ 		\mathcal{P}_1, \mathcal{P}_2, ..., \mathcal{P}_N \Big\}_{i=1}^{N}},
	\end{equation}
	where $\Big\{ \mathcal{T}_k \Big\}_{k=1}^N$, and $\Big\{ \mathcal{P}_k \Big\}_{k=1}^N$ are the 	partial cumulative sums of time and price differences for every LOB level for $N$ samples. 
       \item Mid price deeper levels are equal to:
	\begin{equation}
	DMP = \frac{Ask_l + Bid_l}{2}, \ l = 2:10
	\end{equation}
	where $l$ denotes the depth of the LOB.
        \item Log returns are defined as:
	\begin{equation}
	r(X)_i=X_i-X_{i-1};
	\end{equation}
	where $X_i$ is the logarithmic price
    \end{itemize}

\subsubsection{\textbf{Volatility measures}}
\medskip
 \noindent The features in this category aim to estimate, either the integrated variance (IV), that is the process
    \begin{equation}
    IV_t=\int^t_0\sigma_s^2\mathrm{d}s
    \end{equation}
    or, more generally, the quadratic variation (QV)
    \begin{equation}
    [X,X]_t=\int^t_0\sigma_s^2\mathrm{d}s+\sum_{0<s\leq t}(\zeta_s\mathrm{d}N_s)^2.
    \end{equation}
    Here $X$ is the logarithmic price of some given asset.
    We assume that $X_t$ follows an It\^o semimartingale; that is,
    \begin{equation}
    X_t=X_0+\int^t_0b_s\mathrm{d}s+\int_0^t\sigma_s\mathrm{d}W_s+\int_o^t\zeta_s\mathrm{d}N_s,
    \end{equation}
    where b is locally bounded, $\sigma$ is c\'adl\'ag and predictable, and $W$ is a standard Weiner process, $\zeta$ is a thin (i.e., finite) process mapping the jump size, and $N$ is the counting process associated to the jump times of $X$. We define $\Delta_n$ the time elapsed between two adjacent observations; specifically, if we assume the observations are equidistant in time we have $\Delta_n=\lfloor\frac{t}{n}\rfloor$. As we do not work in calendar time we will have $\Delta_n=\frac{1}{n}$.
    \begin{itemize}
        \item Realized variance\\
        The realized variance \cite{Andersen1998} is the most natural estimator of the quadratic variation process and is equal to:
\begin{equation}
RV_t=\sum^n_{i=1}(r(X)_{i})^2.
\end{equation}
        \item Realized kernel\\
        Realized kernels \cite{Barndorff-Nielsen2008} are used to obtain a noise robust estimate of QV as follows:
\begin{equation}
RK_t=\gamma_0(X_{\Delta_n})+\sum^H_{h=1}k\left (\frac{h}{H}\right )\{\gamma_h(X_{\Delta_n})+\gamma_{-h}(X_{\Delta_n})\},
\end{equation}
with $H$ the kernel bandwidth, $\gamma_h(X_{\Delta_n})$ the autocovariation process, $k$ is the kernel function of choice. In particular we use a non-flat-top Parzen and our implementation follows closely \cite{Barndorff-Nielsen2009}.
        \item Realized pre-averaged variance\\
        The pre-averaged realized variance \cite{Jacod2009} is akin to the realized kernel estimator (in fact they are asymptotically equivalent). As for the realized kernel, the pre-averaged realized variance is used to retrieve a noise-free measurement of the quadratic variation of our price process and it is calculated as follows:
\begin{equation}
PA-RV_t=\frac{\sqrt{\Delta_n}}{\theta\psi_2}\sum^{n-H+1}_{i=0}(\bar{X}^n_i)^2-\frac{\psi_1\Delta_n}{2\theta^2\psi_2}\sum^{N}_{i=0}(r(X))^2.
\end{equation}
As before we have $H$ the kernel bandwidth and $\theta$ the pre-averaging horizon. Further, given a nonzero real-valued function $g:[0,1]\rightarrow\mathbb{R}$ with $g(0)=g(1)=0$ and which is further continous and piecewise continuously differentiable such that its derivative $g^\prime$ is piecewise Lipschitz. Then, we define:
\[
\psi_1=\int_0^1(g^\prime(s))^2\mathrm{d}s,\quad\quad \psi_1=\int_0^1(g(s))^2\mathrm{d}s.
\]
In our application we follow \cite{Christensen2014} and set $H=\theta\sqrt{n}$ and $\theta=1$, $g(x)=x\wedge (1-x)$. Hence we will have $\psi_1=1$ and, $\psi_2=\frac{1}{12}$.
        \item Realized semi-variance ($+$, $-$)\\
        Positive ($+$) and negative ($-$) realized semi-variances \cite{Barndorff-Nielsen2010} measure upside and downside risk respectively, as follows:
\begin{align}
\begin{split}
RSV^{+}(X)_{t} &= \sum_{i=1}^{n} r(X)_{i}^2 \textbf{1}_{(r(X)_{i}>0)}\\
RSV^{-}(X)_{t} &= \sum_{i=1}^{n} r(X)_{i}^2 \textbf{1}_{(r(X)_{i}<0)}
\end{split}
\label{eq: definition of RSV}
\end{align}
where $\textbf{1}$ is a simple indicator function.
        \item Realized bipower variation\\ 
        The realized bipower variation \cite{Barndorff-Nielsen2004} measures the diffusive component of the price process, isolating it from the variation caused by the jump components and it is equal to:
\begin{equation}
BV(X)_{t} := \frac{\pi}{2} \sum_{i=2}^{n} |r(X)_{i}| |r(X)_{i-1}|
\end{equation}
        \item Realized bipower variation (lag 2)
        
        \begin{equation}
BV(X)_{t} := \frac{\pi}{2} \sum_{i=3}^{n} |r(X)_{i}| |r(X)_{i-2}|
\end{equation}
        \item Realized bipower semivariance ($+$, $-$)\\ 
        Realized bipower semivariances \cite{Barndorff-Nielsen2010} are used to measure the upside and downside risk of the diffusive component:
\begin{align}
\begin{split}
BV^+(X)_{t} := \frac{\pi}{2} \sum_{i=2}^{n} |r(X)_{i}| |r(X)_{i-1}|\textbf{1}_{(r(X)_{i}>0)}\\
BV^-(X)_{t} := \frac{\pi}{2} \sum_{i=2}^{n} |r(X)_{i}| |r(X)_{i-1}|\textbf{1}_{(r(X)_{i}<0)}.
\end{split}
\end{align}
        \item Jump variation\\
        We use a modified version of the jump variation estimator \cite{Christensen2014} which is both non-negative and consistent. As hinted by the name, the jump variation estimator provides a measures of the discontinuous variability component:
\begin{equation}
JV(X)_{t} := \max(RV(X)_{t}-BV(X)_{t},0). \label{eq: definition of JV}
\end{equation}
        \item Spot volatility\\
        We only compute the spot volatility (i.e., \cite{Barndorff-Nielsen2002} and \cite{Andersen2010}) estimates on the block. The spot volatility measures the instantaneous volatility. The definition is consistent with the terminology commonly used in the literature on parametric stochastic volatility models in  continuous-time:
\begin{equation}
SV(X)_{t} := \lim_{h\to0}\{\mathbb{E}[([X,X]_{t+h}-[X,X]_t)/h]|\mathcal{F}_t\}. \label{eq: definition of SV}
\end{equation}        
        with $h\to0$ being the time interval upon which the measure is computed.         
        \item Average spot volatility\\
        The average spot volatility provides an historical average of the estimated spot volatilities:
\begin{equation}
\overline{SV}(X)_{t} := \frac{1}{t}\sum^t_{i=0}SV(X)_{i}. \label{eq: definition of avg SV}
\end{equation}                
    \end{itemize}

\subsubsection{\textbf{Noise and uncertainty measures}}
\medskip
\noindent In this category, we incorporate two kinds of measures which are intimately linked to each other. We provide three different estimates for the integrated quarticity and two different estimates for the variance of the contaminating noise process. The integrated quarticity measures the degree of estimation error in the realized variance and can be consistently estimated through the realized quarticity estimators presented below for a fixed window size of 2000 events. The noise variance estimates provide a measure of the intensity of the noise process affecting the underlying price, as follows:
    \begin{equation}
    IQ_t=\int^t_0\sigma_s^4\mathrm{d}s
\end{equation}
 \noindent with the noise variance estimates providing a measure of the contaminating:
    \begin{itemize}
        \item Realized quarticity
        \cite{Barndorff-Nielsen2006}:
        \begin{equation}
        RQ_t=\frac{n}{3}\sum_{i=1}^n(X_i-X_{i-1})^4
        \end{equation}
        \item Realized quarticity Tripower\\
        The tri-power quarticity \cite{Barndorff-Nielsen2006} is a generalization of the realized bipower variation and is a consistent estimator for the integrated quarticity in the presence of jumps:
        \begin{equation}
        RQ_t=n\mu^{-3}_{4/3}\sum_{i=3}^{n}|r(X)_i|^{4/3}|r(X)_{i-1}|^{4/3}|r(X)_{i-2}|^{4/3}
        \end{equation}
        with $\mu_{p}=\mathbb{E}\left (|Z|^{p}\right )$,where $Z$ denotes a standard normally distributed random variable.
        \item Realized quarticity Quadpower\\
        A generalization of multipower variation measures led to the realized quadpower quarticity estimator proposed by \cite{Barndorff-Nielsen2006} and it is equal to:
        \begin{equation}
        RQ_t=n\mu^{-4}_{1}\sum_{i=4}^{n}|r(X)_i||r(X)_{i-1}||r(X)_{i-2}||r(X)_{i-3}|
        \end{equation}
        \item Noise variance
        \cite{Oomen2006}:
        \begin{equation}
        NV_t=-\frac{1}{n-1}\sum^n_{i=2}(r(X)_ir(X)_{i-1}).
        \end{equation}
        \item Noise variance
        \cite{Zhang2005}:
        \begin{equation}
        NV_t=\frac{1}{2n}\sum^n_{i=1}(X_i-X_{i-1})^2.
        \end{equation}
    \end{itemize}

\subsubsection{\textbf{Price discovery features}}
\medskip
    \begin{itemize}
        \item Mid price weighted by order imbalance:
\begin{equation}    
        MidPrice_t=\frac{Ask*V_{Ask}+Bid*V_{Bid}}{V_{Ask}+V_{Bid}}.
\end{equation}
        \item Volume imbalance:
        \begin{equation}
        VolImbalance=\frac{V_{Bid}}{{V_{Ask}+V_{Bid}}}
        \end{equation}
        \item Bid-ask spread:
        \begin{equation}
        BA_{\mathrm{spread}}=Ask-Bid.
        \end{equation}
        \item Normalized bid-ask spread\\ 
        The normalized bid-ask spread expresses the spread as the number of ticks between the bid and the ask price:
        \begin{equation}
        BA_{\mathrm{spread}}=\frac{Ask-Bid}{TickSize}.
		\end{equation}
\end{itemize}

\subsection{Protocol II Results}\label{SS:ProI}

\begin{table}[htb!]
\centering
\scalebox{0.80}{
\begin{tabular}{crcccccc}\hline
\multirow{4}{*}{Model}&\multirow{4}{*}{Stock}& \multicolumn{2}{c}{Econ} & \multicolumn{2}{c}{Tech-Quant}  & \multicolumn{2}{c}{LOB}\\
\cmidrule(lr){3-4} \cmidrule(lr){5-6} \cmidrule(lr){7-8}
 & & UnBal. & Bal. & UnBal. & Bal.& UnBal. & Bal. \\
\midrule
\multirowcell{3}{MLP\_1}&{Amazon} & 0.28&0.26 &0.20 &0.62 &0.31 &0.45 \\
				        &{Google}& 0.28&0.28 &0.18 &0.63 &0.30 &0.55 \\ 
                        &{Joint}& 0.36&0.25 &0.23 & 0.56 &0.27 &0.51\\ 
\midrule
\multirowcell{3}{MLP\_2}&{Amazon} & 0.35&0.26 &0.20&0.58 &0.25 &0.31 \\
                        &{Google}& 0.31&0.35 &0.19 &0.35 &0.32 &0.50 \\
                        &{Joint}& 0.21&0.25 &0.24 &0.49 &0.29 &0.50\\  
\midrule
\multirowcell{3}{MLP\_3}&{Amazon} & 0.25&0.26 &0.26 &0.46 &0.26 &0.44 \\
                        &{Google}& 0.27&0.33 &0.19 &0.47 &0.26 &0.48 \\     
                        &{Joint}& 0.28&0.21 &0.33&0.63 &\underline{\textbf{0.51}} &0.56\\ 
\midrule
\multirowcell{3}{MLP\_4}&{Amazon} & 0.21&0.23 &0.19 &0.53 &0.15 &0.41 \\
                        &{Google}& 0.27&0.27 &0.20&\underline{\textbf{0.65}} &0.27 &0.56 \\ 
                        &{Joint}& 0.21&0.26 &0.27 &\underline{\textbf{0.65}} &0.21&0.59\\ 
\midrule
\multirowcell{3}{MLP\_5}&{Amazon} & 0.31&0.26 &0.21 &0.56 &0.20 &0.39 \\
                        &{Google}& 0.33&0.31 &0.20&0.62 &0.21 &0.56 \\      
                        &{Joint}&0.36&0.30 &0.24 &0.52 &0.35 &0.57\\ 
\midrule
\midrule
\multirowcell{3}{CNN\_1}&{Amazon} & 0.34&0.25 &0.19 &0.16 &0.24 &0.18 \\
                        &{Google}& 0.33&0.34 &0.21 &0.19 &0.27 &0.22 \\ 
                        &{Joint}& 0.35&0.27 &0.19&0.22 &0.29 &0.20\\ 
\midrule
\multirowcell{3}{CNN\_2}&{Amazon} & 0.31&0.26 &0.22 &0.21 &0.26 &0.21 \\
                        &{Google}& 0.35&0.22 &0.25 &0.20 &0.31 &0.22 \\ 
                        &{Joint}& 0.37&0.29 &0.23 &0.24 &0.30 &0.23\\
\midrule
\midrule
\multirowcell{5}{LSTM\_1}&{Amazon} & 0.33&0.22 &0.23 &0.22 &0.28 &0.15 \\
                        &{Google}& 0.31& 0.35 &0.22 &0.23 &0.33 &0.14 \\ 
                        &{Joint}& 0.35&0.23 &0.19 &0.25 &0.21 &0.19\\ 
\midrule
\multirowcell{5}{LSTM\_2}&{Amazon} & 0.34&0.26 &0.21 &0.25 &0.26 &0.21 \\
                        &{Google}& 0.32& 0.37&0.24 &0.26 &\underline{\textbf{0.41}} &0.19 \\ 
                        &{Joint}& 0.37&0.25 &0.20 &0.27 &0.42 &0.22\\   
\bottomrule 
\end{tabular}}
        \caption{Protocol II: f1 scores based on US stocks for the handcrafted features. \\{\tiny \textit{Note: Highlighted text shows the best f1 performance for: \\1) Joint/Unbalanced, 2) Joint/Balanced, \\ 3) Stock-Specific/Unbalanced, and \\4) Stock-Specific/Balanced cases\\}}}
        \label{tab:table_12}
\end{table}


\begin{table}[htb!]
\centering
\scalebox{0.80}{
\begin{tabular}{crcc}\hline
\multirow{4}{*}{Model}&\multirow{4}{*}{Stock}& \multicolumn{2}{c}{LSTM AE} \\
\cmidrule(lr){3-4} 
 & & UnBal. & Bal.  \\
\midrule
\multirowcell{3}{MLP\_1}   &{Amazon} & 0.27&0.28  \\
				         &{Google}& 0.28&\underline{\textbf{0.31}}  \\ 
                        			&{Joint}& 0.27&0.25 \\ 
\midrule
\multirowcell{3}{MLP\_2}  &{Amazon} & 0.11&0.11  \\
                        			&{Google}& 0.15&0.18  \\
                        			&{Joint}& 0.19&0.25 \\  
\midrule
\multirowcell{3}{MLP\_3}  &{Amazon} & 0.22&0.22  \\
                        			&{Google}& 0.27&0.24  \\     
                        			&{Joint}& 0.23&0.21 \\ 
\midrule
\multirowcell{3}{MLP\_4}  &{Amazon} & 0.25&0.25 \\
                        			&{Google}& 0.24&0.23  \\ 
                        			&{Joint}& 0.22&0.26 \\ 
\midrule
\multirowcell{3}{MLP\_5  }&{Amazon} & 0.21&0.21  \\
                        			&{Google}& 0.23&0.28  \\      
                        			&{Joint}&0.24&\underline{\textbf{0.30}} \\ 
\midrule
\midrule
\multirowcell{3}{CNN\_1}  &{Amazon} & 0.27&0.21  \\
                        			&{Google}& 0.24&0.22 \\ 
                        			&{Joint}& 0.25&0.19 \\ 
\midrule
\multirowcell{3}{CNN\_2}&{Amazon} & 0.27&0.25  \\
                        		       &{Google}& 0.29&0.26  \\ 
                        		       &{Joint}& 0.30&0.25 \\
\midrule
\midrule
\multirowcell{5}{LSTM\_1}&{Amazon} & 0.19&0.21  \\
                        		        &{Google}& 0.33& 0.27  \\ 
                        		        &{Joint}& \underline{\textbf{0.33}}&0.22 \\ 
\midrule
\multirowcell{5}{LSTM\_2}&{Amazon} & 0.21&0.23  \\
                        			&{Google}& \underline{\textbf{0.34}}& 0.21 \\ 
                        			&{Joint}& 0.28&0.24 \\   
\bottomrule 
\end{tabular}}
        \caption{Protocol II: f1 scores based on US stocks for the fully automated features. \\{\tiny \textit{Note: Highlighted text shows the best f1 performance for: \\1) Joint/Unbalanced, 2) Joint/Balanced, \\ 3) Stock-Specific/Unbalanced, and \\4) Stock-Specific/Balanced cases\\}}}
        \label{tab:US2}
\end{table}


\begin{table}[htb!]
\centering
\scalebox{0.80}{
\begin{tabular}{crcccccc}\hline
\multirow{4}{*}{Model}&\multirow{4}{*}{Stock}& \multicolumn{2}{c}{Econ} & \multicolumn{2}{c}{Tech-Quant}  & \multicolumn{2}{c}{LOB}\\
\cmidrule(lr){3-4} \cmidrule(lr){5-6} \cmidrule(lr){7-8}
 & & UnBal. & Bal. & UnBal. & Bal.& UnBal. & Bal. \\
\midrule
\multirowcell{6}{MLP\_1}&{Kesko Oyj} & 0.42&0.29 &\underline{\textbf{0.63}} &\underline{\textbf{0.56}} & 0.59 &0.25 \\
				      &{Outokumpu Oyj}&  0.30 & 0.25 & 0.51&0.45 &0.52 &0.40 \\ 
				      &{Sampo Oyj}& 0.32&0.30 &0.56 &0.48 &0.57 &0.36\\ 
   				      &{Rautaruukki}& 0.19 &0.37  &0.42 &0.48 &0.42 &0.41\\ 
				      &{Wartsila Oyj}& 0.30&0.35 &0.43 &0.37 &0.36 &0.41\\ 
				      &{Joint}& 0.37&0.34 &0.48 &0.43 &0.45 &0.42\\ 
\midrule
\multirowcell{6}{MLP\_2}&{Kesko Oyj} & 0.45 & 0.38&0.55 &0.53 &0.51&0.44 \\
				      &{Outokumpu Oyj}& 0.30& 0.31 &0.46 &0.45 &0.51 &0.45 \\ 
				      &{Sampo Oyj}& 0.30&0.30 &0.49 &0.48 &0.55&0.45\\ 
  				      &{Rautaruukki}& 0.31&0.35&0.38 &0.37&0.40 &0.41\\ 
				      &{Wartsila Oyj}& 0.34 &0.36 &0.35 &0.38 &0.36 &0.39\\ 
				      &{Joint}& 0.32&0.34 &0.34 &0.42 &0.37 &0.44\\ 
\midrule
\multirowcell{6}{MLP\_3}&{Kesko Oyj} & 0.44&0.39 &0.56 &0.48 &0.53 &0.43 \\
				&{Outokumpu Oyj}& 0.30& 0.31 &0.47 &0.46&0.50 &0.44 \\ 
				&{Sampo Oyj}& 0.30&0.29&0.50 &0.48 &0.54 &0.52\\ 
				&{Rautaruukki}& 0.34&0.37 &0.40 &0.42 &0.41 &0.41\\ 
				&{Wartsila Oyj}& 0.30&0.37 &0.36&0.41 &0.34 &0.33\\ 
				&{Joint}& 0.33&0.33 &0.49 &0.43 &0.48 &0.43\\ 
\midrule
\multirowcell{6}{MLP\_4}&{Kesko Oyj} & 0.44&0.36 &0.54 &0.52&0.52 &0.41 \\
				&{Outokumpu Oyj}& 0.30&0.29 &0.45 &0.45 &0.52 &0.46 \\ 
				&{Sampo Oyj}& 0.31 &0.31 &0.51&0.49 &0.54 &0.46\\ 
				&{Rautaruukki}& 0.33&0.35 &0.41 &0.40 &0.42 &0.42\\ 
				&{Wartsila Oyj}& 0.33&0.35&0.40 &0.40 &0.38 &0.40\\ 
				&{Joint}& 0.30&0.33 &\underline{\textbf{0.51}} &0.41 &0.46 &0.41\\ 
\midrule
\multirowcell{6}{MLP\_5}&{Kesko Oyj} & 0.45& 0.37&0.49 &0.49 &0.53& 0.39\\
				&{Outokumpu Oyj}& 0.30& 0.30 &0.49 &0.48 &0.51 &0.43 \\ 
				&{Sampo Oyj}& 0.32&0.30&0.51 &0.49 &0.53 &0.50\\ 
				&{Rautaruukki}& 0.32&0.35 &0.43 &0.40 &0.42 &0.41\\ 
				&{Wartsila Oyj}& 0.29&0.35 &0.41 &0.38 &0.37 &0.41\\ 
				&{Joint}& 0.32&0.33 &0.50 &0.44 &0.44 &\underline{\textbf{0.45}}\\
\midrule
\midrule
\multirowcell{6}{CNN\_1}&{Kesko Oyj} & 0.42&0.26 &0.51 &0.46 &0.48 &0.09 \\
			    &{Outokumpu Oyj}& 0.31&0.26 &0.46 &0.29 &0.48 &0.18 \\ 
			    &{Sampo Oyj}& 0.34&0.27 &0.45 &0.38 &0.50 &0.34\\ 
                &{Rautaruukki}& 0.32&0.38 &0.40&0.38&0.40 &0.29\\ 
                &{Wartsila Oyj}& 0.31&0.23 &0.36 &0.24 &0.33 &0.20\\ 
                &{Joint}& 0.32&0.25 &0.44 &0.26 &0.45 &0.29\\ 
\midrule
\multirowcell{6}{CNN\_2}&{Kesko Oyj} & 0.45&0.13 &0.52 &0.37 &0.54 &0.20 \\
                        &{Outokumpu Oyj}& 0.28&0.17 &0.51 &0.30 &0.51 &0.22 \\ 
                        &{Sampo Oyj}& 0.33&0.19 &0.52 &0.26 &0.55 &0.20\\ 
                        &{Rautaruukki}& 0.40&0.16 &0.41 &0.29 &0.40 &0.28\\ 
                        &{Wartsila Oyj}& 0.33&0.30&0.36 &0.26 &0.38 &0.28\\ 
                        &{Joint}& 0.31&0.30 &0.49 &0.26 &0.47 &0.27\\
                        \midrule
                        \midrule
\multirowcell{6}{LSTM\_1}&{Kesko Oyj} & 0.43&0.28 &0.52 &0.50 &0.54 &0.14 \\
                        &{Outokumpu Oyj}& 0.31&0.24 &0.45&0.34 &0.49 &0.24 \\ 
                        &{Sampo Oyj}& 0.32&0.31 &0.50 &0.39 &0.51 &0.30\\ 
                        &{Rautaruukki}& 0.28&0.24 &0.38 &0.31 &0.40 &0.30\\ 
                        &{Wartsila Oyj}& 0.33&0.32 &0.35 &0.29 &0.39 &0.27\\ 
                        &{Joint}& 0.32&0.27 &0.46&0.27 &0.45 &0.29\\ 
\midrule
\multirowcell{6}{LSTM\_2}&{Kesko Oyj} & 0.44&0.21 &0.54 &0.47 &0.48 &0.13 \\
                         &{Outokumpu Oyj}& 0.32&0.23 &0.45&0.19 &0.49 &0.19 \\ 
                         &{Sampo Oyj}& 0.29&0.25 &0.50 &0.40 &0.52 &0.31\\ 
                         &{Rautaruukki}& 0.32&0.30 &0.38 &0.30 &0.39 &0.30\\ 
                         &{Wartsila Oyj}& 0.31&0.31 &0.35 &0.30 &0.38 &0.27\\ 
                         &{Joint}& 0.34& 0.26 &0.46 &0.26 &0.48 &0.27\\ 
\bottomrule 
\end{tabular}}
        \caption{Protocol II: f1 scores based on Nordic stocks for the handcrafted features. \\{\tiny \textit{Note: Highlighted text shows the best f1 performance for: \\1) Joint/Unbalanced, 2) Joint/Balanced, \\ 3) Stock-Specific/Unbalanced, and \\4) Stock-Specific/Balanced cases\\}}}
        \label{tab:table_11}
\end{table}


\begin{table}[htb!]
\centering
\scalebox{0.80}{
\begin{tabular}{crcccccc}\hline
\multirow{4}{*}{Model}&\multirow{4}{*}{Stock}& \multicolumn{2}{c}{LSTM AE} \\
\cmidrule(lr){3-4} 
 & & UnBal. & Bal.  \\
\midrule
\multirowcell{6}{MLP\_1}&{Kesko Oyj} & 0.35&0.30  \\
				      &{Outokumpu Oyj}&  0.20 & 0.19  \\ 
				      &{Sampo Oyj}& 0.28&0.26 \\ 
   				      &{Rautaruukki}& 0.19 &0.21  \\ 
				      &{Wartsila Oyj}& 0.28&0.22 \\ 
				      &{Joint}& \underline{\textbf{0.37}}&0.26 \\ 
\midrule
\multirowcell{6}{MLP\_2}&{Kesko Oyj} & 0.34 & 0.29 \\
				      &{Outokumpu Oyj}& 0.20& 0.18  \\ 
				      &{Sampo Oyj}& 0.27&0.21 \\ 
  				      &{Rautaruukki}& 0.31&0.21\\ 
				      &{Wartsila Oyj}& 0.27 &0.22 \\ 
				      &{Joint}& 0.32&0.20 \\ 
\midrule
\multirowcell{6}{MLP\_3}&{Kesko Oyj} & 0.32&\underline{\textbf{0.33}}  \\
				&{Outokumpu Oyj}& 0.20& 0.17 \\ 
				&{Sampo Oyj}& 0.26&0.29\\ 
				&{Rautaruukki}& 0.26&0.23 \\ 
				&{Wartsila Oyj}& 0.26&0.26 \\ 
				&{Joint}& 0.33&\underline{\textbf{0.29}}\\ 
\midrule
\multirowcell{6}{MLP\_4}&{Kesko Oyj} & 0.30&0.31  \\
				&{Outokumpu Oyj}& 0.20&0.19\\ 
				&{Sampo Oyj}& 0.28 &0.32 \\ 
				&{Rautaruukki}& 0.26&0.28 \\ 
				&{Wartsila Oyj}& 0.33&0.27\\ 
				&{Joint}& 0.30&0.25 \\ 
\midrule
\multirowcell{6}{MLP\_5}&{Kesko Oyj} & 0.30& 0.28\\
				&{Outokumpu Oyj}& 0.20& 0.19 \\ 
				&{Sampo Oyj}& 0.18&0.17\\ 
				&{Rautaruukki}& 0.19&0.18 \\ 
				&{Wartsila Oyj}& 0.18&0.18 \\ 
				&{Joint}& 0.32&0.23 \\
\midrule
\midrule
\multirowcell{6}{CNN\_1}&{Kesko Oyj} & 0.28&0.26 \\
			               &{Outokumpu Oyj}& 0.29&0.27 \\ 
			               &{Sampo Oyj}& 0.26&0.27 \\ 
                                          &{Rautaruukki}& 0.31&0.21 \\ 
                                          &{Wartsila Oyj}& 0.30&0.22 \\ 
                                          &{Joint}& 0.32&0.19  \\ 
\midrule
\multirowcell{6}{CNN\_2}&{Kesko Oyj} & 0.29&0.13  \\
                                          &{Outokumpu Oyj}& 0.27&0.17  \\ 
                                          &{Sampo Oyj}& 0.29&0.19 \\ 
                                          &{Rautaruukki}& \underline{\textbf{0.36}}&0.16 \\ 
                                          &{Wartsila Oyj}& 0.31&0.24\\ 
                                          &{Joint}& 0.31&0.21 \\
                        \midrule
                        \midrule
\multirowcell{6}{LSTM\_1}&{Kesko Oyj} & 0.28&0.28 \\
                                           &{Outokumpu Oyj}& 0.32&0.24  \\ 
                                           &{Sampo Oyj}& 0.31&0.25 \\ 
                                           &{Rautaruukki}& 0.27&0.25 \\ 
                                           &{Wartsila Oyj}& 0.31&0.24 \\ 
                                           &{Joint}& 0.33&0.27 \\ 
\midrule
\multirowcell{6}{LSTM\_2}&{Kesko Oyj} & 0.31&0.21  \\
                                            &{Outokumpu Oyj}& 0.33&0.23  \\ 
                                            &{Sampo Oyj}& 0.33&0.23 \\ 
                                            &{Rautaruukki}& 0.34&0.22 \\ 
                                            &{Wartsila Oyj}& 0.32&0.22 \\ 
                                            &{Joint}& 0.31& 0.25 \\ 
\bottomrule 
\end{tabular}}
        \caption{Protocol II: f1 scores based on Nordic stocks for the fully automated features. \\{\tiny \textit{Note: Highlighted text shows the best f1 performance for: \\1) Joint/Unbalanced, 2) Joint/Balanced, \\ 3) Stock-Specific/Unbalanced, and \\4) Stock-Specific/Balanced cases\\}}}
        \label{tab:NORDIC2}
\end{table}


\clearpage
\section*{Acknowledgment}

\noindent The research leading to these results has received funding from the H2020 Project BigDataFinance MSCA-ITN-ETN 675044 (http://bigdatafinance.eu), Training for Big Data in Financial Research and Risk Management.
\\

\noindent The authors wish to acknowledge CSC-IT Center for Science, Finland, for generous computational resources.

\section*{}
\bibliography{Aarhus_Bib.bib}

\begin{thebibliography}{}

\bibitem[Alberg \& Lipton, 2017]{alberg2017improving}
Alberg, J. \& Lipton, Z.~C. (2017).
\newblock Improving factor-based quantitative investing by forecasting company
  fundamentals.
\newblock {\em arXiv preprint arXiv:1711.04837}.

\bibitem[Andersen \& Bollerslev, 1998]{Andersen1998}
Andersen, T.~G. \& Bollerslev, T. (1998).
\newblock Answering the skeptics: Yes, standard volatility models do provide
  accurate forecasts.
\newblock {\em International Economic Review}, pages 885--905.

\bibitem[Andersen et~al., 2010]{Andersen2010}
Andersen, T.~G., Bollerslev, T., \& Diebold, F.~X. (2010).
\newblock Parametric and nonparametric volatility measurement.
\newblock In {\em Handbook of Financial Econometrics, Vol 1}, chapter~2, pages
  67--137. Elsevier B.V.

\bibitem[Barndorff-Nielsen et~al., 2010]{Barndorff-Nielsen2010}
Barndorff-Nielsen, O., Kinnebrock, S., \& Shephard, N. (2010).
\newblock Measuring downside risk: Realised semivariance.
\newblock {\em Volatility and Time Series Econometrics: Essays in honour of Rob
  Engle}.

\bibitem[Barndorff-Nielsen et~al., 2008]{Barndorff-Nielsen2008}
Barndorff-Nielsen, O.~E., Hansen, P.~R., Lunde, A., \& Shephard, N. (2008).
\newblock Designing realized kernels to measure the ex post variation of equity
  prices in the presence of noise.
\newblock {\em Econometrica}, 76(6):1481--1536.

\bibitem[Barndorff-Nielsen et~al., 2009]{Barndorff-Nielsen2009}
Barndorff-Nielsen, O.~E., Hansen, P.~R., Lunde, A., \& Shephard, N. (2009).
\newblock Realized kernels in practice: Trades and quotes.
\newblock {\em The Econometrics Journal}, 12(3):C1--C32.

\bibitem[Barndorff-Nielsen \& Shephard, 2002]{Barndorff-Nielsen2002}
Barndorff-Nielsen, O.~E. \& Shephard, N. (2002).
\newblock Econometric analysis of realized volatility and its use in estimating
  stochastic volatility models.
\newblock {\em Journal of the Royal Statistical Society: Series B (Statistical
  Methodology)}, 64(2):253--280.

\bibitem[Barndorff-Nielsen \& Shephard, 2004]{Barndorff-Nielsen2004}
Barndorff-Nielsen, O.~E. \& Shephard, N. (2004).
\newblock Power and bipower variation with stochastic volatility and jumps.
\newblock {\em Journal of Financial Econometrics}, 2(1):1--37.

\bibitem[Barndorff-Nielsen \& Shephard, 2006]{Barndorff-Nielsen2006}
Barndorff-Nielsen, O.~E. \& Shephard, N. (2006).
\newblock Econometrics of testing for jumps in financial economics using
  bipower variation.
\newblock {\em Journal of Financial Econometrics}, 4(1):1--30.
\newblock Available from: \url{http://dx.doi.org/10.1093/jjfinec/nbi022}.

\bibitem[Baydin et~al., 2015]{baydin2015automatic}
Baydin, A.~G., Pearlmutter, B.~A., Radul, A.~A., \& Siskind, J.~M. (2015).
\newblock Automatic differentiation in machine learning: a survey.

\bibitem[Bishop et~al., 1995]{bishop1995neural}
Bishop, C.~M. et~al. (1995).
\newblock {\em Neural networks for pattern recognition}.
\newblock Oxford university press.

\bibitem[Brownlees \& Gallo, 2006]{brownlees2006financial}
Brownlees, C.~T. \& Gallo, G.~M. (2006).
\newblock Financial econometric analysis at ultra-high frequency: Data handling
  concerns.
\newblock {\em Computational Statistics \& Data Analysis}, 51(4):2232--2245.

\bibitem[Butler \& Kazakov, 2011]{butler2011effects}
Butler, M. \& Kazakov, D. (2011).
\newblock The effects of variable stationarity in a financial time-series on
  artificial neural networks.
\newblock In {\em 2011 IEEE Symposium on Computational Intelligence for
  Financial Engineering and Economics (CIFEr)}, pages 1--8. IEEE.

\bibitem[Chen et~al., 2018]{chen2018artificial}
Chen, L., Qiao, Z., Wang, M., Wang, C., Du, R., \& Stanley, H.~E. (2018).
\newblock Which artificial intelligence algorithm better predicts the chinese
  stock market?
\newblock {\em IEEE Access}, 6:48625--48633.

\bibitem[Chollet et~al., 2015]{chollet2015keras}
Chollet, F. et~al. (2015).
\newblock Keras.
\newblock \url{https://github.com/fchollet/keras}.

\bibitem[Christensen et~al., 2014]{Christensen2014}
Christensen, K., Oomen, R.~C., \& Podolskij, M. (2014).
\newblock Fact or friction: Jumps at ultra high frequency.
\newblock {\em Journal of Financial Economics}, 114(3):576 -- 599.
\newblock Available from:
  \url{http://www.sciencedirect.com/science/article/pii/S0304405X14001548}.

\bibitem[Chung \& Chuwonganant, 2018]{chung2018market}
Chung, K.~H. \& Chuwonganant, C. (2018).
\newblock Market volatility and stock returns: The role of liquidity providers.
\newblock {\em Journal of Financial Markets}, 37:17--34.

\bibitem[Dacorogna et~al., 2001]{Dacorogna2001}
Dacorogna, M.~M., Gençay, R., Müller, U.~A., Olsen, R.~B., \& Pictet, O.~V.
  (2001).
\newblock 4 - adaptive data cleaning.
\newblock In Dacorogna, M.~M., Gençay, R., Müller, U.~A., Olsen, R.~B., \&
  Pictet, O.~V., editors, {\em An Introduction to High-Frequency Finance},
  pages 82 -- 120. Academic Press, San Diego.
\newblock Available from:
  \url{http://www.sciencedirect.com/science/article/pii/B9780122796715500071}.

\bibitem[Dash \& Dash, 2016]{dash2016hybrid}
Dash, R. \& Dash, P.~K. (2016).
\newblock A hybrid stock trading framework integrating technical analysis with
  machine learning techniques.
\newblock {\em The Journal of Finance and Data Science}, 2(1):42--57.

\bibitem[Dixon, 2018]{dixon2018sequence}
Dixon, M. (2018).
\newblock Sequence classification of the limit order book using recurrent
  neural networks.
\newblock {\em Journal of computational science}, 24:277--286.

\bibitem[Doering et~al., 2017]{doering2017convolutional}
Doering, J., Fairbank, M., \& Markose, S. (2017).
\newblock Convolutional neural networks applied to high-frequency market
  microstructure forecasting.
\newblock In {\em Computer Science and Electronic Engineering (CEEC), 2017},
  pages 31--36. IEEE.

\bibitem[Dozat, 2016]{dozat2016incorporating}
Dozat, T. (2016).
\newblock Incorporating nesterov momentum into adam.

\bibitem[G{\"o}{\c{c}}ken et~al., 2016]{goccken2016integrating}
G{\"o}{\c{c}}ken, M., {\"O}z{\c{c}}al{\i}c{\i}, M., Boru, A., \&
  Dosdo{\u{g}}ru, A.~T. (2016).
\newblock Integrating metaheuristics and artificial neural networks for
  improved stock price prediction.
\newblock {\em Expert Systems with Applications}, 44:320--331.

\bibitem[Goodfellow et~al., 2016]{Goodfellow-et-al-2016}
Goodfellow, I., Bengio, Y., \& Courville, A. (2016).
\newblock {\em Deep Learning}.
\newblock MIT Press.
\newblock \url{http://www.deeplearningbook.org}.

\bibitem[Gudelek et~al., 2017]{gudelek2017deep}
Gudelek, M.~U., Boluk, S.~A., \& Ozbayoglu, A.~M. (2017).
\newblock A deep learning based stock trading model with 2-d cnn trend
  detection.
\newblock In {\em Computational Intelligence (SSCI), 2017 IEEE Symposium Series
  on}, pages 1--8.

\bibitem[Guo, 2004]{guo2004limited}
Guo, H. (2004).
\newblock Limited stock market participation and asset prices in a dynamic
  economy.
\newblock {\em Journal of Financial and Quantitative Analysis}, 39(3):495--516.

\bibitem[Han et~al., 2015]{han2015machine}
Han, J., Hong, J., Sutardja, N., \& Wong, S.~F. (2015).
\newblock Machine learning techniques for price change forecast using the limit
  order book data.
\newblock Technical report, Working Paper.

\bibitem[Hochreiter \& Schmidhuber, 1997]{hochreiter1997long}
Hochreiter, S. \& Schmidhuber, J. (1997).
\newblock Long short-term memory.
\newblock {\em Neural computation}, 9(8):1735--1780.

\bibitem[Jacod et~al., 2009]{Jacod2009}
Jacod, J., Li, Y., Mykland, P.~A., Podolskij, M., \& Vetter, M. (2009).
\newblock Microstructure noise in the continuous case: the pre-averaging
  approach.
\newblock {\em Stochastic Processes and their Applications}, 119(7):2249--2276.

\bibitem[Kanagal et~al., 2017]{kanagal2017market}
Kanagal, K., Wu, Y., \& Chen, K. (2017).
\newblock Market making with machine learning methods.
\newblock {\em Available online here https://web. stanford.
  edu/class/msande448/2017/Final/Reports/gr4. pdf}.

\bibitem[Kercheval \& Zhang, 2015]{doi:10.1080/14697688.2015.1032546}
Kercheval, A.~N. \& Zhang, Y. (2015).
\newblock Modelling high-frequency limit order book dynamics with support
  vector machines.
\newblock {\em Quantitative Finance}, 15(8):1315--1329.
\newblock Available from: \url{https://doi.org/10.1080/14697688.2015.1032546}.

\bibitem[Kim et~al., 2004]{kim2004artificial}
Kim, T.~Y., Oh, K.~J., Kim, C., \& Do, J.~D. (2004).
\newblock Artificial neural networks for non-stationary time series.
\newblock {\em Neurocomputing}, 61:439--447.

\bibitem[Lahmiri, 2014]{lahmiri2014wavelet}
Lahmiri, S. (2014).
\newblock Wavelet low-and high-frequency components as features for predicting
  stock prices with backpropagation neural networks.
\newblock {\em Journal of King Saud University-Computer and Information
  Sciences}, 26(2):218--227.

\bibitem[Lettau \& Ludvigson, 2001]{lettau2001consumption}
Lettau, M. \& Ludvigson, S. (2001).
\newblock Consumption, aggregate wealth, and expected stock returns.
\newblock {\em the Journal of Finance}, 56(3):815--849.

\bibitem[Liou et~al., 2014]{liou2014autoencoder}
Liou, C.-Y., Cheng, W.-C., Liou, J.-W., \& Liou, D.-R. (2014).
\newblock Autoencoder for words.
\newblock {\em Neurocomputing}, 139:84--96.

\bibitem[Liou et~al., 2008]{liou2008modeling}
Liou, C.-Y., Huang, J.-C., \& Yang, W.-C. (2008).
\newblock Modeling word perception using the elman network.
\newblock {\em Neurocomputing}, 71(16-18):3150--3157.

\bibitem[M{\"a}kinen et~al., 2018]{makinen2018predicting}
M{\"a}kinen, M., Iosifidis, A., Gabbouj, M., \& Kanniainen, J. (2018).
\newblock Predicting jump arrivals in stock prices using neural networks with
  limit order book data.

\bibitem[Minh et~al., 2018]{minh2018deep}
Minh, D.~L., Sadeghi-Niaraki, A., Huy, H.~D., Min, K., \& Moon, H. (2018).
\newblock Deep learning approach for short-term stock trends prediction based
  on two-stream gated recurrent unit network.
\newblock {\em IEEE Access}, 6:55392--55404.

\bibitem[Nousi et~al., 2018]{nousi2018machine}
Nousi, P., Tsantekidis, A., Passalis, N., Ntakaris, A., Kanniainen, J., Tefas,
  A., Gabbouj, M., \& Iosifidis, A. (2018).
\newblock Machine learning for forecasting mid price movement using limit order
  book data.
\newblock {\em arXiv preprint arXiv:1809.07861}.

\bibitem[Ntakaris et~al., 2018a]{ntakaris2018mid}
Ntakaris, A., Kanniainen, J., Gabbouj, M., \& Iosifidis, A. (2018a).
\newblock Mid-price prediction based on machine learning methods with technical
  and quantitative indicators.

\bibitem[Ntakaris et~al., 2018b]{ntakaris2018benchmark}
Ntakaris, A., Magris, M., Kanniainen, J., Gabbouj, M., \& Iosifidis, A.
  (2018b).
\newblock Benchmark dataset for mid-price forecasting of limit order book data
  with machine learning methods.
\newblock {\em Journal of Forecasting}.

\bibitem[Oomen, 2006]{Oomen2006}
Oomen, R. C.~A. (2006).
\newblock Properties of realized variance under alternative sampling schemes.
\newblock {\em Journal of Business \& Economic Statistics}, 24(2):219--237.
\newblock Available from: \url{http://www.jstor.org/stable/27638871}.

\bibitem[Passalis et~al., 2017]{passalis2017time}
Passalis, N., Tsantekidis, A., Tefas, A., Kanniainen, J., Gabbouj, M., \&
  Iosifidis, A. (2017).
\newblock Time-series classification using neural bag-of-features.
\newblock In {\em Signal Processing Conference (EUSIPCO), 2017 25th European},
  pages 301--305.

\bibitem[Qi \& Zhang, 2008]{qi2008trend}
Qi, M. \& Zhang, G.~P. (2008).
\newblock Trend time--series modeling and forecasting with neural networks.
\newblock {\em IEEE Transactions on neural networks}, 19(5):808--816.

\bibitem[Qian, 2017]{qian2017financial}
Qian, X.-Y. (2017).
\newblock Financial series prediction: Comparison between precision of time
  series models and machine learning methods.
\newblock {\em arXiv preprint arXiv:1706.00948}.

\bibitem[Sezer et~al., 2017]{sezer2017artificial}
Sezer, O.~B., Ozbayoglu, A.~M., \& Dogdu, E. (2017).
\newblock An artificial neural network-based stock trading system using
  technical analysis and big data framework.
\newblock In {\em Proceedings of the SouthEast Conference}, pages 223--226.
  ACM.

\bibitem[Siami-Namini \& Namin, 2018]{siami2018forecasting}
Siami-Namini, S. \& Namin, A.~S. (2018).
\newblock Forecasting economics and financial time series: Arima vs. lstm.
\newblock {\em arXiv preprint arXiv:1803.06386}.

\bibitem[Sirignano, 2016]{sirignano2016deep}
Sirignano, J. (2016).
\newblock Deep learning for limit order books.

\bibitem[Sirignano \& Cont, 2018]{sirignano2018universal}
Sirignano, J. \& Cont, R. (2018).
\newblock Universal features of price formation in financial markets:
  perspectives from deep learning.

\bibitem[Tran et~al., 2018]{tran2018temporal}
Tran, D.~T., Iosifidis, A., Kanniainen, J., \& Gabbouj, M. (2018).
\newblock Temporal attention-augmented bilinear network for financial
  time-series data analysis.
\newblock {\em IEEE transactions on neural networks and learning systems}.

\bibitem[Tran et~al., 2017]{tran2017tensor}
Tran, D.~T., Magris, M., Kanniainen, J., Gabbouj, M., \& Iosifidis, A. (2017).
\newblock Tensor representation in high-frequency financial data for price
  change prediction.
\newblock In {\em Computational Intelligence (SSCI), 2017 IEEE Symposium Series
  on}, pages 1--7. IEEE.

\bibitem[Tsantekidis et~al., 2017a]{tsantekidis2017forecasting}
Tsantekidis, A., Passalis, N., Tefas, A., Kanniainen, J., Gabbouj, M., \&
  Iosifidis, A. (2017a).
\newblock Forecasting stock prices from the limit order book using
  convolutional neural networks.
\newblock In {\em Business Informatics (CBI), 2017 IEEE 19th Conference on},
  volume~1, pages 7--12.

\bibitem[Tsantekidis et~al., 2017b]{tsantekidis2017using}
Tsantekidis, A., Passalis, N., Tefas, A., Kanniainen, J., Gabbouj, M., \&
  Iosifidis, A. (2017b).
\newblock Using deep learning to detect price change indications in financial
  markets.
\newblock In {\em Signal Processing Conference (EUSIPCO), 2017 25th European},
  pages 2511--2515.

\bibitem[Tsantekidis et~al., 2018]{tsantekidis2018using}
Tsantekidis, A., Passalis, N., Tefas, A., Kanniainen, J., Gabbouj, M., \&
  Iosifidis, A. (2018).
\newblock Using deep learning for price prediction by exploiting stationary
  limit order book features.
\newblock {\em arXiv preprint arXiv:1810.09965}.

\bibitem[Velay \& Daniel, 2018]{velay2018stock}
Velay, M. \& Daniel, F. (2018).
\newblock Stock chart pattern recognition with deep learning.
\newblock {\em arXiv preprint arXiv:1808.00418}.

\bibitem[Wang, 2011]{wang2011pricing}
Wang, P. (2011).
\newblock Pricing currency options with support vector regression and
  stochastic volatility model with jumps.
\newblock {\em Expert Systems with Applications}, 38(1):1--7.

\bibitem[Zhang et~al., 2005]{Zhang2005}
Zhang, L., Mykland, P.~A., \& A{\"i}t-Sahalia, Y. (2005).
\newblock A tale of two time scales: Determining integrated volatility with
  noisy high-frequency data.
\newblock {\em Journal of the American Statistical Association},
  100(472):1394--1411.

\bibitem[Zhang et~al., 2019]{zhang2019comparison}
Zhang, X., Xue, T., \& Stanley, H.~E. (2019).
\newblock Comparison of econometric models and artificial neural networks
  algorithms for the prediction of baltic dry index.
\newblock {\em IEEE Access}, 7:1647--1657.

\bibitem[Zhang et~al., 2018]{zhang2018deeplob}
Zhang, Z., Zohren, S., \& Roberts, S. (2018).
\newblock Deeplob: Deep convolutional neural networks for limit order books.
\newblock {\em arXiv preprint arXiv:1808.03668}.

\bibitem[Zheng et~al., 2012]{zheng2012price}
Zheng, B., Moulines, E., \& Abergel, F. (2012).
\newblock Price jump prediction in limit order book.
\newblock {\em arXiv preprint arXiv:1204.1381}.

\bibitem[Zhou et~al., 2016]{zhou2016attention}
Zhou, P., Shi, W., Tian, J., Qi, Z., Li, B., Hao, H., \& Xu, B. (2016).
\newblock Attention-based bidirectional long short-term memory networks for
  relation classification.
\newblock In {\em Proceedings of the 54th Annual Meeting of the Association for
  Computational Linguistics (Volume 2: Short Papers)}, volume~2, pages
  207--212.

\bibitem[Zieba et~al., 2016]{zikeba2016ensemble}
Zieba, M., Tomczak, S.~K., \& Tomczak, J.~M. (2016).
\newblock Ensemble boosted trees with synthetic features generation in
  application to bankruptcy prediction.
\newblock {\em Expert Systems with Applications}, 58:93--101.

\end{thebibliography}

\end{document}